\documentclass[twocolumn,aps,prx]{revtex4-1}

\usepackage{amsmath}
\usepackage{amsfonts}
\usepackage{amssymb}
\usepackage[colorlinks=true,citecolor=blue,linkcolor=red]{hyperref}
\usepackage{graphicx}
\usepackage{bbold}					
\usepackage[dvipsnames]{xcolor}

\usepackage{array,booktabs,ragged2e}	
\usepackage{soul}						
\usepackage{cancel}						
\usepackage{tabularx}					
\usepackage{multirow}
\usepackage{hhline}

\usepackage{pdfpages}

\usepackage[utf8]{inputenc}
\usepackage[T1]{fontenc}

\def\bs#1{\boldsymbol{#1}}						
\def\e#1{\mathrm{e}^{#1}}           			
\def\imi{\mathrm{i}}							

\newcolumntype{P}[1]{>{\raggedleft\arraybackslash}p{#1}}
\newcolumntype{R}[1]{>{\centering\arraybackslash}p{#1}}

\begin{document}


\title{Nodal-chain metals}

\author{Tom\'{a}\v{s} Bzdu\v{s}ek$^1$}
\author{QuanSheng Wu$^{1,2}$}
\author{Andreas R\"{u}egg$^{1}$}
\author{Manfred Sigrist$^{1}$}
\author{Alexey A. Soluyanov$^{1,2,3}$}

\affiliation{$^{1}$Institut f\"{u}r Theoretische Physik, ETH Zurich, 8093 Zurich, Switzerland}
\affiliation{${^{2}}$Station Q Zurich, ETH Zurich, 8093 Zurich, Switzerland}
\affiliation{${^{3}}$Department of Physics, St. Petersburg State University, St. Petersburg, 199034 Russia}
\date{\today}

\begin{abstract}
The band theory of solids is arguably the most successful theory of condensed matter physics, providing the description of the electronic energy levels in a variety of materials. Electronic wavefunctions obtained from the band theory allow for a topological characterization of the system and the electronic spectrum may host robust, topologically protected fermionic quasiparticles. 
Many of these quasiparticles are analogs of the elementary particles of the Standard Model, but others do not have a counterpart in relativistic high-energy theories. A full list of possible quasiparticles in solids is still unknown, even in the non-interacting case.
Here, we report on a new type of fermionic excitation that appears in metals. This excitation forms a nodal chain -- a chain of connected loops in momentum space -- along which conduction and valence band touch. We prove that the nodal chain is topologically distinct from any other excitation reported before. We discuss the symmetry requirements for the appearance of this novel excitation and predict that it is realized in an existing material IrF$_4$, as well as in other compounds of this material class. Using IrF$_4$ as an example, we provide a detailed discussion of the topological surface states associated with the nodal chain. Furthermore, we argue that the presence of the novel quasiparticles results in anomalous magnetotransport properties, distinct from those of the known materials.  

\end{abstract}
\maketitle


In metallic band structures valence and conduction bands overlap, and a degeneracy can occur between them at points, lines or planes in the Brillouin zone (BZ). In some cases, the degeneracies are stable against perturbations because their existence is protected by a topological invariant. If such degeneracies occur close to the Fermi level, also the low energy excitations of the metal are topologically protected~\cite{Volovik:2003}. This is the case in Weyl (Dirac) metals~\cite{Wang:2012, Wang:2013, Neupane:2014, Liu:2014a,Liu:2014b, Liang:2015, Weng:2015, Huang:2015a, Xu:2015a, Lv:2015, Wehling:2014}, where a topologically protected degeneracy of two (four) bands occurs at isolated points in the BZ~\cite{Murakami:2007,Wan:2011,Young:2012,Yang:2014,Liu:2014c, Matsuura:2013,Bzdusek:2015}. Remarkably, these materials exhibit the chiral anomaly~\cite{Nielsen:1983,Son:2013,Huang:2015b, Xiong:2015, Zhang:2015} and topological surface Fermi arcs~\cite{Wan:2011,Xu:2015,Xu:2015a,Lv:2015,Inoue:2016}, while their low energy quasiparticle excitations correspond to Weyl (Dirac) fermions of the quantum field theory.
\begin{figure}[b]
  \includegraphics[width=0.48\textwidth]{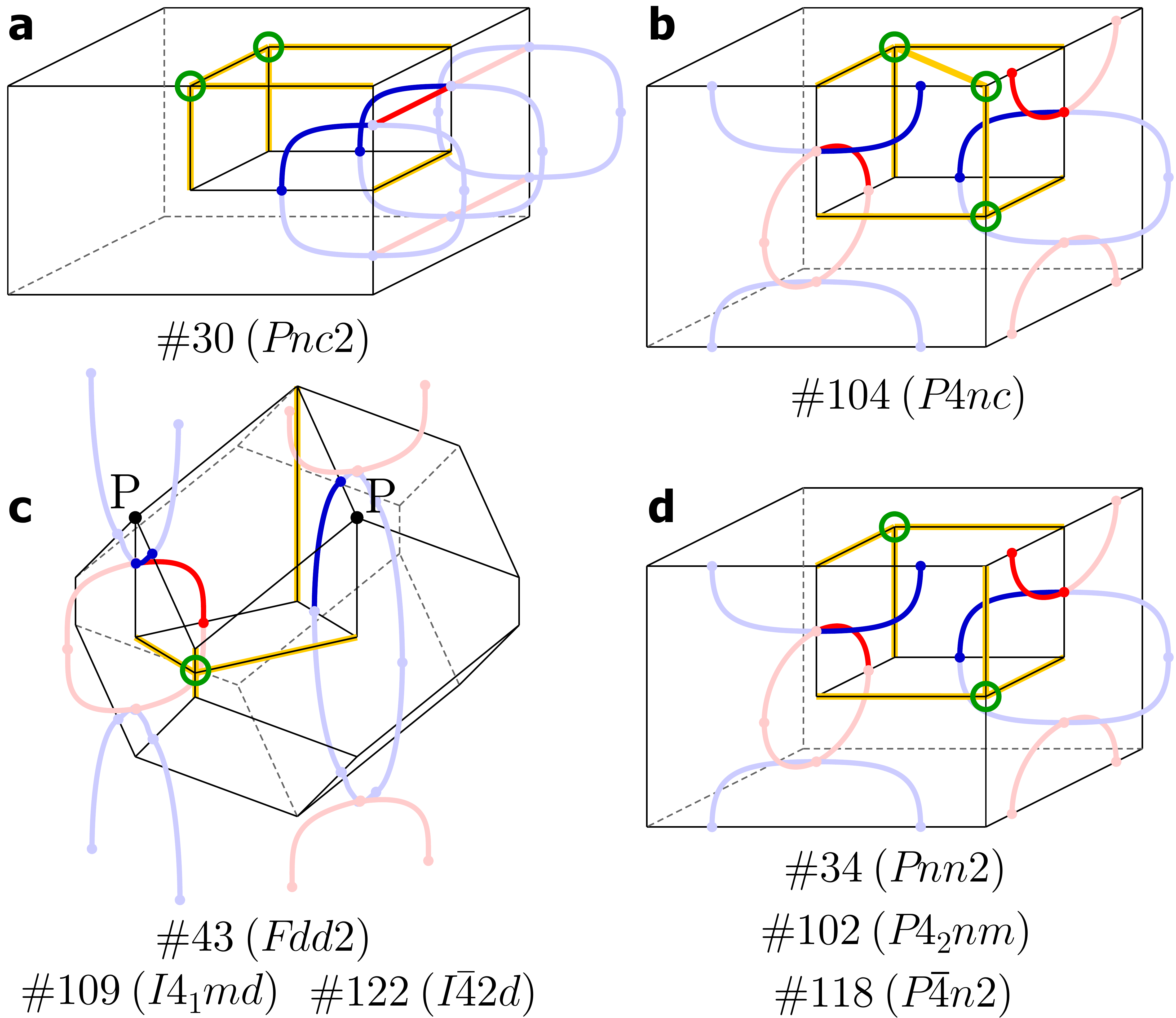}  
  \caption{{\small \textsf{\textbf{Catalog of nodal chain metals.} A nodal chain appears in metals with these space groups whenever there are $4n+2$ electrons per primitive unit cell. The blue and red lines show the nodal lines (NLs) located in mutually orthogonal planes. The additional double Weyl points are marked with green circles. The high-symmetry lines supporting a two-fold degeneracy of valence (conduction) bands are highlighted in orange. In space groups 109 and 122 the NLs touching point is at the point P.}}}
  \label{fig:main1}
\end{figure}

Topological metals can also host elementary excitations that do not have a direct analog in relativistic field theories~\cite{Soluyanov:2015,Volovik:2014,Burkov:2011,Yu:2015, Bian:2015a,Bian:2016,Chen:2015,Wieder:2015,Fang:2015b,Wang:2016,Chang:2016,Liang:2016,Bradlyn:2016}. For example, in Weyl nodal line (NL) metals~\cite{Burkov:2011,Yu:2015,Bian:2015a,Bian:2016}, two bands are degenerate along a line in the BZ. In the presence of spin-orbit coupling, this NL, referred to as an accidental NL (ANL) in the following, occurs on mirror-symmetric planes and is formed by crossing bands of opposite mirror-symmetry eigenvalue. The ANL materials are predicted to host flat topological drumhead states~\cite{Yu:2015,Weng:2015b}, which were argued to provide a route to higher-temperature superconductivity~\cite{Heikkila:2011}. Given the variety of striking physical phenomena emerging from topological excitations in materials, the identification of novel topological phases and their material realizations is of fundamental importance.

In this work, we describe a new topological fermionic excitation -- a nodal chain -- and predict that an existing material IrF$_4$ and a family of related compounds host this novel state of matter. A nodal chain, illustrated in Fig.~\ref{fig:main1}, consists of NLs of a yet not described kind, whose properties are distinct from those of ANLs. The appearance of these NLs is enforced in the vicinity of the Fermi level of certain non-centrosymmetric materials provided their crystal structure has a non-symmorphic glide plane symmetry $g=\{\sigma|\bs{t}\}$ formed by a reflection $\sigma$, followed by a translation by a fraction of a primitive lattice vector ${\bs t}$. For several space groups (see Fig.~\ref{fig:main1}) such non-symmorphic NLs (NSNLs) appear on distinct high-symmetry planes, which are orthogonal to each other. The NSNLs located in different planes touch each other at points on a high-symmetry axis, forming a chain of double degeneracy that goes across the entire BZ. 

To set the stage for the discussion of nodal chains, we first describe their building blocks, NSNLs. For spin-orbit coupled systems $g^2=-\e{-\imi 2\bs{k}\cdot\bs{t}_\parallel}$, where $\bs{k}$ is the electron momentum and $\bs{t}_\parallel$ is the in-plane component of $\bs{t}$. This means the possible eigenvalues of $g$ are $\eta_\pm(\bs{k}) = \pm\imi \e{-\imi \bs{k}\cdot\bs{t}_\parallel}$, becoming $\bs{k}$-dependent whenever $\bs{t}_\parallel\neq 0$~\cite{Young:2015}. 

\begin{figure}[b]
  \includegraphics[width=0.48\textwidth]{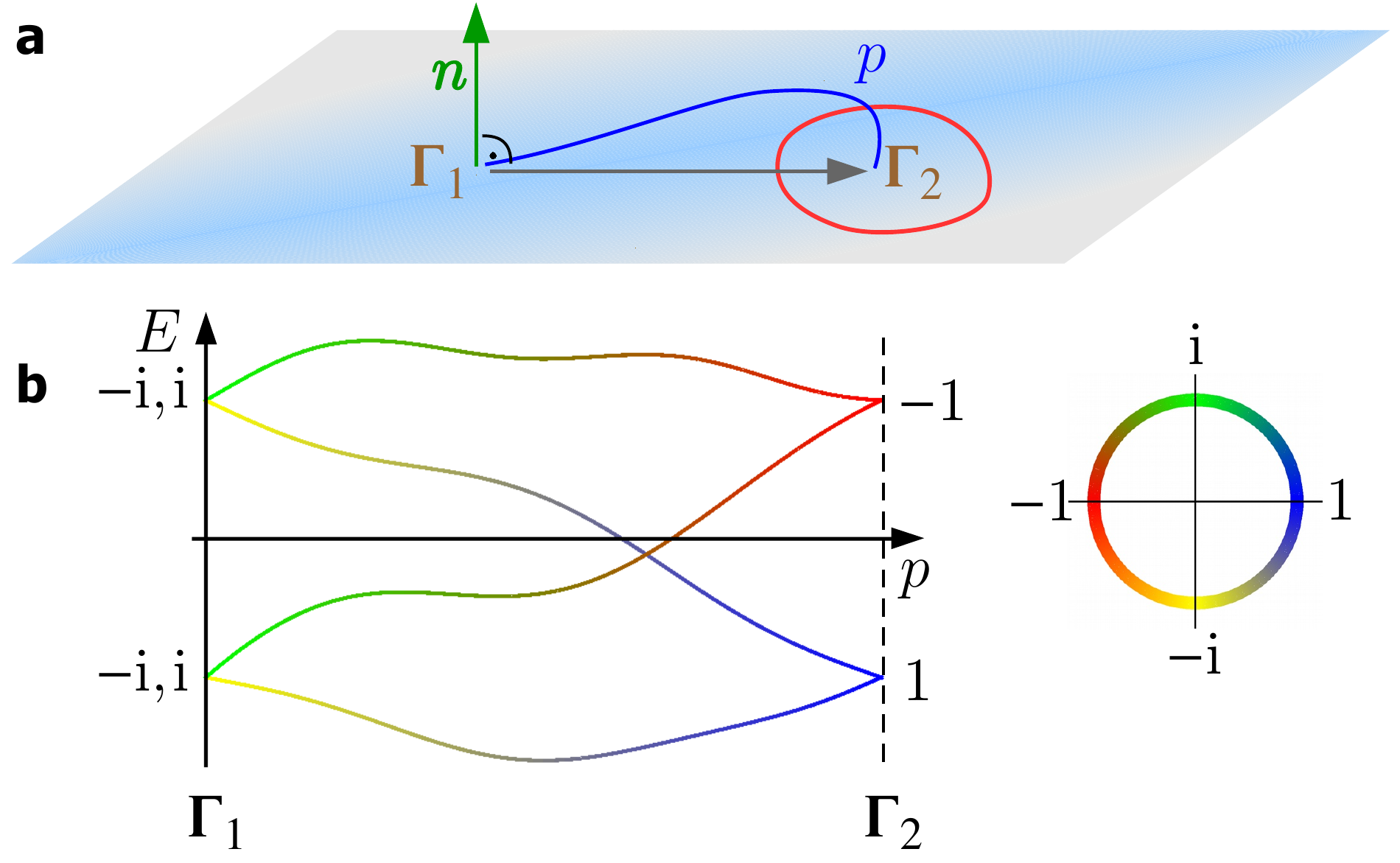}  
  \caption{{\small \textsf{\textbf{Non-symmorphic nodal line.} (a) Any path $p$ connecting a pair of time-reversal invariant momenta $\bs{\Gamma}_{1,2}$ in a glide-invariant plane (blue) fulfilling the criterion of equation~(\ref{eqn:NSNLcriterion}) has to have a gap closing point, which belongs to a non-symmorphic nodal line (red). (b) Two Kramer's pairs shown along any path $p$ connecting $\bs{\Gamma}_{1,2}$. The evolution of the glide eigenvalues along the path is shown in color for all bands.}}}
  \label{fig:1}
\end{figure}

If in addition to a glide plane $g$ time-reversal symmetry $\Theta$ is present in the system [see Supplementary Material (SM) for the generalization of the argument to antiferromagnetic systems], the relation $\bs{\Gamma}_i\cdot\bs{t}_\parallel = 0$ (mod $\frac{\pi}{2}$) holds for any of the four in-plane time-reversal invariant momenta (TRIMs) $\bs{\Gamma}_i$, defined as $\bs{\Gamma}_i=-\bs{\Gamma}_i+{\bs G}$ with ${\bs G}$ a reciprocal lattice vector (see SM). This makes it possible for the two TRIMs $\bs{\Gamma}_{1,2}$ to satisfy the relation
\begin{equation}
\left(\bs{\Gamma}_1-\bs{\Gamma}_2\right)\cdot \bs{t}_\parallel = \frac{\pi}{2}\mod \pi,\label{eqn:NSNLcriterion}
\end{equation}
so that the glide eigenvalues $\eta_\pm(\bs{k})$ are $\pm\imi$ at $\bs{k}=\bs{\Gamma}_1$ and $\pm 1$ at $\bs{k}=\bs{\Gamma}_2$. Hence, along any in-plane path $p$ that connects $\bs{\Gamma}_1$ to $\bs{\Gamma}_2$, the glide eigenvalues $\eta_\pm(\bs{k})$ must smoothly evolve from $(+\imi,-\imi)$ to $(+1,-1)$, as illustrated in Fig.~\ref{fig:1}a. But in $\Theta$-symmetric systems the bands form Kramers pairs, which are degenerate at TRIMs and carry complex conjugate eigenvalues. Since at $\bs{\Gamma}_2$ the eigenvalues are \emph{not} complex conjugate anymore, they belong to different Kramers doublets, meaning that there are several Kramers pairs that switch partners along $p$, as shown in Fig.~\ref{fig:1}b. This argument holds for \emph{any} in-plane path $p$, and thus there exists a nodal line (the NSNL) separating the two TRIMs, shown as a red loop in Fig.~\ref{fig:1}a. 

A simple electron counting suggests that the NSNL is formed between valence and conduction bands whenever there are
\begin{equation}
\nu_\textrm{filled} = 4n+2, \indent n\in \mathbb{N},
\label{eqn:FillingCondition}
\end{equation}
filled bands, irrespective of all further material details. This is one of the differences between NSNLs and ANLs~\cite{Bian:2015a,Bian:2016}, which are accidental and not symmetry-enforced (see SM for material examples of NSNLs formed by valence or conduction bands).

The topological protection of NSNLs is similar to that of ANLs~\cite{Gosablez:2015}. The corresponding topological invariant is defined on a gapped loop that is symmetric under the glide operation. The Berry phase accumulated by the valence bands on such a loop is equal to $\pi$ times the number of NLs enclosed by the loop. 

Due to the periodicity of $\bs{k}$-space, a particular example of such a loop is a one-dimensional straight line threading across the BZ, orthogonal to the glide plane. For 1D insulators with spatial symmetries quantizing the Berry phase to either $0$ or $\pi$, it is known that in the latter case a metallic end mode of half an electron charge has to exist~\cite{King-Smith:1993,Hughes:2011}. Considering a 3D crystal as a collection of coupled 1D chains, one arrives at the conclusion that the projection of the NSNL on a crystal surface has to enclose a topological metallic surface state, which is similar to the drumhead state of ANLs~\cite{Weng:2015b}. 
\begin{figure*}[t!]
  \includegraphics[width=0.99\textwidth]{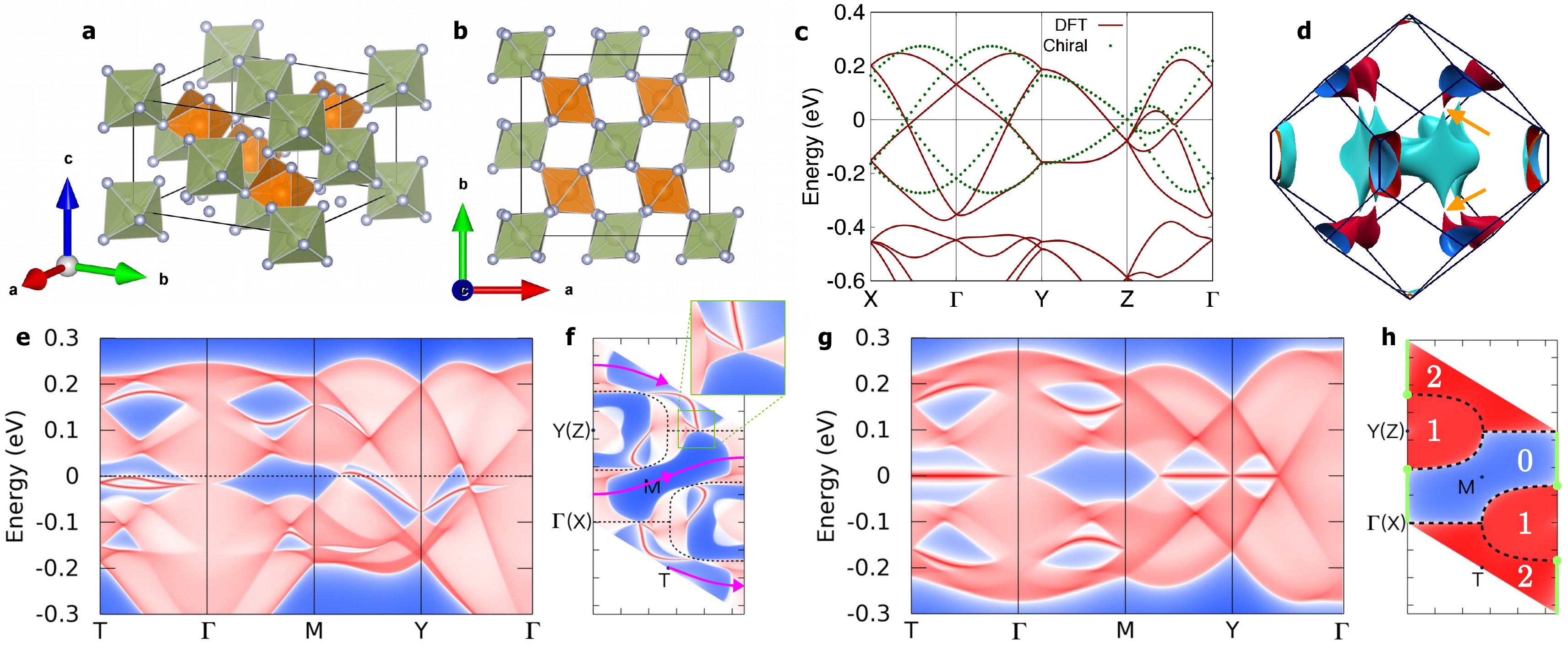}  
  \caption{{\small \textsf{\textbf{Iridium tetrafluoride and its band structure.} (a) The crystal structure of IrF$_4$: Corner-sharing octahedra of F enclose Ir atoms. The color indicates the two sublattices related by an approximate chiral symmetry (CHS). (b) The same structure viewed along the $[001]$ direction. (c) Band structure of paramagnetic IrF$_4$: First-principles (solid red) and tight-binding with CHS (dotted green) bands are shown. (d) Fermi surface of IrF$_4$ consists of electron pockets (cyan) and hole pockets (red) touching (orange arrows) along the nodal rings. (e) The $(100)$ surface density of states shown along the high-symmetry lines of the surface BZ. Topological surface states are clearly visible. (f) The cut of the density of states in (e) at the Fermi energy plotted in the $(100)$ surface BZ. End points of the surface Fermi arcs coincide with the projections of the bulk kissing points of the electron and hole pockets. The dashed black line is the projection of the nodal chain into the surface BZ. The magenta line is the projection of a plane used to calculate the bulk $\mathbb{Z}_2$ invariant. (g,h) are the analogs of (e,f) for a tight-binding model endowed with CHS. The numbers in (h) indicate the number of topological surface bands in that region, and the green lines correspond to the additional nodal loop imposed by CHS projection into the surface BZ.}}}
  \label{fig:3}
\end{figure*}

Despite this similarity, we argue that low energy excitations produced by ANLs and NSNLs are very different. Unlike ANLs, NSNLs in $\Theta$-symmetric non-centrosymmetric systems always enclose a TRIM, and thus a single nodal loop contains a time-reversed image of each Bloch state in addition to the state itself. In fact, if inversion symmetry breaking terms are smoothly tuned to zero in a NSNL Hamiltonian, the NSNL shrinks into a Dirac point~\cite{Young:2012,Yang:2014} (see SM). This leads to immediate consequences in transport.   

In particular, as outlined in SM, application of a magnetic field in the direction orthogonal to the NSNL results in field-driven topological phase transitions. We find that the conduction and valence band Landau levels (LLs) \emph{touch} at certain values $B_{\mathrm c}$ of the magnetic field. Since application of a magnetic field orthogonal to the glide plane preserves the glide-plane, each LL has a well defined glide eigenvalue. At the touchings, the two LLs exchange glide eigenvalues, thus realizing a topological phase transition. Since momentum $k_B$ along the field direction is still a good quantum number and the glide plane quantizes the Berry phase of the effectively 1D system to either $0$ or $\pi$, the field-driven band touchings 
change the Berry phase of the system by $\pi$. This results in pumping a charge of $e/2$ per area covered by a magnetic flux quantum to the surface of the sample parallel to the plane of the NSNL. Hence, a step change in the Hall response of the metallic surface state is expected at magnetic field values $B_{\mathrm c}$. 

The response of the NSNLs to the mirror symmetry breaking in-plane magnetic field is also distinct from that of the ANLs. While for the latter the Landau spectrum is gapped, for NSNLs it is always gapless~\cite{Rhim:2015}. The crossing of the two LLs is protected by the the product symmetry $\Theta \circ g$ that survives the application of the in-plane field. The gapless structure of the LLs suggests unusual transport properties for materials hosting NSNLs when an electric field is aligned with the in-plane magnetic field, similar to case of the chiral anomaly in Weyl and Dirac semimetals~\cite{Son:2013,Huang:2015b, Xiong:2015}.

Having established the NSNLs, we are now in a position to tackle systems with \emph{two} glide planes. Such systems can accommodate \emph{nodal chains} formed by a pair of touching NSNLs located in mutually orthogonal planes, while the bands at the touching point are still only doubly degenerate. The nodal chain continues indefinitely in the extended $\bs{k}$-space.

The critera for the occurrence of a nodal chain are: (1) The system has to be symmetric under \emph{two} inequivalent glide planes $g_{1,2}=\left\{\sigma_{1,2}|\bs{t}_{1,2}\right\}$ such that the criterion of equation~(\ref{eqn:NSNLcriterion}) is fulfilled for the two TRIMs $\bs{\Gamma}_{1,2}$, located on the intersection of the two glide-invariant planes, for both translation vectors $\bs{t}_{1,2}$, and (2) the two bands forming the chain belong to two-dimensional representations at $\bs{\Gamma}_{1,2}$, which split into one-dimensional ones on the high-symmetry line connecting $\bs{\Gamma}_1$ and  $\bs{\Gamma}_2$.

Out of the $230$ space groups~\cite{Bradley:1972,Aroyo:2006}, the ones satisfying the above criteria for two \emph{mutually orthogonal} glide planes are listed in the catalog of Fig.~\ref{fig:main1}. 
The space group $\# 110$ ($I4_1cd$) is discussed separately in SM. 
All the listed space groups must contain a nodal chain between the valence and conduction bands whenever the condition of equation~(\ref{eqn:FillingCondition}) is satisfied. The bands forming the nodal chain are only doubly degenerate along the whole chain, including the touching points of the NSNLs. In all the cases shown in the catalog we find that at least one additional point of 4-fold degeneracy, formed by two Weyl points of opposite chirality, is present at a high-symmetry point on the BZ boundary~\cite{Gao:2015}.

A nodal chain represents a novel topological excitation, distinct from a collection of NSNLs. To see this, first note that the two NSNLs that form a nodal chain cannot be separated. Indeed, the argument provided above for the appearance of the NSNL guarantees that there must be an odd number of band crossings along the high-symmetry line connecting $\bs{\Gamma}_1$ and $\bs{\Gamma}_2$. 

While a detailed study will be reported elsewhere~\cite{Bzdusek:2016}, the non-trivial transport properties of the nodal chain can be inferred from the above analysis of NSNLs in magnetic fields, suggesting several distinct scenarios for the LL spectrum. Here we proceed with the analysis of the topological surface states of nodal chains that we illustrate using a particular real material example.

We found the nodal chain state in iridium tetrafluoride (IrF$_4$). The orthorhombic crystal structure of this compound belongs to space group $\# 43$ ($Fdd2$). The primitive unit cell contains two formula units~\cite{Rao:1976} so that the number of electrons satisfies the constraint of equation~(\ref{eqn:FillingCondition}). Each Ir site is surrounded by an octahedron of six F atoms, four of which are shared between the neighboring octahedra. The octahedra form a bipartite lattice as shown in Fig.~\ref{fig:3}a-b (see SM for a more detailed description of the crystal structure of IrF$_4$). The space group contains two mutually orthogonal glide planes: $g_1$ ($g_2$) formed by a reflection about the $(100)$ ($(010)$) plane, followed by a translation of $(1/4,1/4,1/4)$ in the reduced coordinates. 

Possible antiferromagnetic ordering with $T_\textrm{N}\lesssim 100$~K was reported for IrF$_4$ in magnetic susceptibility measurements~\cite{Rao:1976}. A paramagnetic phase is expected to occur at  temperatures above $T_\textrm{N}$, still accessible for an angle resolved photoemission spectroscopy (ARPES). We first discuss the paramagnetic phase, in which the crystal symmetries and  band filling guarantee the presence of a nodal chain corresponding to Fig.~\ref{fig:main1}c. 

\begin{figure}[t]
  \includegraphics[width=0.48\textwidth]{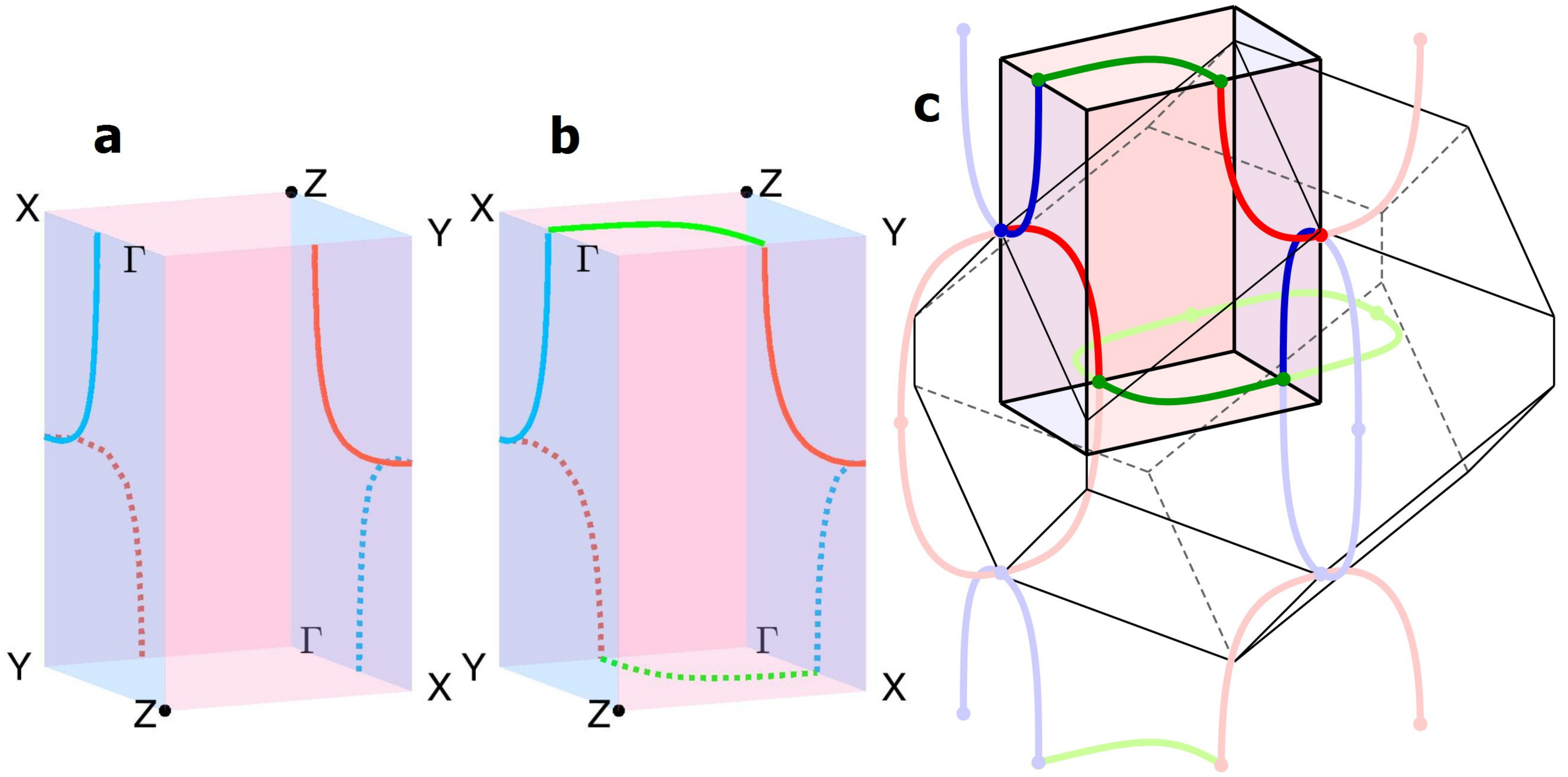}  
  \caption{{\small \textsf{\textbf{Nodal chain and nodal net of IrF$_4$.} (a) Nodal chain structure of
 IrF$_4$. (b) Nodal net in the chiral-symmetric model of IrF$_4$. (c) Nodal net in the extended $\bs{k}$-space. Different colors correspond to different orientations of the nodal lines.}}}
  \label{fig:4}
\end{figure}

To study paramagnetic IrF$_4$ we performed first-principle calculations as detailed in SM. The obtained band structure is shown in Fig.~\ref{fig:3}c. We indeed find a nodal chain, plotted in Fig.~\ref{fig:4}a, consisting of two NSNLs in the $(100)$ and $(010)$ planes. Both NSNLs cross the chemical potential four times resulting in topologically protected touching points between electron and hole pockets shown with arrows in Fig.~\ref{fig:3}d. Similar touchings of carrier pockets, although of different topological origin, were predicted for type-II Weyl semimetals~\cite{Soluyanov:2015} and ANLs~\cite{Burkov:2011,Weng:2015b,Heikkila:2015}. These Fermi surface touching points can be probed with the soft X-ray ARPES~\cite{Strocov:2010}, and were recently argued to be important for potential higher temperature superconducting phases~\cite{Yudin:2014}.  

The nodal chain produces non-trivial topological surface states on the $(100)$ surface of IrF$_4$ as shown in Fig.~\ref{fig:3}e-f. On this surface, the projection of $(010)$ ($(100)$) NSNL is a line (oval) shown with dashed lines in Fig.~\ref{fig:3}f. Fermi arcs arise from the touching points of the Fermi pockets. For the projection of the $(100)$ NSNL (also shown as region 1 in Fig.~\ref{fig:3}h) a single such arc produced by the drumhead state emerges from the touching point. The touching points that appear on a linear projection of the $(010)$ NSNL, however, produce two Fermi arcs, consistent with the fact that there are two such Fermi pocket touchings that project onto the same point in the surface BZ.

The arcs originating on different NSNLs are connected either directly or through a carrier pocket. 
Moreover, the $\mathbb{Z}_2$ invariant computed along the gapped time-reversal symmetric plane projected onto the magenta line in Fig.~\ref{fig:3}f is non-trivial. Hence, such path corresponds to an edge of a 2D topological insulator, and has to host an odd number of Kramer's pairs of edge states~\cite{Kane:2005}, in accord with the observed connectivity of Fermi arcs -- there is a single Kramer's pair of such edge states (see SM).    

To understand why both Fermi arcs of the $(010)$ NSNL appear on the same side of its projection onto the $(100)$ surface, we need to expose the approximate chiral symmetry (CHS) present in the material. We constructed a tight-binding model for the $j=\tfrac{1}{2}$ orbitals located on the $\textrm{Ir}$ sites representing the two sublattices of the IrF$_4$ structure, and fitted the parameters to reproduce the first-principles results (see SM). We found that the avoided crossings along the $\textrm{Z}-\Gamma$ line in Fig.~\ref{fig:3}c originates from the hoppings within the sublattices. The amplitudes of these hoppings are more than three times smaller than those of the inter-sublattice ones, meaning that in IrF$_4$ there exists a weakly broken CHS, relating the two sublattices of the crystal structure. The CHS can be restored in the model by setting the intra-sublattice hoppings to zero. The corresponding band structure is shown in Fig.~\ref{fig:3}c, and one can see that now the gap along the $\textrm{Z}-\Gamma$ vanishes, and an additional NL appears. It connects to the nodal chain, thus creating a {\it nodal net}, shown in Fig.~\ref{fig:4}. The projection of the additional NL onto the $(100)$ surface is shown in green in Fig.~\ref{fig:3}h. 

Endowed with the CHS, the Hamiltonian allows for an additional topological classification (see SM), which predicts two/one/zero surface modes to exist in regions marked as 2/1/0 in Fig.~\ref{fig:3}h. In the presence of the CHS, all these regions are topologically distinct and separated with NLs. When the CHS is weakly broken in real IrF$_4$,
only the \emph{parity} of the number of surface states remains topologially protected and the additional NL gets gapped. However, since the breaking of the CHS is weak, the location of surface modes in the surface BZ of IrF$_4$ is inherited from the chiral symmetric structure.      

The possible antiferromagnetic ordering in IrF$_4$ below $T_c$ preserves the nodal chain structure if the magnetic moment is aligned with the $[001]$ axis. In fact, the nodal chain survives weak breaking of time-reversal symmetry, but not the breaking of glide planes. If magnetic ordering preserves only one of the two glide planes, the nodal chain state transforms into the NSNL state. 

We looked for other possible nodal chain candidates, which would be nonmagnetic and less prone to correlations than IrF$_4$. Although no exhaustive crystallographic data exist, several reports~\cite{Rao:1976,McCarley:1963,Habermehl:2010} (some with partial crystallographic data) of the stability of XY$_4$ crystals (X=Ir, Ta, Re; Y=F, Cl, Br, I) with octahedra-formed lattices similar to that of IrF$_4$ exist. Assuming these compounds crystallize in the same space group as IrF$_4$, and hence satisfy the criterion of equation~(\ref{eqn:FillingCondition}), we carried out an exhaustive first-principles study (see SM) of each of them. We predict a nodal chain structure for all of them. We find the particular shape of the chain and its position relative to the Fermi level to depend on the lattice constants of the unit cell, suggesting possible fine tuning with uniaxial or hydrostatic strains.  

The prediction of the novel nodal chain state of matter in IrF$_4$ material class opens interesting venues for further study of novel physical properties associated with these compounds. The presence of both strongly and weakly correlated compounds in this family allows one to study the interplay between the nodal chain topology and electron-electron interactions, as well as magnetism. Application of strains that break one of the glide planes in these compounds provides a route for a similar study of the NSNL phase, as well as for experimental probe of the anomalous magnetoelectric response predicted here for NSNLs.

{\it Acknowledgments.} The crystal structures in Fig.~\ref{fig:3} were plotted using VESTA 3~\cite{Momma:2011}. We would like to thank A. Bouhon, C. L. Kane, G. E. Volovik, B. A. Bernevig and R. J. Cava for useful discussions. T.B., A.R., M.S. acknowledge financial support through an ETH research grant and the Swiss National Science Foundation. Q.S.W., A.A.S. acknowledge the support of Microsoft Research and Swiss National Science Foundation through the National Competence Centers in Research MARVEL and QSIT.

{\it Author Contributions.} T.B. had the initial idea, carried out the theoretical analysis, and determined the suitable space groups. Q.W. performed the first-principle studies and determined $\textrm{IrF}_4$ to be a nodal chain metal. T.B. and A.A.S. wrote the manuscript. All authors discussed and commented on the manuscript.


\bibliography{paper}

\begin{thebibliography}{58}%
\makeatletter
\providecommand \@ifxundefined [1]{%
 \@ifx{#1\undefined}
}%
\providecommand \@ifnum [1]{%
 \ifnum #1\expandafter \@firstoftwo
 \else \expandafter \@secondoftwo
 \fi
}%
\providecommand \@ifx [1]{%
 \ifx #1\expandafter \@firstoftwo
 \else \expandafter \@secondoftwo
 \fi
}%
\providecommand \natexlab [1]{#1}%
\providecommand \enquote  [1]{``#1''}%
\providecommand \bibnamefont  [1]{#1}%
\providecommand \bibfnamefont [1]{#1}%
\providecommand \citenamefont [1]{#1}%
\providecommand \href@noop [0]{\@secondoftwo}%
\providecommand \href [0]{\begingroup \@sanitize@url \@href}%
\providecommand \@href[1]{\@@startlink{#1}\@@href}%
\providecommand \@@href[1]{\endgroup#1\@@endlink}%
\providecommand \@sanitize@url [0]{\catcode `\\12\catcode `\$12\catcode
  `\&12\catcode `\#12\catcode `\^12\catcode `\_12\catcode `\%12\relax}%
\providecommand \@@startlink[1]{}%
\providecommand \@@endlink[0]{}%
\providecommand \url  [0]{\begingroup\@sanitize@url \@url }%
\providecommand \@url [1]{\endgroup\@href {#1}{\urlprefix }}%
\providecommand \urlprefix  [0]{URL }%
\providecommand \Eprint [0]{\href }%
\providecommand \doibase [0]{http://dx.doi.org/}%
\providecommand \selectlanguage [0]{\@gobble}%
\providecommand \bibinfo  [0]{\@secondoftwo}%
\providecommand \bibfield  [0]{\@secondoftwo}%
\providecommand \translation [1]{[#1]}%
\providecommand \BibitemOpen [0]{}%
\providecommand \bibitemStop [0]{}%
\providecommand \bibitemNoStop [0]{.\EOS\space}%
\providecommand \EOS [0]{\spacefactor3000\relax}%
\providecommand \BibitemShut  [1]{\csname bibitem#1\endcsname}%
\let\auto@bib@innerbib\@empty
\bibitem [{\citenamefont {Volovik}(2003)}]{Volovik:2003}%
  \BibitemOpen
  \bibfield  {author} {\bibinfo {author} {\bibfnamefont {G.~E.}\ \bibnamefont
  {Volovik}},\ }\href@noop {} {\emph {\bibinfo {title} {The {U}niverse in a
  {H}elium {D}roplet}}}\ (\bibinfo  {publisher} {Oxford University Press},\
  \bibinfo {address} {New York},\ \bibinfo {year} {2003})\BibitemShut {NoStop}%
\bibitem [{\citenamefont {Wang}\ \emph {et~al.}(2012)\citenamefont {Wang},
  \citenamefont {Sun}, \citenamefont {Chen}, \citenamefont {Franchini},
  \citenamefont {Xu}, \citenamefont {Weng}, \citenamefont {Dai},\ and\
  \citenamefont {Fang}}]{Wang:2012}%
  \BibitemOpen
  \bibfield  {author} {\bibinfo {author} {\bibfnamefont {Z.}~\bibnamefont
  {Wang}}, \bibinfo {author} {\bibfnamefont {Y.}~\bibnamefont {Sun}}, \bibinfo
  {author} {\bibfnamefont {X.-Q.}\ \bibnamefont {Chen}}, \bibinfo {author}
  {\bibfnamefont {C.}~\bibnamefont {Franchini}}, \bibinfo {author}
  {\bibfnamefont {G.}~\bibnamefont {Xu}}, \bibinfo {author} {\bibfnamefont
  {H.}~\bibnamefont {Weng}}, \bibinfo {author} {\bibfnamefont {X.}~\bibnamefont
  {Dai}}, \ and\ \bibinfo {author} {\bibfnamefont {Z.}~\bibnamefont {Fang}},\
  }\href {\doibase 10.1103/PhysRevB.85.195320} {\bibfield  {journal} {\bibinfo
  {journal} {Phys. Rev. B}\ }\textbf {\bibinfo {volume} {85}},\ \bibinfo
  {pages} {195320} (\bibinfo {year} {2012})}\BibitemShut {NoStop}%
\bibitem [{\citenamefont {Wang}\ \emph {et~al.}(2013)\citenamefont {Wang},
  \citenamefont {Weng}, \citenamefont {Wu}, \citenamefont {Dai},\ and\
  \citenamefont {Fang}}]{Wang:2013}%
  \BibitemOpen
  \bibfield  {author} {\bibinfo {author} {\bibfnamefont {Z.}~\bibnamefont
  {Wang}}, \bibinfo {author} {\bibfnamefont {H.}~\bibnamefont {Weng}}, \bibinfo
  {author} {\bibfnamefont {Q.}~\bibnamefont {Wu}}, \bibinfo {author}
  {\bibfnamefont {X.}~\bibnamefont {Dai}}, \ and\ \bibinfo {author}
  {\bibfnamefont {Z.}~\bibnamefont {Fang}},\ }\href {\doibase
  10.1103/PhysRevB.88.125427} {\bibfield  {journal} {\bibinfo  {journal} {Phys.
  Rev. B}\ }\textbf {\bibinfo {volume} {88}},\ \bibinfo {pages} {125427}
  (\bibinfo {year} {2013})}\BibitemShut {NoStop}%
\bibitem [{\citenamefont {Neupane}\ \emph {et~al.}(2014)\citenamefont
  {Neupane}, \citenamefont {Xu}, \citenamefont {Sankar}, \citenamefont
  {Alidoust}, \citenamefont {Bian}, \citenamefont {Liu}, \citenamefont
  {Belopolski}, \citenamefont {Chang}, \citenamefont {Jeng}, \citenamefont
  {Lin}, \citenamefont {Bansil}, \citenamefont {Chou},\ and\ \citenamefont
  {Hasan}}]{Neupane:2014}%
  \BibitemOpen
  \bibfield  {author} {\bibinfo {author} {\bibfnamefont {M.}~\bibnamefont
  {Neupane}}, \bibinfo {author} {\bibfnamefont {S.-Y.}\ \bibnamefont {Xu}},
  \bibinfo {author} {\bibfnamefont {R.}~\bibnamefont {Sankar}}, \bibinfo
  {author} {\bibfnamefont {N.}~\bibnamefont {Alidoust}}, \bibinfo {author}
  {\bibfnamefont {G.}~\bibnamefont {Bian}}, \bibinfo {author} {\bibfnamefont
  {C.}~\bibnamefont {Liu}}, \bibinfo {author} {\bibfnamefont {I.}~\bibnamefont
  {Belopolski}}, \bibinfo {author} {\bibfnamefont {T.-R.}\ \bibnamefont
  {Chang}}, \bibinfo {author} {\bibfnamefont {H.-T.}\ \bibnamefont {Jeng}},
  \bibinfo {author} {\bibfnamefont {H.}~\bibnamefont {Lin}}, \bibinfo {author}
  {\bibfnamefont {A.}~\bibnamefont {Bansil}}, \bibinfo {author} {\bibfnamefont
  {F.}~\bibnamefont {Chou}}, \ and\ \bibinfo {author} {\bibfnamefont
  {Z.}~\bibnamefont {Hasan}},\ }\href {\doibase 10.1038/ncomms4786} {\bibfield
  {journal} {\bibinfo  {journal} {Nat. Comm.}\ }\textbf {\bibinfo {volume}
  {5}},\ \bibinfo {pages} {4786} (\bibinfo {year} {2014})}\BibitemShut
  {NoStop}%
\bibitem [{\citenamefont {Liu}\ \emph {et~al.}(2014{\natexlab{a}})\citenamefont
  {Liu}, \citenamefont {Zhou}, \citenamefont {Zhang}, \citenamefont {Wang},
  \citenamefont {Weng}, \citenamefont {Prabhakaran}, \citenamefont {Mo},
  \citenamefont {Shen}, \citenamefont {Fang}, \citenamefont {Dai},
  \citenamefont {Hussain},\ and\ \citenamefont {Chen}}]{Liu:2014a}%
  \BibitemOpen
  \bibfield  {author} {\bibinfo {author} {\bibfnamefont {Z.~K.}\ \bibnamefont
  {Liu}}, \bibinfo {author} {\bibfnamefont {B.}~\bibnamefont {Zhou}}, \bibinfo
  {author} {\bibfnamefont {Y.}~\bibnamefont {Zhang}}, \bibinfo {author}
  {\bibfnamefont {Z.~J.}\ \bibnamefont {Wang}}, \bibinfo {author}
  {\bibfnamefont {H.~M.}\ \bibnamefont {Weng}}, \bibinfo {author}
  {\bibfnamefont {D.}~\bibnamefont {Prabhakaran}}, \bibinfo {author}
  {\bibfnamefont {S.-K.}\ \bibnamefont {Mo}}, \bibinfo {author} {\bibfnamefont
  {Z.~X.}\ \bibnamefont {Shen}}, \bibinfo {author} {\bibfnamefont
  {Z.}~\bibnamefont {Fang}}, \bibinfo {author} {\bibfnamefont {X.}~\bibnamefont
  {Dai}}, \bibinfo {author} {\bibfnamefont {Z.}~\bibnamefont {Hussain}}, \ and\
  \bibinfo {author} {\bibfnamefont {Y.}~\bibnamefont {Chen}},\ }\href {\doibase
  10.1126/science.1245085} {\bibfield  {journal} {\bibinfo  {journal}
  {Science}\ }\textbf {\bibinfo {volume} {343}},\ \bibinfo {pages} {864}
  (\bibinfo {year} {2014}{\natexlab{a}})}\BibitemShut {NoStop}%
\bibitem [{\citenamefont {Liu}\ \emph {et~al.}(2014{\natexlab{b}})\citenamefont
  {Liu}, \citenamefont {Jiang}, \citenamefont {Zhou}, \citenamefont {Wang},
  \citenamefont {Zhang}, \citenamefont {Weng}, \citenamefont {Prabhakaran},
  \citenamefont {Mo}, \citenamefont {Peng}, \citenamefont {Dudin},
  \citenamefont {Kim}, \citenamefont {M.}, \citenamefont {Fang}, \citenamefont
  {Dai}, \citenamefont {Shen}, \citenamefont {Hussain},\ and\ \citenamefont
  {Chen}}]{Liu:2014b}%
  \BibitemOpen
  \bibfield  {author} {\bibinfo {author} {\bibfnamefont {Z.~K.}\ \bibnamefont
  {Liu}}, \bibinfo {author} {\bibfnamefont {J.}~\bibnamefont {Jiang}}, \bibinfo
  {author} {\bibfnamefont {B.}~\bibnamefont {Zhou}}, \bibinfo {author}
  {\bibfnamefont {Z.~J.}\ \bibnamefont {Wang}}, \bibinfo {author}
  {\bibfnamefont {Y.}~\bibnamefont {Zhang}}, \bibinfo {author} {\bibfnamefont
  {H.~M.}\ \bibnamefont {Weng}}, \bibinfo {author} {\bibfnamefont
  {D.}~\bibnamefont {Prabhakaran}}, \bibinfo {author} {\bibfnamefont {S.-K.}\
  \bibnamefont {Mo}}, \bibinfo {author} {\bibfnamefont {H.}~\bibnamefont
  {Peng}}, \bibinfo {author} {\bibfnamefont {P.}~\bibnamefont {Dudin}},
  \bibinfo {author} {\bibfnamefont {T.}~\bibnamefont {Kim}}, \bibinfo {author}
  {\bibfnamefont {H.}~\bibnamefont {M.}}, \bibinfo {author} {\bibfnamefont
  {Z.}~\bibnamefont {Fang}}, \bibinfo {author} {\bibfnamefont {X.}~\bibnamefont
  {Dai}}, \bibinfo {author} {\bibfnamefont {D.~L.}\ \bibnamefont {Shen},
  \bibfnamefont {Z.~X.~Feng}}, \bibinfo {author} {\bibfnamefont
  {Z.}~\bibnamefont {Hussain}}, \ and\ \bibinfo {author} {\bibfnamefont
  {Y.}~\bibnamefont {Chen}},\ }\href {\doibase 10.1038/nmat3990} {\bibfield
  {journal} {\bibinfo  {journal} {Nature Mat.}\ }\textbf {\bibinfo {volume}
  {13}},\ \bibinfo {pages} {677} (\bibinfo {year}
  {2014}{\natexlab{b}})}\BibitemShut {NoStop}%
\bibitem [{\citenamefont {Liang}\ \emph {et~al.}(2015)\citenamefont {Liang},
  \citenamefont {Gibson}, \citenamefont {Ali}, \citenamefont {Liu},
  \citenamefont {Cava},\ and\ \citenamefont {Ong}}]{Liang:2015}%
  \BibitemOpen
  \bibfield  {author} {\bibinfo {author} {\bibfnamefont {T.}~\bibnamefont
  {Liang}}, \bibinfo {author} {\bibfnamefont {Q.}~\bibnamefont {Gibson}},
  \bibinfo {author} {\bibfnamefont {M.~N.}\ \bibnamefont {Ali}}, \bibinfo
  {author} {\bibfnamefont {M.}~\bibnamefont {Liu}}, \bibinfo {author}
  {\bibfnamefont {R.~J.}\ \bibnamefont {Cava}}, \ and\ \bibinfo {author}
  {\bibfnamefont {N.~P.}\ \bibnamefont {Ong}},\ }\href {\doibase
  10.1038/nmat4143} {\bibfield  {journal} {\bibinfo  {journal} {Nature
  materials}\ }\textbf {\bibinfo {volume} {14}},\ \bibinfo {pages} {280}
  (\bibinfo {year} {2015})}\BibitemShut {NoStop}%
\bibitem [{\citenamefont {Weng}\ \emph
  {et~al.}(2015{\natexlab{a}})\citenamefont {Weng}, \citenamefont {Fang},
  \citenamefont {Fang}, \citenamefont {Bernevig},\ and\ \citenamefont
  {Dai}}]{Weng:2015}%
  \BibitemOpen
  \bibfield  {author} {\bibinfo {author} {\bibfnamefont {H.}~\bibnamefont
  {Weng}}, \bibinfo {author} {\bibfnamefont {C.~F.}\ \bibnamefont {Fang}},
  \bibinfo {author} {\bibfnamefont {Z.}~\bibnamefont {Fang}}, \bibinfo {author}
  {\bibfnamefont {B.~A.}\ \bibnamefont {Bernevig}}, \ and\ \bibinfo {author}
  {\bibfnamefont {X.}~\bibnamefont {Dai}},\ }\href {\doibase
  10.1103/PhysRevX.5.011029} {\bibfield  {journal} {\bibinfo  {journal} {Phys.
  Rev. X}\ }\textbf {\bibinfo {volume} {5}},\ \bibinfo {pages} {011029}
  (\bibinfo {year} {2015}{\natexlab{a}})}\BibitemShut {NoStop}%
\bibitem [{\citenamefont {Huang}\ \emph
  {et~al.}(2015{\natexlab{a}})\citenamefont {Huang}, \citenamefont {Xu},
  \citenamefont {Belopolski}, \citenamefont {Lee}, \citenamefont {Chang},
  \citenamefont {Wang}, \citenamefont {Alidoust}, \citenamefont {Bian},
  \citenamefont {Neupane}, \citenamefont {Zhang}, \citenamefont {Jia},
  \citenamefont {Bansil}, \citenamefont {Lin},\ and\ \citenamefont
  {Hasan}}]{Huang:2015a}%
  \BibitemOpen
  \bibfield  {author} {\bibinfo {author} {\bibfnamefont {S.-M.}\ \bibnamefont
  {Huang}}, \bibinfo {author} {\bibfnamefont {S.-Y.}\ \bibnamefont {Xu}},
  \bibinfo {author} {\bibfnamefont {I.}~\bibnamefont {Belopolski}}, \bibinfo
  {author} {\bibfnamefont {C.-C.}\ \bibnamefont {Lee}}, \bibinfo {author}
  {\bibfnamefont {G.}~\bibnamefont {Chang}}, \bibinfo {author} {\bibfnamefont
  {B.}~\bibnamefont {Wang}}, \bibinfo {author} {\bibfnamefont {N.}~\bibnamefont
  {Alidoust}}, \bibinfo {author} {\bibfnamefont {G.}~\bibnamefont {Bian}},
  \bibinfo {author} {\bibfnamefont {M.}~\bibnamefont {Neupane}}, \bibinfo
  {author} {\bibfnamefont {C.}~\bibnamefont {Zhang}}, \bibinfo {author}
  {\bibfnamefont {S.}~\bibnamefont {Jia}}, \bibinfo {author} {\bibfnamefont
  {A.}~\bibnamefont {Bansil}}, \bibinfo {author} {\bibfnamefont
  {H.}~\bibnamefont {Lin}}, \ and\ \bibinfo {author} {\bibfnamefont {M.~Z.}\
  \bibnamefont {Hasan}},\ }\href {\doibase 10.1038/ncomms8373} {\bibfield
  {journal} {\bibinfo  {journal} {Nat. Commun.}\ }\textbf {\bibinfo {volume}
  {6}},\ \bibinfo {pages} {7373} (\bibinfo {year}
  {2015}{\natexlab{a}})}\BibitemShut {NoStop}%
\bibitem [{\citenamefont {Xu}\ \emph {et~al.}(2015{\natexlab{a}})\citenamefont
  {Xu}, \citenamefont {Belopolski}, \citenamefont {Alidoust}, \citenamefont
  {Neupane}, \citenamefont {Bian}, \citenamefont {Zhang}, \citenamefont
  {Sankar}, \citenamefont {Chang}, \citenamefont {Yuan}, \citenamefont {Lee},
  \citenamefont {Huang}, \citenamefont {Zheng}, \citenamefont {Ma},
  \citenamefont {Sanchez}, \citenamefont {Wang}, \citenamefont {Bansil},
  \citenamefont {Chou}, \citenamefont {Shibayev}, \citenamefont {Lin},
  \citenamefont {Jia},\ and\ \citenamefont {Hasan}}]{Xu:2015a}%
  \BibitemOpen
  \bibfield  {author} {\bibinfo {author} {\bibfnamefont {S.-Y.}\ \bibnamefont
  {Xu}}, \bibinfo {author} {\bibfnamefont {I.}~\bibnamefont {Belopolski}},
  \bibinfo {author} {\bibfnamefont {N.}~\bibnamefont {Alidoust}}, \bibinfo
  {author} {\bibfnamefont {M.}~\bibnamefont {Neupane}}, \bibinfo {author}
  {\bibfnamefont {G.}~\bibnamefont {Bian}}, \bibinfo {author} {\bibfnamefont
  {C.}~\bibnamefont {Zhang}}, \bibinfo {author} {\bibfnamefont
  {R.}~\bibnamefont {Sankar}}, \bibinfo {author} {\bibfnamefont
  {G.}~\bibnamefont {Chang}}, \bibinfo {author} {\bibfnamefont
  {Z.}~\bibnamefont {Yuan}}, \bibinfo {author} {\bibfnamefont {C.-C.}\
  \bibnamefont {Lee}}, \bibinfo {author} {\bibfnamefont {S.-M.}\ \bibnamefont
  {Huang}}, \bibinfo {author} {\bibfnamefont {H.}~\bibnamefont {Zheng}},
  \bibinfo {author} {\bibfnamefont {J.}~\bibnamefont {Ma}}, \bibinfo {author}
  {\bibfnamefont {D.~S.}\ \bibnamefont {Sanchez}}, \bibinfo {author}
  {\bibfnamefont {B.}~\bibnamefont {Wang}}, \bibinfo {author} {\bibfnamefont
  {A.}~\bibnamefont {Bansil}}, \bibinfo {author} {\bibfnamefont
  {F.}~\bibnamefont {Chou}}, \bibinfo {author} {\bibfnamefont {P.~P.}\
  \bibnamefont {Shibayev}}, \bibinfo {author} {\bibfnamefont {H.}~\bibnamefont
  {Lin}}, \bibinfo {author} {\bibfnamefont {S.}~\bibnamefont {Jia}}, \ and\
  \bibinfo {author} {\bibfnamefont {M.~Z.}\ \bibnamefont {Hasan}},\ }\href
  {\doibase 10.1126/science.aaa9297} {\bibfield  {journal} {\bibinfo  {journal}
  {Science}\ }\textbf {\bibinfo {volume} {349}},\ \bibinfo {pages} {613}
  (\bibinfo {year} {2015}{\natexlab{a}})}\BibitemShut {NoStop}%
\bibitem [{\citenamefont {Lv}\ \emph {et~al.}(2015)\citenamefont {Lv},
  \citenamefont {Weng}, \citenamefont {Fu}, \citenamefont {Wang}, \citenamefont
  {Miao}, \citenamefont {Ma}, \citenamefont {Richard}, \citenamefont {Huang},
  \citenamefont {Zhao}, \citenamefont {Chen}, \citenamefont {Fang},
  \citenamefont {Dai}, \citenamefont {Qian},\ and\ \citenamefont
  {Ding}}]{Lv:2015}%
  \BibitemOpen
  \bibfield  {author} {\bibinfo {author} {\bibfnamefont {B.}~\bibnamefont
  {Lv}}, \bibinfo {author} {\bibfnamefont {H.}~\bibnamefont {Weng}}, \bibinfo
  {author} {\bibfnamefont {B.}~\bibnamefont {Fu}}, \bibinfo {author}
  {\bibfnamefont {X.}~\bibnamefont {Wang}}, \bibinfo {author} {\bibfnamefont
  {H.}~\bibnamefont {Miao}}, \bibinfo {author} {\bibfnamefont {J.}~\bibnamefont
  {Ma}}, \bibinfo {author} {\bibfnamefont {P.}~\bibnamefont {Richard}},
  \bibinfo {author} {\bibfnamefont {X.}~\bibnamefont {Huang}}, \bibinfo
  {author} {\bibfnamefont {L.}~\bibnamefont {Zhao}}, \bibinfo {author}
  {\bibfnamefont {G.}~\bibnamefont {Chen}}, \bibinfo {author} {\bibfnamefont
  {Z.}~\bibnamefont {Fang}}, \bibinfo {author} {\bibfnamefont {X.}~\bibnamefont
  {Dai}}, \bibinfo {author} {\bibfnamefont {T.}~\bibnamefont {Qian}}, \ and\
  \bibinfo {author} {\bibfnamefont {H.}~\bibnamefont {Ding}},\ }\href {\doibase
  10.1103/PhysRevX.5.031013} {\bibfield  {journal} {\bibinfo  {journal} {Phys.
  Rev. X}\ }\textbf {\bibinfo {volume} {5}},\ \bibinfo {pages} {031013}
  (\bibinfo {year} {2015})}\BibitemShut {NoStop}%
\bibitem [{\citenamefont {Wehling}\ \emph {et~al.}(2014)\citenamefont
  {Wehling}, \citenamefont {Black-Schaffer},\ and\ \citenamefont
  {Balatsky}}]{Wehling:2014}%
  \BibitemOpen
  \bibfield  {author} {\bibinfo {author} {\bibfnamefont {T.~O.}\ \bibnamefont
  {Wehling}}, \bibinfo {author} {\bibfnamefont {A.~M.}\ \bibnamefont
  {Black-Schaffer}}, \ and\ \bibinfo {author} {\bibfnamefont {A.~V.}\
  \bibnamefont {Balatsky}},\ }\href {\doibase 10.1080/00018732.2014.927109}
  {\bibfield  {journal} {\bibinfo  {journal} {Adv. Phys.}\ }\textbf {\bibinfo
  {volume} {63}},\ \bibinfo {pages} {1} (\bibinfo {year} {2014})}\BibitemShut
  {NoStop}%
\bibitem [{\citenamefont {Murakami}(2007)}]{Murakami:2007}%
  \BibitemOpen
  \bibfield  {author} {\bibinfo {author} {\bibfnamefont {S.}~\bibnamefont
  {Murakami}},\ }\href {\doibase 10.1088/1367-2630/9/9/356} {\bibfield
  {journal} {\bibinfo  {journal} {New J. Phys.}\ }\textbf {\bibinfo {volume}
  {9}},\ \bibinfo {pages} {356} (\bibinfo {year} {2007})}\BibitemShut {NoStop}%
\bibitem [{\citenamefont {Wan}\ \emph {et~al.}(2011)\citenamefont {Wan},
  \citenamefont {Turner}, \citenamefont {Vishwanath},\ and\ \citenamefont
  {Savrasov}}]{Wan:2011}%
  \BibitemOpen
  \bibfield  {author} {\bibinfo {author} {\bibfnamefont {X.}~\bibnamefont
  {Wan}}, \bibinfo {author} {\bibfnamefont {A.~M.}\ \bibnamefont {Turner}},
  \bibinfo {author} {\bibfnamefont {A.}~\bibnamefont {Vishwanath}}, \ and\
  \bibinfo {author} {\bibfnamefont {S.~Y.}\ \bibnamefont {Savrasov}},\ }\href
  {\doibase 10.1103/PhysRevB.83.205101} {\bibfield  {journal} {\bibinfo
  {journal} {Phys. Rev. B}\ }\textbf {\bibinfo {volume} {83}},\ \bibinfo
  {pages} {205101} (\bibinfo {year} {2011})}\BibitemShut {NoStop}%
\bibitem [{\citenamefont {Young}\ \emph {et~al.}(2012)\citenamefont {Young},
  \citenamefont {Zaheer}, \citenamefont {Teo}, \citenamefont {Kane},
  \citenamefont {Mele},\ and\ \citenamefont {Rappe}}]{Young:2012}%
  \BibitemOpen
  \bibfield  {author} {\bibinfo {author} {\bibfnamefont {S.~M.}\ \bibnamefont
  {Young}}, \bibinfo {author} {\bibfnamefont {S.}~\bibnamefont {Zaheer}},
  \bibinfo {author} {\bibfnamefont {J.~C.~Y.}\ \bibnamefont {Teo}}, \bibinfo
  {author} {\bibfnamefont {C.~L.}\ \bibnamefont {Kane}}, \bibinfo {author}
  {\bibfnamefont {E.~J.}\ \bibnamefont {Mele}}, \ and\ \bibinfo {author}
  {\bibfnamefont {A.~M.}\ \bibnamefont {Rappe}},\ }\href {\doibase
  10.1103/PhysRevLett.108.140405} {\bibfield  {journal} {\bibinfo  {journal}
  {Phys. Rev. Lett.}\ }\textbf {\bibinfo {volume} {108}},\ \bibinfo {pages}
  {140405} (\bibinfo {year} {2012})}\BibitemShut {NoStop}%
\bibitem [{\citenamefont {Yang}\ and\ \citenamefont
  {Nagaosa}(2014)}]{Yang:2014}%
  \BibitemOpen
  \bibfield  {author} {\bibinfo {author} {\bibfnamefont {B.-J.}\ \bibnamefont
  {Yang}}\ and\ \bibinfo {author} {\bibfnamefont {N.}~\bibnamefont {Nagaosa}},\
  }\href {\doibase 10.1038/ncomms5898} {\bibfield  {journal} {\bibinfo
  {journal} {Nat. Commun.}\ }\textbf {\bibinfo {volume} {5}},\ \bibinfo {pages}
  {4898} (\bibinfo {year} {2014})}\BibitemShut {NoStop}%
\bibitem [{\citenamefont {Liu}\ and\ \citenamefont
  {Vanderbilt}(2014)}]{Liu:2014c}%
  \BibitemOpen
  \bibfield  {author} {\bibinfo {author} {\bibfnamefont {J.}~\bibnamefont
  {Liu}}\ and\ \bibinfo {author} {\bibfnamefont {D.}~\bibnamefont
  {Vanderbilt}},\ }\href {\doibase 10.1103/PhysRevB.90.155316} {\bibfield
  {journal} {\bibinfo  {journal} {Physical Review B}\ }\textbf {\bibinfo
  {volume} {90}},\ \bibinfo {pages} {155316} (\bibinfo {year}
  {2014})}\BibitemShut {NoStop}%
\bibitem [{\citenamefont {Matsuura}\ \emph {et~al.}(2013)\citenamefont
  {Matsuura}, \citenamefont {Chang}, \citenamefont {Schnyder},\ and\
  \citenamefont {Ryu}}]{Matsuura:2013}%
  \BibitemOpen
  \bibfield  {author} {\bibinfo {author} {\bibfnamefont {S.}~\bibnamefont
  {Matsuura}}, \bibinfo {author} {\bibfnamefont {P.-Y.}\ \bibnamefont {Chang}},
  \bibinfo {author} {\bibfnamefont {A.}~\bibnamefont {Schnyder}}, \ and\
  \bibinfo {author} {\bibfnamefont {S.}~\bibnamefont {Ryu}},\ }\href {\doibase
  10.1088/1367-2630/15/6/065001} {\bibfield  {journal} {\bibinfo  {journal}
  {New J. Phys.}\ }\textbf {\bibinfo {volume} {15}},\ \bibinfo {pages} {065001}
  (\bibinfo {year} {2013})}\BibitemShut {NoStop}%
\bibitem [{\citenamefont {Bzdu\v{s}ek}\ \emph {et~al.}(2015)\citenamefont
  {Bzdu\v{s}ek}, \citenamefont {R\"{u}egg},\ and\ \citenamefont
  {Sigrist}}]{Bzdusek:2015}%
  \BibitemOpen
  \bibfield  {author} {\bibinfo {author} {\bibfnamefont {T.}~\bibnamefont
  {Bzdu\v{s}ek}}, \bibinfo {author} {\bibfnamefont {A.}~\bibnamefont
  {R\"{u}egg}}, \ and\ \bibinfo {author} {\bibfnamefont {M.}~\bibnamefont
  {Sigrist}},\ }\href {\doibase 10.1103/PhysRevB.91.165105} {\bibfield
  {journal} {\bibinfo  {journal} {Phys. Rev. B}\ }\textbf {\bibinfo {volume}
  {91}},\ \bibinfo {pages} {165105} (\bibinfo {year} {2015})}\BibitemShut
  {NoStop}%
\bibitem [{\citenamefont {Nielsen}\ and\ \citenamefont
  {Ninomiya}(1983)}]{Nielsen:1983}%
  \BibitemOpen
  \bibfield  {author} {\bibinfo {author} {\bibfnamefont {H.~B.}\ \bibnamefont
  {Nielsen}}\ and\ \bibinfo {author} {\bibfnamefont {M.}~\bibnamefont
  {Ninomiya}},\ }\href {\doibase 10.1016/0370-2693(83)91529-0} {\bibfield
  {journal} {\bibinfo  {journal} {Physics Letters B}\ }\textbf {\bibinfo
  {volume} {130}},\ \bibinfo {pages} {389} (\bibinfo {year}
  {1983})}\BibitemShut {NoStop}%
\bibitem [{\citenamefont {Son}\ and\ \citenamefont {Spivak}(2013)}]{Son:2013}%
  \BibitemOpen
  \bibfield  {author} {\bibinfo {author} {\bibfnamefont {D.~T.}\ \bibnamefont
  {Son}}\ and\ \bibinfo {author} {\bibfnamefont {B.~Z.}\ \bibnamefont
  {Spivak}},\ }\href {\doibase 10.1103/PhysRevB.88.104412} {\bibfield
  {journal} {\bibinfo  {journal} {Phys. Rev. B}\ }\textbf {\bibinfo {volume}
  {88}},\ \bibinfo {pages} {104412} (\bibinfo {year} {2013})}\BibitemShut
  {NoStop}%
\bibitem [{\citenamefont {Huang}\ \emph
  {et~al.}(2015{\natexlab{b}})\citenamefont {Huang}, \citenamefont {Zhao},
  \citenamefont {Long}, \citenamefont {Wang}, \citenamefont {Chen},
  \citenamefont {Yang}, \citenamefont {Liang}, \citenamefont {Xue},
  \citenamefont {Weng}, \citenamefont {Fang}, \citenamefont {Dai},\ and\
  \citenamefont {Chen}}]{Huang:2015b}%
  \BibitemOpen
  \bibfield  {author} {\bibinfo {author} {\bibfnamefont {X.}~\bibnamefont
  {Huang}}, \bibinfo {author} {\bibfnamefont {L.}~\bibnamefont {Zhao}},
  \bibinfo {author} {\bibfnamefont {Y.}~\bibnamefont {Long}}, \bibinfo {author}
  {\bibfnamefont {P.}~\bibnamefont {Wang}}, \bibinfo {author} {\bibfnamefont
  {D.}~\bibnamefont {Chen}}, \bibinfo {author} {\bibfnamefont {Z.}~\bibnamefont
  {Yang}}, \bibinfo {author} {\bibfnamefont {H.}~\bibnamefont {Liang}},
  \bibinfo {author} {\bibfnamefont {M.}~\bibnamefont {Xue}}, \bibinfo {author}
  {\bibfnamefont {H.}~\bibnamefont {Weng}}, \bibinfo {author} {\bibfnamefont
  {Z.}~\bibnamefont {Fang}}, \bibinfo {author} {\bibfnamefont {X.}~\bibnamefont
  {Dai}}, \ and\ \bibinfo {author} {\bibfnamefont {G.}~\bibnamefont {Chen}},\
  }\href {\doibase 10.1103/PhysRevX.5.031023} {\bibfield  {journal} {\bibinfo
  {journal} {Phys. Rev. X}\ }\textbf {\bibinfo {volume} {5}},\ \bibinfo {pages}
  {031023} (\bibinfo {year} {2015}{\natexlab{b}})}\BibitemShut {NoStop}%
\bibitem [{\citenamefont {Xiong}\ \emph {et~al.}(2015)\citenamefont {Xiong},
  \citenamefont {Kushwaha}, \citenamefont {Liang}, \citenamefont {Krizan},
  \citenamefont {Wang}, \citenamefont {Cava},\ and\ \citenamefont
  {Ong}}]{Xiong:2015}%
  \BibitemOpen
  \bibfield  {author} {\bibinfo {author} {\bibfnamefont {J.}~\bibnamefont
  {Xiong}}, \bibinfo {author} {\bibfnamefont {S.~K.}\ \bibnamefont {Kushwaha}},
  \bibinfo {author} {\bibfnamefont {T.}~\bibnamefont {Liang}}, \bibinfo
  {author} {\bibfnamefont {J.~W.}\ \bibnamefont {Krizan}}, \bibinfo {author}
  {\bibfnamefont {W.}~\bibnamefont {Wang}}, \bibinfo {author} {\bibfnamefont
  {R.}~\bibnamefont {Cava}}, \ and\ \bibinfo {author} {\bibfnamefont
  {N.}~\bibnamefont {Ong}},\ }\href@noop {} {\bibfield  {journal} {\bibinfo
  {journal} {ArXiv e-prints}\ } (\bibinfo {year} {2015})},\ \Eprint
  {http://arxiv.org/abs/1503.08179} {arXiv:1503.08179} \BibitemShut {NoStop}%
\bibitem [{\citenamefont {Zhang}\ \emph {et~al.}(2015)\citenamefont {Zhang},
  \citenamefont {Xu}, \citenamefont {Belopolski}, \citenamefont {Yuan},
  \citenamefont {Lin}, \citenamefont {Tong}, \citenamefont {Alidoust},
  \citenamefont {Lee}, \citenamefont {Huang}, \citenamefont {Lin} \emph
  {et~al.}}]{Zhang:2015}%
  \BibitemOpen
  \bibfield  {author} {\bibinfo {author} {\bibfnamefont {C.}~\bibnamefont
  {Zhang}}, \bibinfo {author} {\bibfnamefont {S.-Y.}\ \bibnamefont {Xu}},
  \bibinfo {author} {\bibfnamefont {I.}~\bibnamefont {Belopolski}}, \bibinfo
  {author} {\bibfnamefont {Z.}~\bibnamefont {Yuan}}, \bibinfo {author}
  {\bibfnamefont {Z.}~\bibnamefont {Lin}}, \bibinfo {author} {\bibfnamefont
  {B.}~\bibnamefont {Tong}}, \bibinfo {author} {\bibfnamefont {N.}~\bibnamefont
  {Alidoust}}, \bibinfo {author} {\bibfnamefont {C.-C.}\ \bibnamefont {Lee}},
  \bibinfo {author} {\bibfnamefont {S.-M.}\ \bibnamefont {Huang}}, \bibinfo
  {author} {\bibfnamefont {H.}~\bibnamefont {Lin}},  \emph {et~al.},\
  }\href@noop {} {\bibfield  {journal} {\bibinfo  {journal} {ArXiv e-prints}\ }
  (\bibinfo {year} {2015})},\ \Eprint {http://arxiv.org/abs/1503.02630}
  {arXiv:1503.02630} \BibitemShut {NoStop}%
\bibitem [{\citenamefont {Xu}\ \emph {et~al.}(2015{\natexlab{b}})\citenamefont
  {Xu}, \citenamefont {Liu}, \citenamefont {Kushwaha}, \citenamefont {Sankar},
  \citenamefont {Krizan}, \citenamefont {Belopolski}, \citenamefont {Neupane},
  \citenamefont {Bian}, \citenamefont {Alidoust}, \citenamefont {Chang},
  \citenamefont {Jeng}, \citenamefont {Huang}, \citenamefont {Tsai},
  \citenamefont {Lin}, \citenamefont {Chou}, \citenamefont {Shibayev},
  \citenamefont {Cava},\ and\ \citenamefont {Hasan}}]{Xu:2015}%
  \BibitemOpen
  \bibfield  {author} {\bibinfo {author} {\bibfnamefont {S.-Y.}\ \bibnamefont
  {Xu}}, \bibinfo {author} {\bibfnamefont {C.}~\bibnamefont {Liu}}, \bibinfo
  {author} {\bibfnamefont {S.~K.}\ \bibnamefont {Kushwaha}}, \bibinfo {author}
  {\bibfnamefont {R.}~\bibnamefont {Sankar}}, \bibinfo {author} {\bibfnamefont
  {J.~W.}\ \bibnamefont {Krizan}}, \bibinfo {author} {\bibfnamefont
  {I.}~\bibnamefont {Belopolski}}, \bibinfo {author} {\bibfnamefont
  {M.}~\bibnamefont {Neupane}}, \bibinfo {author} {\bibfnamefont
  {G.}~\bibnamefont {Bian}}, \bibinfo {author} {\bibfnamefont {N.}~\bibnamefont
  {Alidoust}}, \bibinfo {author} {\bibfnamefont {T.-R.}\ \bibnamefont {Chang}},
  \bibinfo {author} {\bibfnamefont {H.-T.}\ \bibnamefont {Jeng}}, \bibinfo
  {author} {\bibfnamefont {C.-Y.}\ \bibnamefont {Huang}}, \bibinfo {author}
  {\bibfnamefont {W.-F.}\ \bibnamefont {Tsai}}, \bibinfo {author}
  {\bibfnamefont {H.}~\bibnamefont {Lin}}, \bibinfo {author} {\bibfnamefont
  {F.}~\bibnamefont {Chou}}, \bibinfo {author} {\bibfnamefont {P.~P.}\
  \bibnamefont {Shibayev}}, \bibinfo {author} {\bibfnamefont {R.~J.}\
  \bibnamefont {Cava}}, \ and\ \bibinfo {author} {\bibfnamefont {M.~Z.}\
  \bibnamefont {Hasan}},\ }\href {\doibase 10.1126/science.1256742} {\bibfield
  {journal} {\bibinfo  {journal} {Science}\ }\textbf {\bibinfo {volume}
  {347}},\ \bibinfo {pages} {6219} (\bibinfo {year}
  {2015}{\natexlab{b}})}\BibitemShut {NoStop}%
\bibitem [{\citenamefont {Inoue}\ \emph {et~al.}(2016)\citenamefont {Inoue},
  \citenamefont {Gyenis}, \citenamefont {Wang}, \citenamefont {Li},
  \citenamefont {Oh}, \citenamefont {Jiang}, \citenamefont {Ni}, \citenamefont
  {Bernevig},\ and\ \citenamefont {Yazdani}}]{Inoue:2016}%
  \BibitemOpen
  \bibfield  {author} {\bibinfo {author} {\bibfnamefont {H.}~\bibnamefont
  {Inoue}}, \bibinfo {author} {\bibfnamefont {A.}~\bibnamefont {Gyenis}},
  \bibinfo {author} {\bibfnamefont {Z.}~\bibnamefont {Wang}}, \bibinfo {author}
  {\bibfnamefont {J.}~\bibnamefont {Li}}, \bibinfo {author} {\bibfnamefont
  {S.~W.}\ \bibnamefont {Oh}}, \bibinfo {author} {\bibfnamefont
  {S.}~\bibnamefont {Jiang}}, \bibinfo {author} {\bibfnamefont
  {N.}~\bibnamefont {Ni}}, \bibinfo {author} {\bibfnamefont {B.~A.}\
  \bibnamefont {Bernevig}}, \ and\ \bibinfo {author} {\bibfnamefont
  {A.}~\bibnamefont {Yazdani}},\ }\href {\doibase 10.1126/science.aad8766}
  {\bibfield  {journal} {\bibinfo  {journal} {Science}\ }\textbf {\bibinfo
  {volume} {351}},\ \bibinfo {pages} {1184} (\bibinfo {year}
  {2016})}\BibitemShut {NoStop}%
\bibitem [{\citenamefont {Soluyanov}\ \emph {et~al.}(2015)\citenamefont
  {Soluyanov}, \citenamefont {Gresch}, \citenamefont {Wang}, \citenamefont
  {Wu}, \citenamefont {Troyer}, \citenamefont {Dai},\ and\ \citenamefont
  {Bernevig}}]{Soluyanov:2015}%
  \BibitemOpen
  \bibfield  {author} {\bibinfo {author} {\bibfnamefont {A.~A.}\ \bibnamefont
  {Soluyanov}}, \bibinfo {author} {\bibfnamefont {D.}~\bibnamefont {Gresch}},
  \bibinfo {author} {\bibfnamefont {Z.}~\bibnamefont {Wang}}, \bibinfo {author}
  {\bibfnamefont {Q.}~\bibnamefont {Wu}}, \bibinfo {author} {\bibfnamefont
  {M.}~\bibnamefont {Troyer}}, \bibinfo {author} {\bibfnamefont
  {X.}~\bibnamefont {Dai}}, \ and\ \bibinfo {author} {\bibfnamefont {B.~A.}\
  \bibnamefont {Bernevig}},\ }\href@noop {} {\bibfield  {journal} {\bibinfo
  {journal} {Nature}\ }\textbf {\bibinfo {volume} {527}},\ \bibinfo {pages}
  {495} (\bibinfo {year} {2015})}\BibitemShut {NoStop}%
\bibitem [{\citenamefont {Volovik}\ and\ \citenamefont
  {Zubkov}(2014)}]{Volovik:2014}%
  \BibitemOpen
  \bibfield  {author} {\bibinfo {author} {\bibfnamefont {G.~E.}\ \bibnamefont
  {Volovik}}\ and\ \bibinfo {author} {\bibfnamefont {M.~A.}\ \bibnamefont
  {Zubkov}},\ }\href {\doibase 10.1016/j.nuclphysb.2014.02.018} {\bibfield
  {journal} {\bibinfo  {journal} {Nucl. Phys. B}\ }\textbf {\bibinfo {volume}
  {881}},\ \bibinfo {pages} {514} (\bibinfo {year} {2014})}\BibitemShut
  {NoStop}%
\bibitem [{\citenamefont {Burkov}\ \emph {et~al.}(2012)\citenamefont {Burkov},
  \citenamefont {Hook},\ and\ \citenamefont {Balents}}]{Burkov:2011}%
  \BibitemOpen
  \bibfield  {author} {\bibinfo {author} {\bibfnamefont {A.~A.}\ \bibnamefont
  {Burkov}}, \bibinfo {author} {\bibfnamefont {M.~D.}\ \bibnamefont {Hook}}, \
  and\ \bibinfo {author} {\bibfnamefont {L.}~\bibnamefont {Balents}},\ }\href
  {\doibase 10.1103/PhysRevB.84.235126} {\bibfield  {journal} {\bibinfo
  {journal} {Phys. Rev. B}\ }\textbf {\bibinfo {volume} {84}},\ \bibinfo
  {pages} {235126} (\bibinfo {year} {2012})}\BibitemShut {NoStop}%
\bibitem [{\citenamefont {Yu}\ \emph {et~al.}(2015)\citenamefont {Yu},
  \citenamefont {Weng}, \citenamefont {Fang}, \citenamefont {Dai},\ and\
  \citenamefont {Hu}}]{Yu:2015}%
  \BibitemOpen
  \bibfield  {author} {\bibinfo {author} {\bibfnamefont {R.}~\bibnamefont
  {Yu}}, \bibinfo {author} {\bibfnamefont {H.}~\bibnamefont {Weng}}, \bibinfo
  {author} {\bibfnamefont {Z.}~\bibnamefont {Fang}}, \bibinfo {author}
  {\bibfnamefont {X.}~\bibnamefont {Dai}}, \ and\ \bibinfo {author}
  {\bibfnamefont {X.}~\bibnamefont {Hu}},\ }\href {\doibase
  10.1103/PhysRevLett.115.036807} {\bibfield  {journal} {\bibinfo  {journal}
  {Phys. Rev. Lett.}\ }\textbf {\bibinfo {volume} {115}},\ \bibinfo {pages}
  {036807} (\bibinfo {year} {2015})}\BibitemShut {NoStop}%
\bibitem [{\citenamefont {Bian}\ \emph {et~al.}(2015)\citenamefont {Bian},
  \citenamefont {Chang}, \citenamefont {Sankar}, \citenamefont {Xu},
  \citenamefont {Zheng}, \citenamefont {Neupert}, \citenamefont {Chiu},
  \citenamefont {Huang}, \citenamefont {Chang}, \citenamefont {Belopolski},
  \citenamefont {Sanchez}, \citenamefont {Neupane}, \citenamefont {Alidoust},
  \citenamefont {Liu}, \citenamefont {Wang}, \citenamefont {Lee}, \citenamefont
  {Jeng}, \citenamefont {Bansil}, \citenamefont {Chou}, \citenamefont {Lin},\
  and\ \citenamefont {Hasan}}]{Bian:2015a}%
  \BibitemOpen
  \bibfield  {author} {\bibinfo {author} {\bibfnamefont {G.}~\bibnamefont
  {Bian}}, \bibinfo {author} {\bibfnamefont {T.-R.}\ \bibnamefont {Chang}},
  \bibinfo {author} {\bibfnamefont {R.}~\bibnamefont {Sankar}}, \bibinfo
  {author} {\bibfnamefont {S.-Y.}\ \bibnamefont {Xu}}, \bibinfo {author}
  {\bibfnamefont {H.}~\bibnamefont {Zheng}}, \bibinfo {author} {\bibfnamefont
  {T.}~\bibnamefont {Neupert}}, \bibinfo {author} {\bibfnamefont {C.-K.}\
  \bibnamefont {Chiu}}, \bibinfo {author} {\bibfnamefont {S.-M.}\ \bibnamefont
  {Huang}}, \bibinfo {author} {\bibfnamefont {G.}~\bibnamefont {Chang}},
  \bibinfo {author} {\bibfnamefont {I.}~\bibnamefont {Belopolski}}, \bibinfo
  {author} {\bibfnamefont {D.~S.}\ \bibnamefont {Sanchez}}, \bibinfo {author}
  {\bibfnamefont {M.}~\bibnamefont {Neupane}}, \bibinfo {author} {\bibfnamefont
  {N.}~\bibnamefont {Alidoust}}, \bibinfo {author} {\bibfnamefont
  {C.}~\bibnamefont {Liu}}, \bibinfo {author} {\bibfnamefont {B.}~\bibnamefont
  {Wang}}, \bibinfo {author} {\bibfnamefont {C.-C.}\ \bibnamefont {Lee}},
  \bibinfo {author} {\bibfnamefont {H.-T.}\ \bibnamefont {Jeng}}, \bibinfo
  {author} {\bibfnamefont {A.}~\bibnamefont {Bansil}}, \bibinfo {author}
  {\bibfnamefont {F.}~\bibnamefont {Chou}}, \bibinfo {author} {\bibfnamefont
  {H.}~\bibnamefont {Lin}}, \ and\ \bibinfo {author} {\bibfnamefont {M.~Z.}\
  \bibnamefont {Hasan}},\ }\href@noop {} {\bibfield  {journal} {\bibinfo
  {journal} {ArXiv e-prints}\ } (\bibinfo {year} {2015})},\ \Eprint
  {http://arxiv.org/abs/1505.03069} {arXiv:1505.03069} \BibitemShut {NoStop}%
\bibitem [{\citenamefont {Bian}\ \emph {et~al.}(2016)\citenamefont {Bian},
  \citenamefont {Chang}, \citenamefont {Zheng}, \citenamefont {Velury},
  \citenamefont {Xu}, \citenamefont {Neupert}, \citenamefont {Chiu},
  \citenamefont {Huang}, \citenamefont {Sanchez}, \citenamefont {Belopolski},
  \citenamefont {Alidoust}, \citenamefont {Chen}, \citenamefont {Chang},
  \citenamefont {Bansil}, \citenamefont {Jeng}, \citenamefont {Lin},\ and\
  \citenamefont {Hasan}}]{Bian:2016}%
  \BibitemOpen
  \bibfield  {author} {\bibinfo {author} {\bibfnamefont {G.}~\bibnamefont
  {Bian}}, \bibinfo {author} {\bibfnamefont {T.-R.}\ \bibnamefont {Chang}},
  \bibinfo {author} {\bibfnamefont {H.}~\bibnamefont {Zheng}}, \bibinfo
  {author} {\bibfnamefont {S.}~\bibnamefont {Velury}}, \bibinfo {author}
  {\bibfnamefont {S.-Y.}\ \bibnamefont {Xu}}, \bibinfo {author} {\bibfnamefont
  {T.}~\bibnamefont {Neupert}}, \bibinfo {author} {\bibfnamefont {C.-K.}\
  \bibnamefont {Chiu}}, \bibinfo {author} {\bibfnamefont {S.-M.}\ \bibnamefont
  {Huang}}, \bibinfo {author} {\bibfnamefont {D.~S.}\ \bibnamefont {Sanchez}},
  \bibinfo {author} {\bibfnamefont {I.}~\bibnamefont {Belopolski}}, \bibinfo
  {author} {\bibfnamefont {N.}~\bibnamefont {Alidoust}}, \bibinfo {author}
  {\bibfnamefont {P.-J.}\ \bibnamefont {Chen}}, \bibinfo {author}
  {\bibfnamefont {G.}~\bibnamefont {Chang}}, \bibinfo {author} {\bibfnamefont
  {A.}~\bibnamefont {Bansil}}, \bibinfo {author} {\bibfnamefont {H.-T.}\
  \bibnamefont {Jeng}}, \bibinfo {author} {\bibfnamefont {H.}~\bibnamefont
  {Lin}}, \ and\ \bibinfo {author} {\bibfnamefont {M.~Z.}\ \bibnamefont
  {Hasan}},\ }\href {\doibase 10.1103/PhysRevB.93.121113} {\bibfield  {journal}
  {\bibinfo  {journal} {Phys. Rev. B}\ }\textbf {\bibinfo {volume} {93}},\
  \bibinfo {pages} {121113} (\bibinfo {year} {2016})}\BibitemShut {NoStop}%
\bibitem [{\citenamefont {Chen}\ \emph {et~al.}(2015)\citenamefont {Chen},
  \citenamefont {Lu},\ and\ \citenamefont {Kee}}]{Chen:2015}%
  \BibitemOpen
  \bibfield  {author} {\bibinfo {author} {\bibfnamefont {Y.}~\bibnamefont
  {Chen}}, \bibinfo {author} {\bibfnamefont {Y.-M.}\ \bibnamefont {Lu}}, \ and\
  \bibinfo {author} {\bibfnamefont {H.-Y.}\ \bibnamefont {Kee}},\ }\href
  {\doibase 10.1038/ncomms7593} {\bibfield  {journal} {\bibinfo  {journal}
  {Nat. Commun.}\ }\textbf {\bibinfo {volume} {6}},\ \bibinfo {pages} {6593}
  (\bibinfo {year} {2015})}\BibitemShut {NoStop}%
\bibitem [{\citenamefont {Wieder}\ \emph {et~al.}(2015)\citenamefont {Wieder},
  \citenamefont {Kim}, \citenamefont {Rappe},\ and\ \citenamefont
  {Kane}}]{Wieder:2015}%
  \BibitemOpen
  \bibfield  {author} {\bibinfo {author} {\bibfnamefont {B.~J.}\ \bibnamefont
  {Wieder}}, \bibinfo {author} {\bibfnamefont {Y.}~\bibnamefont {Kim}},
  \bibinfo {author} {\bibfnamefont {A.~M.}\ \bibnamefont {Rappe}}, \ and\
  \bibinfo {author} {\bibfnamefont {C.~L.}\ \bibnamefont {Kane}},\ }\href@noop
  {} {\bibfield  {journal} {\bibinfo  {journal} {ArXiv e-prints}\ } (\bibinfo
  {year} {2015})},\ \Eprint {http://arxiv.org/abs/1512.00074}
  {arXiv:1512.00074} \BibitemShut {NoStop}%
\bibitem [{\citenamefont {Fang}\ \emph {et~al.}(2015)\citenamefont {Fang},
  \citenamefont {Chen}, \citenamefont {Kee},\ and\ \citenamefont
  {Fu}}]{Fang:2015b}%
  \BibitemOpen
  \bibfield  {author} {\bibinfo {author} {\bibfnamefont {C.}~\bibnamefont
  {Fang}}, \bibinfo {author} {\bibfnamefont {Y.}~\bibnamefont {Chen}}, \bibinfo
  {author} {\bibfnamefont {H.-Y.}\ \bibnamefont {Kee}}, \ and\ \bibinfo
  {author} {\bibfnamefont {L.}~\bibnamefont {Fu}},\ }\href {\doibase
  10.1103/PhysRevB.92.081201} {\bibfield  {journal} {\bibinfo  {journal} {Phys.
  Rev. B}\ }\textbf {\bibinfo {volume} {92}},\ \bibinfo {pages} {081201(R)}
  (\bibinfo {year} {2015})}\BibitemShut {NoStop}%
\bibitem [{\citenamefont {Wang}\ \emph {et~al.}(2016)\citenamefont {Wang},
  \citenamefont {Alexandradinata}, \citenamefont {Cava},\ and\ \citenamefont
  {Bernevig}}]{Wang:2016}%
  \BibitemOpen
  \bibfield  {author} {\bibinfo {author} {\bibfnamefont {Z.}~\bibnamefont
  {Wang}}, \bibinfo {author} {\bibfnamefont {A.}~\bibnamefont
  {Alexandradinata}}, \bibinfo {author} {\bibfnamefont {R.~J.}\ \bibnamefont
  {Cava}}, \ and\ \bibinfo {author} {\bibfnamefont {B.~A.}\ \bibnamefont
  {Bernevig}},\ }\href@noop {} {\bibfield  {journal} {\bibinfo  {journal}
  {ArXiv e-prints}\ } (\bibinfo {year} {2016})},\ \Eprint
  {http://arxiv.org/abs/1602.05585} {arXiv:1602.05585} \BibitemShut {NoStop}%
\bibitem [{\citenamefont {Chang}\ \emph {et~al.}(2016)\citenamefont {Chang},
  \citenamefont {Erten},\ and\ \citenamefont {Coleman}}]{Chang:2016}%
  \BibitemOpen
  \bibfield  {author} {\bibinfo {author} {\bibfnamefont {P.-Y.}\ \bibnamefont
  {Chang}}, \bibinfo {author} {\bibfnamefont {O.}~\bibnamefont {Erten}}, \ and\
  \bibinfo {author} {\bibfnamefont {P.}~\bibnamefont {Coleman}},\ }\href@noop
  {} {\bibfield  {journal} {\bibinfo  {journal} {ArXiv e-prints}\ } (\bibinfo
  {year} {2016})},\ \Eprint {http://arxiv.org/abs/1603.03435}
  {arXiv:1603.03435} \BibitemShut {NoStop}%
\bibitem [{\citenamefont {Liang}\ \emph {et~al.}(2016)\citenamefont {Liang},
  \citenamefont {Zhou}, \citenamefont {Yu}, \citenamefont {Wang},\ and\
  \citenamefont {Weng}}]{Liang:2016}%
  \BibitemOpen
  \bibfield  {author} {\bibinfo {author} {\bibfnamefont {Q.-F.}\ \bibnamefont
  {Liang}}, \bibinfo {author} {\bibfnamefont {J.}~\bibnamefont {Zhou}},
  \bibinfo {author} {\bibfnamefont {R.}~\bibnamefont {Yu}}, \bibinfo {author}
  {\bibfnamefont {Z.}~\bibnamefont {Wang}}, \ and\ \bibinfo {author}
  {\bibfnamefont {H.}~\bibnamefont {Weng}},\ }\href {\doibase
  10.1103/PhysRevB.93.085427} {\bibfield  {journal} {\bibinfo  {journal} {Phys.
  Rev. B}\ }\textbf {\bibinfo {volume} {93}},\ \bibinfo {pages} {085427}
  (\bibinfo {year} {2016})}\BibitemShut {NoStop}%
\bibitem [{\citenamefont {Bradlyn}\ \emph {et~al.}(2016)\citenamefont
  {Bradlyn}, \citenamefont {Cano}, \citenamefont {Wang}, \citenamefont {Cava},\
  and\ \citenamefont {Bernevig}}]{Bradlyn:2016}%
  \BibitemOpen
  \bibfield  {author} {\bibinfo {author} {\bibfnamefont {B.}~\bibnamefont
  {Bradlyn}}, \bibinfo {author} {\bibfnamefont {J.}~\bibnamefont {Cano}},
  \bibinfo {author} {\bibfnamefont {Z.}~\bibnamefont {Wang}}, \bibinfo {author}
  {\bibfnamefont {R.~J.}\ \bibnamefont {Cava}}, \ and\ \bibinfo {author}
  {\bibfnamefont {B.~A.}\ \bibnamefont {Bernevig}},\ }\href@noop {} {\bibfield
  {journal} {\bibinfo  {journal} {ArXiv e-prints}\ } (\bibinfo {year}
  {2016})},\ \Eprint {http://arxiv.org/abs/1603.03093} {arXiv:1603.03093}
  \BibitemShut {NoStop}%
\bibitem [{\citenamefont {Weng}\ \emph
  {et~al.}(2015{\natexlab{b}})\citenamefont {Weng}, \citenamefont {Liang},
  \citenamefont {Xu}, \citenamefont {Yu}, \citenamefont {Fang}, \citenamefont
  {Dai},\ and\ \citenamefont {Kawazoe}}]{Weng:2015b}%
  \BibitemOpen
  \bibfield  {author} {\bibinfo {author} {\bibfnamefont {H.}~\bibnamefont
  {Weng}}, \bibinfo {author} {\bibfnamefont {Y.~L.}\ \bibnamefont {Liang}},
  \bibinfo {author} {\bibfnamefont {Q.}~\bibnamefont {Xu}}, \bibinfo {author}
  {\bibfnamefont {R.}~\bibnamefont {Yu}}, \bibinfo {author} {\bibfnamefont
  {Z.}~\bibnamefont {Fang}}, \bibinfo {author} {\bibfnamefont {X.}~\bibnamefont
  {Dai}}, \ and\ \bibinfo {author} {\bibfnamefont {Y.}~\bibnamefont
  {Kawazoe}},\ }\href {\doibase 10.1103/PhysRevB.92.045108} {\bibfield
  {journal} {\bibinfo  {journal} {Phys. Rev. B}\ }\textbf {\bibinfo {volume}
  {92}},\ \bibinfo {pages} {045108} (\bibinfo {year}
  {2015}{\natexlab{b}})}\BibitemShut {NoStop}%
\bibitem [{\citenamefont {Heikkil\"{a}}\ \emph {et~al.}(2011)\citenamefont
  {Heikkil\"{a}}, \citenamefont {Kopnin},\ and\ \citenamefont
  {Volovik}}]{Heikkila:2011}%
  \BibitemOpen
  \bibfield  {author} {\bibinfo {author} {\bibfnamefont {T.~T.}\ \bibnamefont
  {Heikkil\"{a}}}, \bibinfo {author} {\bibfnamefont {N.~B.}\ \bibnamefont
  {Kopnin}}, \ and\ \bibinfo {author} {\bibfnamefont {G.~E.}\ \bibnamefont
  {Volovik}},\ }\href {\doibase 10.1134/S0021364011150045} {\bibfield
  {journal} {\bibinfo  {journal} {JETP Letters}\ }\textbf {\bibinfo {volume}
  {94}},\ \bibinfo {pages} {233} (\bibinfo {year} {2011})}\BibitemShut
  {NoStop}%
\bibitem [{\citenamefont {Young}\ and\ \citenamefont
  {Kane}(2015)}]{Young:2015}%
  \BibitemOpen
  \bibfield  {author} {\bibinfo {author} {\bibfnamefont {S.~M.}\ \bibnamefont
  {Young}}\ and\ \bibinfo {author} {\bibfnamefont {C.~L.}\ \bibnamefont
  {Kane}},\ }\href {\doibase 10.1103/PhysRevLett.115.126803} {\bibfield
  {journal} {\bibinfo  {journal} {Phys. Rev. Lett.}\ }\textbf {\bibinfo
  {volume} {115}},\ \bibinfo {pages} {126803} (\bibinfo {year}
  {2015})}\BibitemShut {NoStop}%
\bibitem [{\citenamefont {Gosálbez-Martínez}\ \emph
  {et~al.}(2015)\citenamefont {Gosálbez-Martínez}, \citenamefont {Souza},\
  and\ \citenamefont {Vanderbilt}}]{Gosablez:2015}%
  \BibitemOpen
  \bibfield  {author} {\bibinfo {author} {\bibfnamefont {D.}~\bibnamefont
  {Gosálbez-Martínez}}, \bibinfo {author} {\bibfnamefont {I.}~\bibnamefont
  {Souza}}, \ and\ \bibinfo {author} {\bibfnamefont {D.}~\bibnamefont
  {Vanderbilt}},\ }\href {\doibase 10.1103/PhysRevB.92.085138} {\bibfield
  {journal} {\bibinfo  {journal} {Phys. Rev. B}\ }\textbf {\bibinfo {volume}
  {92}},\ \bibinfo {pages} {085138} (\bibinfo {year} {2015})}\BibitemShut
  {NoStop}%
\bibitem [{\citenamefont {King-Smith}\ and\ \citenamefont
  {Vanderbilt}(1993)}]{King-Smith:1993}%
  \BibitemOpen
  \bibfield  {author} {\bibinfo {author} {\bibfnamefont {R.~D.}\ \bibnamefont
  {King-Smith}}\ and\ \bibinfo {author} {\bibfnamefont {D.}~\bibnamefont
  {Vanderbilt}},\ }\href {\doibase 10.1103/PhysRevB.47.1651} {\bibfield
  {journal} {\bibinfo  {journal} {Phys. Rev. B}\ }\textbf {\bibinfo {volume}
  {47}},\ \bibinfo {pages} {1651(R)} (\bibinfo {year} {1993})}\BibitemShut
  {NoStop}%
\bibitem [{\citenamefont {Hughes}\ \emph {et~al.}(2011)\citenamefont {Hughes},
  \citenamefont {Prodan},\ and\ \citenamefont {Bernevig}}]{Hughes:2011}%
  \BibitemOpen
  \bibfield  {author} {\bibinfo {author} {\bibfnamefont {T.~L.}\ \bibnamefont
  {Hughes}}, \bibinfo {author} {\bibfnamefont {E.}~\bibnamefont {Prodan}}, \
  and\ \bibinfo {author} {\bibfnamefont {B.~A.}\ \bibnamefont {Bernevig}},\
  }\href {\doibase 10.1103/PhysRevB.83.245132} {\bibfield  {journal} {\bibinfo
  {journal} {Phys. Rev. B}\ }\textbf {\bibinfo {volume} {83}},\ \bibinfo
  {pages} {245132} (\bibinfo {year} {2011})}\BibitemShut {NoStop}%
\bibitem [{\citenamefont {Rhim}\ and\ \citenamefont {Kim}(2015)}]{Rhim:2015}%
  \BibitemOpen
  \bibfield  {author} {\bibinfo {author} {\bibfnamefont {J.-W.}\ \bibnamefont
  {Rhim}}\ and\ \bibinfo {author} {\bibfnamefont {Y.~B.}\ \bibnamefont {Kim}},\
  }\href {\doibase 10.1103/PhysRevB.92.045126} {\bibfield  {journal} {\bibinfo
  {journal} {Phys. Rev. B}\ }\textbf {\bibinfo {volume} {92}},\ \bibinfo
  {pages} {045126} (\bibinfo {year} {2015})}\BibitemShut {NoStop}%
\bibitem [{\citenamefont {Bradley}\ and\ \citenamefont
  {Cracknell}(1972)}]{Bradley:1972}%
  \BibitemOpen
  \bibfield  {author} {\bibinfo {author} {\bibfnamefont {C.~J.}\ \bibnamefont
  {Bradley}}\ and\ \bibinfo {author} {\bibfnamefont {A.~P.}\ \bibnamefont
  {Cracknell}},\ }\href@noop {} {\emph {\bibinfo {title} {The {M}athematical
  {T}heory of {S}ymmetry in {S}olids}}}\ (\bibinfo  {publisher} {Clarendon
  Press},\ \bibinfo {address} {Oxford},\ \bibinfo {year} {1972})\BibitemShut
  {NoStop}%
\bibitem [{\citenamefont {Aroyo}\ \emph {et~al.}(2006)\citenamefont {Aroyo},
  \citenamefont {Perez-Mato}, \citenamefont {Capillas}, \citenamefont
  {Kroumova}, \citenamefont {Ivantchev}, \citenamefont {Madariaga},
  \citenamefont {Kirov},\ and\ \citenamefont {Wondratschek}}]{Aroyo:2006}%
  \BibitemOpen
  \bibfield  {author} {\bibinfo {author} {\bibfnamefont {M.~I.}\ \bibnamefont
  {Aroyo}}, \bibinfo {author} {\bibfnamefont {J.~M.}\ \bibnamefont
  {Perez-Mato}}, \bibinfo {author} {\bibfnamefont {C.}~\bibnamefont
  {Capillas}}, \bibinfo {author} {\bibfnamefont {E.}~\bibnamefont {Kroumova}},
  \bibinfo {author} {\bibfnamefont {S.}~\bibnamefont {Ivantchev}}, \bibinfo
  {author} {\bibfnamefont {G.}~\bibnamefont {Madariaga}}, \bibinfo {author}
  {\bibfnamefont {A.}~\bibnamefont {Kirov}}, \ and\ \bibinfo {author}
  {\bibfnamefont {H.}~\bibnamefont {Wondratschek}},\ }\href {\doibase
  10.1524/zkri.2006.221.1.15} {\bibfield  {journal} {\bibinfo  {journal} {Z.
  Krist}\ }\textbf {\bibinfo {volume} {221}},\ \bibinfo {pages} {15} (\bibinfo
  {year} {2006})}\BibitemShut {NoStop}%
\bibitem [{\citenamefont {Gao}\ \emph {et~al.}(2015)\citenamefont {Gao},
  \citenamefont {Hua}, \citenamefont {Zhang},\ and\ \citenamefont
  {Zhang}}]{Gao:2015}%
  \BibitemOpen
  \bibfield  {author} {\bibinfo {author} {\bibfnamefont {Z.}~\bibnamefont
  {Gao}}, \bibinfo {author} {\bibfnamefont {M.}~\bibnamefont {Hua}}, \bibinfo
  {author} {\bibfnamefont {H.}~\bibnamefont {Zhang}}, \ and\ \bibinfo {author}
  {\bibfnamefont {X.}~\bibnamefont {Zhang}},\ }\href@noop {} {\bibfield
  {journal} {\bibinfo  {journal} {ArXiv e-prints}\ } (\bibinfo {year}
  {2015})},\ \Eprint {http://arxiv.org/abs/1507.07504} {arXiv:1507.07504}
  \BibitemShut {NoStop}%
\bibitem [{\citenamefont {Bzdu\v{s}ek}\ \emph {et~al.}(2016)\citenamefont
  {Bzdu\v{s}ek}, \citenamefont {Wu},\ and\ \citenamefont
  {Soluyanov}}]{Bzdusek:2016}%
  \BibitemOpen
  \bibfield  {author} {\bibinfo {author} {\bibfnamefont {T.}~\bibnamefont
  {Bzdu\v{s}ek}}, \bibinfo {author} {\bibfnamefont {Q.}~\bibnamefont {Wu}}, \
  and\ \bibinfo {author} {\bibfnamefont {A.~A.}\ \bibnamefont {Soluyanov}},\
  }\href@noop {} {\bibfield  {journal} {\bibinfo  {journal} {in preparation}\ }
  (\bibinfo {year} {2016})}\BibitemShut {NoStop}%
\bibitem [{\citenamefont {Rao}\ \emph {et~al.}(1976)\citenamefont {Rao},
  \citenamefont {Tressaud},\ and\ \citenamefont {Bartlett}}]{Rao:1976}%
  \BibitemOpen
  \bibfield  {author} {\bibinfo {author} {\bibfnamefont {P.~R.}\ \bibnamefont
  {Rao}}, \bibinfo {author} {\bibfnamefont {A.}~\bibnamefont {Tressaud}}, \
  and\ \bibinfo {author} {\bibfnamefont {N.}~\bibnamefont {Bartlett}},\ }\href
  {\doibase 10.1016/0022-1902(76)80588-X} {\bibfield  {journal} {\bibinfo
  {journal} {Inorg. Nucl. Chem.}\ }\textbf {\bibinfo {volume} {28}},\ \bibinfo
  {pages} {23} (\bibinfo {year} {1976})}\BibitemShut {NoStop}%
\bibitem [{\citenamefont {Heikkil\"{a}}\ and\ \citenamefont
  {Volovik}(2015)}]{Heikkila:2015}%
  \BibitemOpen
  \bibfield  {author} {\bibinfo {author} {\bibfnamefont {T.~T.}\ \bibnamefont
  {Heikkil\"{a}}}\ and\ \bibinfo {author} {\bibfnamefont {G.~E.}\ \bibnamefont
  {Volovik}},\ }\href@noop {} {\bibfield  {journal} {\bibinfo  {journal} {ArXiv
  e-prints}\ } (\bibinfo {year} {2015})},\ \Eprint
  {http://arxiv.org/abs/1504.05824} {arXiv:1504.05824} \BibitemShut {NoStop}%
\bibitem [{\citenamefont {Strocov}\ \emph {et~al.}(2010)\citenamefont
  {Strocov}, \citenamefont {Schmitt}, \citenamefont {Flechsig}, \citenamefont
  {Schmidt}, \citenamefont {Imhof}, \citenamefont {Chen}, \citenamefont
  {Raabe}, \citenamefont {Betemps}, \citenamefont {Zimoch}, \citenamefont
  {Krempasky}, \citenamefont {Wang}, \citenamefont {Grioni}, \citenamefont
  {Piazzalunga},\ and\ \citenamefont {Patthey}}]{Strocov:2010}%
  \BibitemOpen
  \bibfield  {author} {\bibinfo {author} {\bibfnamefont {V.~N.}\ \bibnamefont
  {Strocov}}, \bibinfo {author} {\bibfnamefont {T.}~\bibnamefont {Schmitt}},
  \bibinfo {author} {\bibfnamefont {U.}~\bibnamefont {Flechsig}}, \bibinfo
  {author} {\bibfnamefont {T.}~\bibnamefont {Schmidt}}, \bibinfo {author}
  {\bibfnamefont {A.}~\bibnamefont {Imhof}}, \bibinfo {author} {\bibfnamefont
  {Q.}~\bibnamefont {Chen}}, \bibinfo {author} {\bibfnamefont {J.}~\bibnamefont
  {Raabe}}, \bibinfo {author} {\bibfnamefont {R.}~\bibnamefont {Betemps}},
  \bibinfo {author} {\bibfnamefont {D.}~\bibnamefont {Zimoch}}, \bibinfo
  {author} {\bibfnamefont {J.}~\bibnamefont {Krempasky}}, \bibinfo {author}
  {\bibfnamefont {X.}~\bibnamefont {Wang}}, \bibinfo {author} {\bibfnamefont
  {M.}~\bibnamefont {Grioni}}, \bibinfo {author} {\bibfnamefont
  {A.}~\bibnamefont {Piazzalunga}}, \ and\ \bibinfo {author} {\bibfnamefont
  {L.}~\bibnamefont {Patthey}},\ }\href {\doibase 10.1107/S0909049510019862}
  {\bibfield  {journal} {\bibinfo  {journal} {J. Synchrotron Rad.}\ }\textbf
  {\bibinfo {volume} {17}},\ \bibinfo {pages} {631} (\bibinfo {year}
  {2010})}\BibitemShut {NoStop}%
\bibitem [{\citenamefont {Yudin}\ \emph {et~al.}(2014)\citenamefont {Yudin},
  \citenamefont {Hirschmeier}, \citenamefont {Hafermann}, \citenamefont
  {Eriksson}, \citenamefont {Lichtenstein},\ and\ \citenamefont
  {Katsnelson}}]{Yudin:2014}%
  \BibitemOpen
  \bibfield  {author} {\bibinfo {author} {\bibfnamefont {D.}~\bibnamefont
  {Yudin}}, \bibinfo {author} {\bibfnamefont {D.}~\bibnamefont {Hirschmeier}},
  \bibinfo {author} {\bibfnamefont {H.}~\bibnamefont {Hafermann}}, \bibinfo
  {author} {\bibfnamefont {O.}~\bibnamefont {Eriksson}}, \bibinfo {author}
  {\bibfnamefont {A.~I.}\ \bibnamefont {Lichtenstein}}, \ and\ \bibinfo
  {author} {\bibfnamefont {M.~I.~K.}\ \bibnamefont {Katsnelson}},\ }\href
  {\doibase 10.1103/PhysRevLett.112.070403} {\bibfield  {journal} {\bibinfo
  {journal} {Phys. Rev. Lett.}\ }\textbf {\bibinfo {volume} {112}},\ \bibinfo
  {pages} {070403} (\bibinfo {year} {2014})}\BibitemShut {NoStop}%
\bibitem [{\citenamefont {Kane}\ and\ \citenamefont {Mele}(2005)}]{Kane:2005}%
  \BibitemOpen
  \bibfield  {author} {\bibinfo {author} {\bibfnamefont {C.~L.}\ \bibnamefont
  {Kane}}\ and\ \bibinfo {author} {\bibfnamefont {E.~J.}\ \bibnamefont
  {Mele}},\ }\href {\doibase 10.1103/PhysRevLett.95.146802} {\bibfield
  {journal} {\bibinfo  {journal} {Phys. Rev. Lett.}\ }\textbf {\bibinfo
  {volume} {95}},\ \bibinfo {pages} {146802} (\bibinfo {year}
  {2005})}\BibitemShut {NoStop}%
\bibitem [{\citenamefont {McCarley}\ and\ \citenamefont
  {Boatman}(1963)}]{McCarley:1963}%
  \BibitemOpen
  \bibfield  {author} {\bibinfo {author} {\bibfnamefont {R.~E.}\ \bibnamefont
  {McCarley}}\ and\ \bibinfo {author} {\bibfnamefont {J.~C.}\ \bibnamefont
  {Boatman}},\ }\href {\doibase 10.1021/ic50007a030} {\bibfield  {journal}
  {\bibinfo  {journal} {Inorg. Chem.}\ }\textbf {\bibinfo {volume} {2}},\
  \bibinfo {pages} {547} (\bibinfo {year} {1963})}\BibitemShut {NoStop}%
\bibitem [{\citenamefont {Habermehl}(2010)}]{Habermehl:2010}%
  \BibitemOpen
  \bibfield  {author} {\bibinfo {author} {\bibfnamefont {K.}~\bibnamefont
  {Habermehl}},\ }\emph {\bibinfo {title} {Neue {U}ntersuchungen an
  {H}alogeniden des {N}iobs und {T}antals}},\ \href@noop {} {Ph.D. thesis},\
  \bibinfo  {school} {Universit\"{a}t zu K\"{o}ln} (\bibinfo {year}
  {2010})\BibitemShut {NoStop}%
\bibitem [{\citenamefont {Momma}\ and\ \citenamefont
  {Izumi}(2011)}]{Momma:2011}%
  \BibitemOpen
  \bibfield  {author} {\bibinfo {author} {\bibfnamefont {K.}~\bibnamefont
  {Momma}}\ and\ \bibinfo {author} {\bibfnamefont {F.}~\bibnamefont {Izumi}},\
  }\href {\doibase 10.1107/S0021889811038970} {\bibfield  {journal} {\bibinfo
  {journal} {J. Appl. Cryst.}\ }\textbf {\bibinfo {volume} {44}},\ \bibinfo
  {pages} {1272} (\bibinfo {year} {2011})}\BibitemShut {NoStop}%
\end{thebibliography}%
\bibliographystyle{apsrev4-1}

\newpage
\onecolumngrid



\section*{Supplementary material}

# Supplementary Information file for "Nodal Chain Metals"

Tomáš Bzdušek[1], QuanSheng Wu[1,2], Andreas Rüegg[1], Manfred Sigrist[1], and Alexey A. Soluyanov[1,2,3]

[1]*Institut für Theoretische Physik, ETH Zurich, 8093 Zurich, Switzerland*

[2]*Station Q Zurich, ETH Zurich, 8093 Zurich, Switzerland and*

[3]*Department of Physics, St. Petersburg State University, St. Petersburg, 199034 Russia*

(Dated: August 25, 2016)

## CONTENTS

## A. SYMMETRY CRITERIA FOR THE APPEARANCE OF NON-SYMMORPHIC NODAL LOOPS AND NODAL CHAINS

Here we provide a derivation of the criteria for the appearance of the non-symmorphic nodal loops (NSNLs) and nodal chains presented in the main text. We also discuss the appearance of NSNLs formed by valence or conduction bands, providing real materials where such a situation occurs. Finally, we discuss the NSNLs in antiferromagnetic systems.

### 1. Non-symmorphic nodal loops

We first recall a few definitions. Reciprocal lattice vectors $\boldsymbol{G}$ fulfill

$$\boldsymbol{G} \cdot \boldsymbol{R} = 0 \mod 2\pi \qquad (1)$$

for every Bravais vector $\boldsymbol{R}$. Two momenta are called equivalent if they differ by a reciprocal lattice vector $\boldsymbol{G}$. Time-reversal invariant momenta (TRIMs) $\boldsymbol{\Gamma}$ are equivalent to their time-reversed images, $\boldsymbol{\Gamma} = -\boldsymbol{\Gamma} + \boldsymbol{G}$. Combining this relation with equation (1), we obtain

$$\boldsymbol{\Gamma} \cdot \boldsymbol{R} = 0 \mod \pi. \qquad (2)$$

Similarly, in a mirror-symmetric crystal, a mirror-invariant momentum $\boldsymbol{k}$ is equivalent to its mirror image $\sigma_{\boldsymbol{n}} \boldsymbol{k}$ if

$$2\boldsymbol{n} \left(\boldsymbol{k} \cdot \boldsymbol{n}\right) = \boldsymbol{0} \mod \boldsymbol{G}, \qquad (3)$$

where $\boldsymbol{n}$ is the normal vector of the mirror plane.

A mirror invariant plane contains four inequivalent TRIMs. To see this, pick up two arbitrary reciprocal lattice vectors $\boldsymbol{G}_{1,2}$ such that they both have non-zero in-plane components ($\boldsymbol{G}_{i,\parallel} \neq \boldsymbol{0}$) that are mutually non-collinear ($\boldsymbol{G}_{1,\parallel} \nparallel \boldsymbol{G}_{2,\parallel}$). By symmetry, the mirror images $\sigma_{\boldsymbol{n}} \boldsymbol{G}_i$ are also reciprocal lattice vectors, and hence $2\boldsymbol{G}_{i,\parallel} = \boldsymbol{G}_i + \sigma_{\boldsymbol{n}} \boldsymbol{G}_i$ are *in-plane* reciprocal lattice vectors. Let $\widetilde{\boldsymbol{G}}_i$ be the *shortest* reciprocal lattice vector in the direction $\boldsymbol{G}_{i,\parallel}$. Then, by taking all linear combinations $n_1 \widetilde{\boldsymbol{G}}_1 + n_2 \widetilde{\boldsymbol{G}}_2, n_{1,2} \in \mathbb{Z}$, we construct *all* in-plane reciprocal lattice vectors. These vectors form a grid of equivalent $\boldsymbol{k}$-points. The midpoints of these vectors form a four-times denser mesh of all the in-plane TRIMs. Thus, there are four inequivalent in-plane TRIMs.

We now consider a spin-orbit coupled (SOC) system with only two symmetries: Time-reversal $\Theta$ and a glide plane $g = \{\sigma_{\boldsymbol{n}} | \boldsymbol{t}\}$, i.e. a mirror symmetry $\sigma_{\boldsymbol{n}}$ with respect to normal vector $\boldsymbol{n}$ followed by a translation by $\boldsymbol{t}$. Such a system supports only one-dimensional irreducible representations (IRs) everywhere apart from TRIMs. The square of the glide plane is $g^2 = \left\{ \overline{E} | 2\boldsymbol{t}_{\parallel} \right\}$ where $\boldsymbol{t}_{\parallel} = \frac{1}{2} \left( \boldsymbol{t} + \sigma_{\boldsymbol{n}} \boldsymbol{t} \right)$ is the in-plane component of the translation vector, and $\overline{E}$ is a $2\pi$-rotation, represented by $-\mathbb{1}$ in SOC systems.

It can be seen that $g^2$ corresponds to a pure translation, and hence $2\boldsymbol{t}_{\parallel}$ must be a Bravais vector. Consequently, $g^2$ is represented at $\boldsymbol{k}$ by $-\mathrm{e}^{-i\boldsymbol{k} \cdot 2\boldsymbol{t}_{\parallel}}$ from which we conclude that the possible glide eigenvalues are [1]

$$\eta_{\pm}(\boldsymbol{k}) = \pm i \mathrm{e}^{-i\boldsymbol{k} \cdot \boldsymbol{t}_{\parallel}}. \qquad (4)$$

Since $2\boldsymbol{t}_{\parallel}$ is a Bravais vector, we find from (2) that

$$\boldsymbol{\Gamma} \cdot \boldsymbol{t}_{\parallel} = 0 \mod \frac{\pi}{2}. \qquad (5)$$

| Bravais lattice | $P$ | non-centrosymmetric + glide plane |
|---|---|---|
| Monoclinic P | $C_s$ | #7 |
| Monoclinic C | $C_s$ | #9 |
| Orthorhombic P | $C_{2v}$ | #26, #27, #28, #29, #30, #31, #32, #33, #34 |
| Orthorhombic A,C | | #36, #37, #39, #40, #41 |
| Orthorhombic F | | #43 |
| Orthorhombic I | | #45, #46 |
| Tetragonal P | $C_{4v}$ | #100, #101, #102, #103, #104, #105, #106 |
| | $D_{2d}$ | #112, #114, #116, #117, #118 |
| | | #120, #122 |
| Tetragonal I | $C_{4v}$ | #108, #109, #110 |
| Trigonal R | $C_{3v}$ | #161 |
| Trigonal P | $C_{3v}$ | #158, #159 |
| Hexagonal | $C_{6v}$ | #184, #185, #186 |
| | $D_{3h}$ | #188, #190 |
| Cubic P | $T_d$ | #218 |
| Cubic F | | #219 |
| Cubic I | | #220 |

TABLE I. **Non-centrosymmetric space groups (SGs) with a suitable glide plane.** For every Bravais lattice we identified all those SGs that are non-centrosymmetric and contain a glide plane that fulfills the criterion of Eq. (6). Column $P$ indicates the corresponding point group. The color scheme indicates the number of such glide planes: One = green, two = blue, three = violet, four = red, six = orange.

This leads to two possibilities. Either $\boldsymbol{\Gamma} \cdot \boldsymbol{t}_{\parallel} = 0 \;(\mathrm{mod}\,\pi)$ at all four mirror-invariant TRIMs, in which case the glide eigenvalues (4) are imaginary at all TRIMs and no band crossings are enforced. The second possibility is

$$\boldsymbol{\Gamma}_1 \cdot \boldsymbol{t}_{\parallel} = 0 \quad \mathrm{mod}\ \pi \quad \text{and} \quad \boldsymbol{\Gamma}_2 \cdot \boldsymbol{t}_{\parallel} = \frac{\pi}{2} \quad \mathrm{mod}\ \pi \quad (6)$$

for $\boldsymbol{\Gamma}_{1,2}$ within the mirror-invariant plane. We argue that in such a case $\boldsymbol{\Gamma}_1$ and $\boldsymbol{\Gamma}_2$ have to be separated by a nodal line (NL) located within the miror-invariant plane.

Consider any in-plane path $p$ that connects $\boldsymbol{\Gamma}_1$ to $\boldsymbol{\Gamma}_2$. The Kramer's theorem forces the bands to be doubly degenerate at $\boldsymbol{\Gamma}_{1,2}$, and the trivial commutator $[g, \Theta] = 0$ makes the glide eigenvalues of a Kramer's doublet complex conjugate. According to equations (4) and (6), the glide eigenvalues of a Kramer's doublet at $\boldsymbol{k} = \boldsymbol{\Gamma}_1$ are $(+i, -i)$. As one moves along $p$, the phase of the eigenvalues has to change by

$$(\boldsymbol{\Gamma}_1 - \boldsymbol{\Gamma}_2) \cdot \boldsymbol{t}_{\parallel} = \frac{\pi}{2} \quad \mathrm{mod}\ \pi, \quad (7)$$

so they evolve into $(+1, -1)$ at $\boldsymbol{\Gamma}_2$. However, these are not complex conjugate anymore and hence belong to *different* Kramer's doublets. We infer that Kramer's doublets have to switch partners along $p$, and a band crossing has to appear in the glide-invariant plane. Since the argument holds for *any* in-plane path $p$, there is a *line* of band crossings separating $\boldsymbol{\Gamma}_{1,2}$ within the plane: the NSNL.

In the absence of additional symmetries, the band structure is built up of the quadruplets illustrated in Fig. 2b of the main text. Simple electron counting suggests that the NSNL if formed by occupied and unoccupied whenever there are

$$\nu_{\text{filled}} = 4n + 2, \quad n \in \mathbb{N}, \quad (8)$$

filled bands (see Section A 4 for the discussion of the stability of NSNLs at the Fermi level). Additional symmetries may lead to additional spectrum degeneracies and change the above filling criterion. For example, in the presence of inversion symmetry $\mathcal{I}$ the whole argument becomes not applicable, since $\Theta \circ \mathcal{I}$ leads to spin degeneracy of all bands throughout the Brillouin zone (BZ). Table I lists all 47 non-centrosymmetric space groups (SGs) that contain a glide plane fulfilling criterion of Eq. (6).

### 2. Drumhead surface states

It is known [2] that the presence of accidental NL (ANL) leads to the existence of a drumhead surface state that resides within the projection of the ANL to the surface. Here we argue that this is also the case for NSNLs.

To see this, note that in the presence of spin-orbit coupling ANLs appear in mirror-invariant planes of the Brillouin zone, while, as argued above, the NSNLs reside in glide-symmetric planes. Let us assume that ANL or NSNL is formed by bands $N$ and $N+1$. A topological invariant defining the presence of a NL exists for both ANLs [3] and NSNLs. Indeed, a mirror (glide) symmetry leads to the quantization of the Berry phase accumulated by $N$ lower lying bands around any mirror- (glide-) symmetric loop enclosing the ANL (NSNL) into values 0 or $\pi$. More precisely, the Berry phase on such a loop is equal to $\pi$ times the number of NLs enclosed by the loop.

Due to the periodicity of $\boldsymbol{k}$-space, a particular example of such a loop is a one-dimensional straight line threading across the BZ, orthogonal to the glide plane. For 1D insulators with spatial symmetries quantizing the Berry phase to either 0 or $\pi$ (like mirror or glide), it is known that in the latter case a metallic end mode of half an electron charge has to exist [4, 5]. Considering a 3D crystal as a collection of coupled 1D chains, one arrives at the conclusion that the projection of the NSNL on a crystal surface has to enclose a topological metallic surface state, which is similar to the drumhead state of ANLs [2].

### 3. Nodal chains

We further focus on systems with (1) a *pair* of mutually orthogonal glide planes $g_{1,2} = \{\sigma_{1,2}|\boldsymbol{t}_{1,2}\}$. According to the discussion above, each of them may create a NSNL in a mirror-invariant plane. These NSNLs *necessarily touch* if (2) the suitable choice of the TRIMs $\boldsymbol{\Gamma}_{1,2}$ fulfilling Eq. (6) for both $\boldsymbol{t}_{1,2}$ coincides with the two TRIMs along an intersection of the two mirror-symmetric planes. Moreover, (3) the high-symmetry line connecting

$\mathbf{\Gamma}_1$ and $\mathbf{\Gamma}_2$ should support *at least two* 1D IRs in the $\Theta$-symmetric case for the presented eigenvalue argument to be valid. Of the SGs listed in Table I, only nine satisfy criteria (1-3). Eight of them preserve the filling condition of Eq. (8) and are listed in Fig. 1 of the main text. The additional space group #110 ($I4_1cd$) is more complicated since the minimal number of detached bands is eight rather than four [6, 7], and because it supports only 2D IRs at $\mathbf{P} = (\mathbf{\Gamma}_1 + \mathbf{\Gamma}_2)/2$ [8]. This leads to a rather complicated nodal line structure illustrated in Fig. 1b.

## 4. Tuning non-symmorphic nodal loops and nodal chains to the Fermi energy

While the appearance of the NSNLs formed by conduction and valence bands requires the filling of Eq. (8), the experimental observation of the transport properties of NSNLs and nodal chains would require their proximity to the Fermi level. Notice that in general in metals the condition of Eq. (8) does not guarantee this proximity. Indeed, additional carrier pockets can provide the necessary charge compensation in the undoped materials even when the topologically protected degeneracy is away from the Fermi level.

For example, in nodal chain metals, such pockets would generally appear around the double Weyl point, accompanying the nodal chain. Thus, candidate materials for experimental probes of nodal chains and NSNLs should be chosen such that the nodal loops are realized reasonably close to the Fermi level, like in the case of IrF$_4$. In other materials, some additional fine tuning of the nodal loops location relative to $E_F$ may be required. This fine tuning can be done by straining, applying pressure or doping the compound.

Moreover, in materials with additional Fermi pockets a nodal chain or a NSNL can be probed experimentally even when the condition of Eq. (8) is not fulfilled. For $\nu_{\text{filled}} = 4n$ a NSNL or a nodal chain formed by conduction *or* valence bands can still have drastic effects on transport provided that it appears sufficiently close to the Fermi level. Examples of such materials are provided by the CuTlSe$_2$ material class of space group #122 ($I\bar{4}2md$). These materials were recently predicted to host Weyl points occurring between the valence and conduction bands [9], but the presence of the NSNLs formed by conduction or valence bands was missed. Since in many of these compounds the NSNLs are located close to the Fermi level, we expect both the Weyl points and the NSNLs to be relevant for transport properties of these materials.

## 5. Generalization to antiferromagnets

Let us finally generalize the argument for the appearance of a NSNL to antiferomagnetic (AFM) systems. For such systems the time reversal $\Theta$ is not a good symmetry anymore, but its "non-symmorphic" analogue $\mathcal{T} = \{\Theta|\boldsymbol{t}_\Theta\}$ *is*, where $\boldsymbol{t}_\Theta$ is a vector connecting the two spin-polarized sublattices of the AFM. We denote the glide plane as $g = \{\sigma_{\boldsymbol{n}}|\boldsymbol{t}_\sigma\}$. The two operators square to $\mathcal{T}^2 = \{\bar{E}|2\boldsymbol{t}_\Theta\}$ and $g^2 = \{\bar{E}|2\boldsymbol{t}_{\sigma,\|}\}$, thus being equivalent to translations by Bravais vectors $2\boldsymbol{t}_\Theta$ and $2\boldsymbol{t}_{\sigma,\|}$ correspondingly. The action of the squares of these operators on Bloch wave functions at $\boldsymbol{k}$ is

$$\mathcal{T}^2(\boldsymbol{k}) = -\mathrm{e}^{-2\mathrm{i}\boldsymbol{k}\cdot\boldsymbol{t}_\Theta}\mathbb{1} \quad \text{and} \quad g^2(\boldsymbol{k}) = -\mathrm{e}^{-2\mathrm{i}\boldsymbol{k}\cdot\boldsymbol{t}_\sigma}\mathbb{1}, \quad (9)$$

being equal to $\pm\mathbb{1}$ at mirror-invariant TRIMs. The commutation relation between the operators is

$$\mathcal{T} \circ g = \lambda(\boldsymbol{k})g \circ \mathcal{T} \qquad (10)$$

where

$$\lambda(\boldsymbol{k}) = \mathrm{e}^{-2\mathrm{i}\boldsymbol{k}_\perp \cdot \boldsymbol{t}_\Theta} \qquad (11)$$

and $\boldsymbol{k}_\perp$ denotes the momentum component orthogonal to the mirror plane. We observe from (11) that at TRIMs the operators either commute or anticommute.

To guarantee a two-fold degeneracy at a TRIM $\mathbf{\Gamma}$, its little co-group has to contain an antiunitary operator commuting with $\mathcal{H}(\boldsymbol{k})$ and squaring to $-\mathbb{1}$, i.e. either $\mathcal{T}^2(\mathbf{\Gamma}) = -\mathbb{1}$ or

$$(g \circ \mathcal{T})^2 = \lambda(\mathbf{\Gamma})g^2(\mathbf{\Gamma})\mathcal{T}^2(\mathbf{\Gamma}) = -\mathbb{1}. \qquad (12)$$

We say that a TRIM $\mathbf{\Gamma}$ is type-$n$ if it contains $n$ such operators. All the possibilities are summarized in Table II. The possible glide eigenvalues at a TRIM $\mathbf{\Gamma}$ are

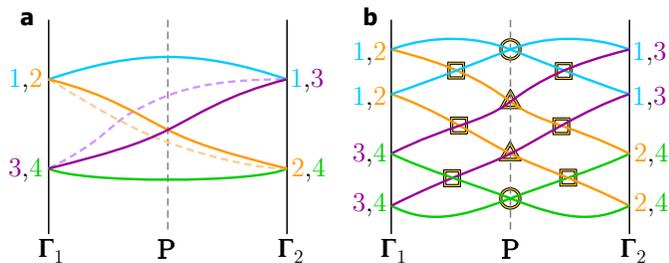

FIG. 1. **Nodal chains along the high-symmetry line.** (a) A generic spectrum along the high-symmetry line connecting $\mathbf{\Gamma}_1$ and $\mathbf{\Gamma}_2$ in nodal chain metals. The labels 1 to 4 (also distinguished by colors) correspond to four different irreducible representations (IRs) characterized by pairs of glide eigenvalues. Bands appear in quadruplets, and Kramer's doublets at $\mathbf{\Gamma}_{1,2}$ switch partners along the way. The imposed band crossing is the touching point of the two NSNLs. In groups #109 and #122, the crossing is pinned to $\mathbf{P} = (\mathbf{\Gamma}_1 + \mathbf{\Gamma}_2)/2$. (b) Space group #110 also supports only one-dimensional IRs along the high-symmetry line, but forces pairs of IRs $1 + 1$, $2 + 3$ and $4 + 4$ to become degenerate at $\mathbf{P}$, leading to band octuplets. There are many imposed nodal lines: The squares indicate single NLs, the circles correspond to crossing of straight NLs along high-symmetry lines, and triangles correspond to a pair of NSNLs crossed by a pair of straight NLs along high-symmetry lines.

$\eta_\pm(\mathbf{\Gamma}) = \pm\sqrt{g^2(\mathbf{\Gamma})}$, and the (anti)commutation relation of Eq. (10) forces the Kramer's doublets to carry eigenvalues

$$(\eta_1, \eta_2)_\pm = \pm\left(\sqrt{g^2(\mathbf{\Gamma})}, \lambda(\mathbf{\Gamma})\sqrt{g^2(\mathbf{\Gamma})}^*\right). \quad (13)$$

Since all glide eigenvalues change their phase along an in-plane path $p$ connecting $\mathbf{\Gamma}_1$ and $\mathbf{\Gamma}_2$ by the *same* increment $\Delta\varphi = (\mathbf{\Gamma}_1 - \mathbf{\Gamma}_2) \cdot \mathbf{t}_{\sigma,\parallel}$, a NSNL has to appear if $\eta_1 = \eta_2$ at one and $\eta_1 = -\eta_2$ at the other of the TRIMs. This is equivalent to $g^2(\mathbf{\Gamma})\lambda(\mathbf{\Gamma})$ being positive at one and negative at the other TRIM. We observe in Table II that this occurs if one TRIM is type-1 and the other type-2.

## B. A RELATION BETWEEN NON-SYMMORPHIC NODAL LOOPS AND NON-SYMMORPHIC DIRAC POINTS

Introducing inversion symmetry $\mathcal{I}$ into a time-reversal symmetric SOC system results in spin degeneracy of all bands. Thus, if the inversion-breaking terms in the Hamiltonian of a NSNL are taken to zero, the NSNL shrinks into a four-fold degeneracy at either $\mathbf{\Gamma}_1$ or $\mathbf{\Gamma}_2$. In a system with time-reversal and glide-plane symmetries this four-fold degeneracy corresponds to one of the non-symmorphic Dirac points (NSDP) classified in Ref. [10] [or to a Dirac line (that is, a line of four-fold degeneracy) along a high-symmetry axis], uncovering a relation between NSNLs and NSDPs.

Here we consider the following question: Is it *always* possible to obtain a NSNL by a suitable distortion of a system possessing a NSDP? We show in the following that there are situations when it is *not* possible to distort a NSDP into a NSNL.

As shown in the work of Ref. [10], a NSDP located at a TRIM on the BZ boundary requires a simultaneous presence of time-reversal $\Theta$, inversion $\mathcal{I}$, and an

| $\mathbf{\Gamma}$-type | $\mathcal{T}^2$ | $g^2$ | $\lambda$ | $(g \circ \mathcal{T})^2$ | eigenvalue pairs |
|---|---|---|---|---|---|
| type-0 | + | + | + | + | no pairs |
|  | + | − | − | + |  |
| type-1 | + | + | − | − | $(1, -1)$ |
|  | − | + | − | + |  |
|  | − | − | + | + | $(\mathrm{i}, -\mathrm{i})$ |
|  | − | − | + | + |  |
| type-2 | − | + | + | − | $(1,1)$ and $(-1,-1)$ |
|  | − | + | + | − | $(\mathrm{i},\mathrm{i})$ and $(-\mathrm{i},-\mathrm{i})$ |

TABLE II. **Non-symmorphic nodal loops (NSNLs) in antiferromagnets.** We classify a time-reversal invariant momentum (TRIM) $\mathbf{\Gamma}$ as being of type-$n$ if its little co-group contains $n$ antiunitary operators that square to $-1$ and commute with $\mathcal{H}(\mathbf{\Gamma})$. Two TRIMs $\mathbf{\Gamma}_{1,2}$ located within the same glide-invariant plane are separated by NSNL for filling factors (8) provided that one of them is type-1 and the other type-2. The paramagnetic situation considered previously corresponds to combining rows 6 and 7 of the table.

even $n$-fold rotational symmetry $C_{nz}$ (set along the $z$-axis by convention). For low order expansion in $\mathbf{k}$, the non-symmorphic symmetry matrices can be considered to be $\mathbf{k}$-independent, so that rotation and screw axes become identical for low order $\mathbf{k} \cdot \mathbf{p}$ expansions. The point group corresponding to these symmetries then contains a twofold rotational symmetry

$$C_{2z} = (C_{nz})^{n/2} \quad (14)$$

and a mirror symmetry

$$\sigma_z = \mathcal{I} \circ C_{2z}. \quad (15)$$

The NSDP appears in the mirror-invariant plane (again, there is no difference between a mirror and a glide for low order $\mathbf{k} \cdot \mathbf{p}$ expansions, in contrast to tight-binding models). Breaking $C_{2z}$ while keeping $\sigma_z$ allows for the appearance of a NSNL (note that breaking $C_{2z}$ also breaks inversion and lifts the spin degeneracy of bands). To find how such a crystal distortion modifies the spectrum, we consider additional terms appearing in the Dirac Hamiltonian that are consistent with the symmetries left.

The Dirac Hamiltonian is

$$\mathcal{H}_{\text{Dirac}}^{(s)} = \hbar\left[v_x\frac{(k_-^s + k_+^s)}{2}\gamma_1 + v_y\frac{\mathrm{i}(k_-^s - k_+^s)}{2}\gamma_2 + v_z k_z\gamma_3\right] \quad (16)$$

where $k_\pm^s = (k_x \pm \mathrm{i}k_y)^s$, and $\gamma_1$ to $\gamma_5$ are mutually anti-commuting Hermitian matrices squaring to $\mathbb{1}$. Together with the commutators $\gamma_{ij} = -\frac{1}{2}[\gamma_i, \gamma_j]$ they form a basis of traceless $4 \times 4$ Hermitian matrices. The value of the parameter $s = 1$ ($s = 3$) corresponds to *linear* (*cubic*) Dirac Hamiltonians. Terms in the Hamiltonian that are independent of $\mathbf{k}$ and consistent with $\Theta$ and $\sigma_z$ can only be proportional to those of the 15 Dirac matrices that commute with both of the symmetries. For $\Theta$ we find [3]

$$[\Theta, \gamma_a] = 0 \quad \text{for} \quad a \in \{4, 14, 24, 34, 45\} \quad (17)$$

and $\{\Theta, \gamma_a\} = 0$ for the remaining Dirac matrices. The classification developed in Ref. [10] allows us to check the anti-/commutation relations of the operator $\sigma_z$ for all possible NSDPs, which we present in Table III.

We find that for a NSDP located along a 2-fold or 6-fold rotation axes

$$\mathcal{H}_{\text{pert.}}^{(n=2,6)} = w\gamma_{34}, \quad (18)$$

leading to a NSNL in the mirror-invariant plane [3]. The bands forming the NSNL have different eigenvalues of $\sigma_z$, hence the nodal loop is stable against including higher-order symmetry-preserving perturbations to the Hamiltonian. This situation corresponds to, both, the NSDPs in $\beta$-cristobalite $BiO_2$ [11] and in distorted spinels $BaXSiO_4$ [12]. Such a distortion might be induced by a structural phase transition at low temperature [13] or by strain.

For NSDPs that occur on 4-fold rotational axes, the corresponding perturbation is

$$\mathcal{H}_{\text{pert.}}^{(n=4)} = m\gamma_4 + u_x\gamma_{14} + u_y\gamma_{24} \quad (19)$$

| $C_{nz}$ | $u_{A\uparrow}$ | type | anti-/commutes with $\sigma_z$ | | | | | | | | |
|---|---|---|---|---|---|---|---|---|---|---|---|
| | | | $\gamma_1$ | $\gamma_2$ | $\gamma_3$ | $\gamma_4$ | $\gamma_5$ | $\gamma_{14}$ | $\gamma_{24}$ | $\gamma_{34}$ | $\gamma_{45}$ |
| $C_{2z}$ | $\pm\exp\left[\mathrm{i}\frac{\pi}{2}\right]$ | linear | C | C | A | A | C | A | A | C | A |
| $C_{4z}$ | $\pm\exp\left[\pm\mathrm{i}\frac{\pi}{4}\right]$ | linear | C | C | A | C | A | C | C | A | A |
| $C_{6z}$ | $\pm\exp\left[\pm\mathrm{i}\frac{\pi}{6}\right]$ | linear | C | C | A | A | C | A | A | C | A |
| $C_{6z}$ | $\pm\exp\left[\mathrm{i}\frac{\pi}{2}\right]$ | cubic | C | C | A | A | C | A | A | C | A |

TABLE III. **Relation of non-symmorphic nodal loops to Dirac points.** The first three columns reproduce the classification of non-symmorphic Dirac points developed in Ref. [10], where $u_{A\uparrow}$ encodes $C_{nz}$ eigenvalues of the four states degenerate at the nodal point and the third column indicates the type of dispersion around the nodal point. We used the $\boldsymbol{k} \cdot \boldsymbol{p}$ models developed in Ref. [10] to check which of the $\Theta$-symmetric matrices (17) are also $\sigma_z$-symmetric. Those labelled C (A) in the orange columns (anti)commute with the mirror and are therefore (not) present in the non-centrosymmetric $\boldsymbol{k} \cdot \boldsymbol{p}$ expansions (18) and (19).

and the spectrum is gapped whenever $m \neq 0$, therefore not leading to a NSNL in general [3]. In the presence of an additional mirror plane *parallel* to the rotation axis, like in the case of $\beta$-cristobalite $\mathrm{BiO}_2$ [11], a NSNL can still appear in this plane. However, the existence of such a mirror is not in general guaranteed, examples being the A and Z points of the simple tetragonal space group #84.

Combining Eqs. (16) and (18) we find a $\boldsymbol{k} \cdot \boldsymbol{p}$ expansion for a NSNL to be

$$\mathcal{H}_{\mathrm{NSNL}}(\boldsymbol{k}) = \hbar\left[v_x k_x \gamma_1 + v_y k_y \gamma_2 + v_z k_z \gamma_3\right] + w\gamma_{34}. \quad (20)$$

Using this Hamiltonian we can analytically compute the topological $\mathbb{Z}_2$ charge of the NSNL defined as

$$z_2(c) = \frac{1}{\pi} \oint_c \mathrm{d}\boldsymbol{k} \cdot \boldsymbol{\mathcal{A}}(\boldsymbol{k}) \mod 2. \quad (21)$$

Here $c$ is a mirror-symmetric path enclosing the NSNL and $\boldsymbol{\mathcal{A}}(\boldsymbol{k})$ is the Berry connection of the occupied bands. Choosing a specific set of Dirac matrices

$$\gamma_1 = -\tau^z \sigma^y \qquad \gamma_2 = \tau^z \sigma^x \qquad \gamma_3 = \tau^y$$
$$\gamma_4 = \tau^x \qquad \gamma_5 = \tau^z \sigma^z. \quad (22)$$

consistent with $\Theta = \mathrm{i}\sigma^y \mathcal{K}$ (where $\mathcal{K}$ is the complex conjugation), and a path enclosing the NSNL

$$c = \left\{ \left( \frac{w + u\cos\alpha}{\hbar v_x}, 0, \frac{u\sin\alpha}{\hbar v_z} \right) \bigg| \alpha \in [0, 2\pi) \right\}. \quad (23)$$

We find smoothly gauged valence band wave functions

$$|\psi(\alpha)\rangle = \frac{\mathrm{e}^{-\mathrm{i}\alpha/2}}{\sqrt{2}} \left( -\sin\tfrac{\alpha}{2}, \mathrm{i}\sin\tfrac{\alpha}{2}, \mathrm{i}\cos\tfrac{\alpha}{2}, \cos\tfrac{\alpha}{2} \right)^\top. \quad (24)$$

The corresponding Berry connection is

$$\mathcal{A}_\alpha = \mathrm{i} \langle\psi(\alpha)| \partial_\alpha |\psi(\alpha)\rangle = \frac{1}{2} \quad (25)$$

which, upon integration over $\alpha$, yields the topologically non-trivial value $z_2(c) = 1$, as expected for the NSNL.

## C. LANDAU LEVELS OF NON-SYMMORPHIC NODAL LOOPS

We consider the effects of magnetic field on the $\boldsymbol{k} \cdot \boldsymbol{p}$ Hamiltonian of a NSNL presented in Eq. (20).

### 1. Out-of-plane field

We first consider a *perpendicular* magnetic field $\boldsymbol{B} = (0, 0, B)$ with $B > 0$. The Peierls substitution gives

$$(\hbar k_x, \hbar k_y, \hbar k_z) \mapsto (-\mathrm{i}\hbar\partial_x + eA_x, -\mathrm{i}\hbar\partial_y + eA_y, \hbar k_z)$$
$$\equiv (\Pi_x, \Pi_y, \hbar k_z) \quad (26)$$

so that the commutator

$$[\Pi_x, \Pi_y] = \mathrm{i}\hbar e\left(\partial_y A_x - \partial_x A_y\right) = -\mathrm{i}\hbar eB. \quad (27)$$

This allows us to express the $\Pi$-operators using the usual ladder operators $[a, a^\dagger] = 1$ as

$$\Pi_x = \sqrt{\frac{e\hbar v_y B}{2v_x}}\left(a + a^\dagger\right), \ \ \Pi_y = \mathrm{i}\sqrt{\frac{e\hbar v_x B}{2v_y}}\left(a - a^\dagger\right). \quad (28)$$

Adopting Dirac matrices of Eq. (22) and representing $\boldsymbol{\tau}$-matrices as $2 \times 2$ blocks and $\boldsymbol{\sigma}$-matrices as elements in the blocks, we obtain the matrix Hamiltonian

$$\mathcal{H}(k_z) = \begin{pmatrix} -w & \mathrm{i}\sqrt{b}\,a & -\mathrm{i}\hbar v_z k_z & 0 \\ -\mathrm{i}\sqrt{b}a^\dagger & -w & 0 & -\mathrm{i}\hbar v_z k_z \\ \mathrm{i}\hbar v_z k_z & 0 & w & -\mathrm{i}\sqrt{b}\,a \\ 0 & \mathrm{i}\hbar v_z k_z & \mathrm{i}\sqrt{b}a^\dagger & w \end{pmatrix} \quad (29)$$

where we introduced a rescaled field $b = 2v_y v_z e\hbar B$. When squared, this Hamiltonian has a block-diagonal form

$$\widetilde{\mathcal{H}}^2 = \begin{pmatrix} b\left(a^\dagger a + 1\right) & -2\mathrm{i}w\sqrt{b}a & 0 & 0 \\ 2\mathrm{i}w\sqrt{b}a^\dagger & ba^\dagger a & 0 & 0 \\ 0 & 0 & b\left(a^\dagger a + 1\right) & -2\mathrm{i}w\sqrt{b}a \\ 0 & 0 & 2\mathrm{i}w\sqrt{b}a^\dagger & ba^\dagger a \end{pmatrix}$$

where $\widetilde{\mathcal{H}}^2 = \mathcal{H}^2(k_z) - \left(w^2 + \hbar^2 v_z^2 k_z^2\right)\mathbb{1}_\tau \otimes \mathbb{1}_\sigma$. Clearly, the eigenvalues of $\mathcal{H}(k_z)$ and of $\widetilde{\mathcal{H}}^2$ are related through $\varepsilon^2(k_z) = \tilde{\varepsilon}^2 + w^2 - \hbar^2 v_z^2 k_z^2$.

The eigenfunctions (that is, Landau levels) of $\mathcal{H}$ are found in the form

$$|\widetilde{\psi}\rangle = \begin{pmatrix} \sum_{n=0}^{+\infty} \widetilde{\psi}_n^{+,+} |n\rangle \\ \sum_{n=0}^{+\infty} \widetilde{\psi}_n^{+,-} |n\rangle \\ \sum_{n=0}^{+\infty} \widetilde{\psi}_n^{-,+} |n\rangle \\ \sum_{n=0}^{+\infty} \widetilde{\psi}_n^{-,-} |n\rangle \end{pmatrix}.. \quad (30)$$

where $|n\rangle$ is the eigenstate of a number operator $a^\dagger a|n\rangle = n|n\rangle$. Using the standard algebra of the ladder operators

$$a|n\rangle = \sqrt{n}|n-1\rangle, \quad a^\dagger|n\rangle = \sqrt{n+1}|n+1\rangle, \quad (31)$$

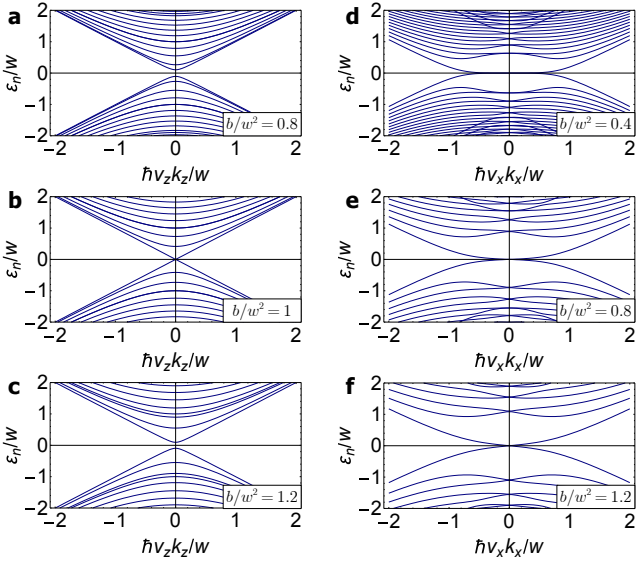

FIG. 2. **Landau levels of a non-symmorphic nodal loop.** (a-c) Landau levels of Hamiltonian (20) for a perpendicular magnetic field are always gapped except when condition (39) is fulfilled, which is the case for (b). (d-f) Gapless Landau levels for an in-plane magnetic field. The zero energy crossing is protected by symmetry as explained in the text.

we find the eigenstates of $\widetilde{\mathcal{H}}^2$ corresponding to the eigenvalue $\widetilde{\varepsilon}^2$ to satisfy conditions

$$b(n+1)\,\widetilde{\psi}_n^{s,+} - 2\mathrm{i}w\sqrt{b}\sqrt{n+1}\,\widetilde{\psi}_{n+1}^{s,-} = \widetilde{\varepsilon}^2\,\widetilde{\psi}_n^{s,+} \quad (32)$$

$$2\mathrm{i}w\sqrt{b}\sqrt{n}\,\widetilde{\psi}_{n-1}^{s,+} + bn\widetilde{\psi}_n^{s,-w} = \widetilde{\varepsilon}^2\,\widetilde{\psi}_n^{s,-} \quad (33)$$

where $n \geq 0$ and $s = +, -$. This leads to solutions of the form

$$\begin{pmatrix} 0 & |0\rangle & 0 & 0 \end{pmatrix}^\top \quad \text{and} \quad \begin{pmatrix} 0 & 0 & 0 & |0\rangle \end{pmatrix}^\top \quad (34)$$

for the lowest eigenvalue $\widetilde{\varepsilon}^2 = 0$, and to solutions

$$\frac{1}{\sqrt{2}}\begin{pmatrix} |n\rangle \\ \pm\mathrm{i}\,|n+1\rangle \\ 0 \\ 0 \end{pmatrix} \quad \text{and} \quad \frac{1}{\sqrt{2}}\begin{pmatrix} 0 \\ 0 \\ |n\rangle \\ \pm\mathrm{i}\,|n+1\rangle \end{pmatrix} \quad (35)$$

with eigenvalues $\widetilde{\varepsilon}^2 = b(n+1) \pm 2w\sqrt{b}\sqrt{n+1}$, where $n$ is any non-negative integer. It is now easy to check that suitable linear combinations of states in Eq. (34) form eigenstates of $\mathcal{H}$ with the corresponding Landau level energies

$$\varepsilon_{0,\pm} = \pm\sqrt{(\hbar v_z k_z)^2 + w^2} \quad (36)$$

while linear combinations of states in Eq. (35) gives the eigenstates of $\mathcal{H}$ with the Landau level energies

$$\varepsilon_{\pm n,\pm} = \pm\sqrt{(\hbar v_z k_z)^2 + \left(w \pm \sqrt{n}\sqrt{b}\right)^2} \quad (37)$$

where $n$ is any positive interger. Results of Eqs. (36)-(37) can be written using a single equation

$$\varepsilon_{n,\pm}(k_z) = \pm\left[(\hbar v_z k_z)^2 + \left(w \pm \mathrm{sign}\,n\sqrt{|n|\,b}\right)^2\right]^{\frac{1}{2}} \quad (38)$$

where $n \in \mathbb{Z}$. The resultant spectrum is gapped unless $w^2/b \in \mathbb{Z}$, i.e. for magnetic fields

$$B_c = \frac{w^2}{2e\hbar v_x v_y n}, \quad n \in \mathbb{Z}\backslash\{0\}. \quad (39)$$

As mentioned in the main text, these gap closings correspond to topological phase transitions, in which a charge $e/2$ is topologically pumped to the surface of the sample, orthogonal to the direction of the field. To see this, notice that the perturbation of Eq. (18) is proportional to the mirror operator $\sigma_z = \mathrm{i}\gamma_{34}$. As a consequence, tuning the magnitude of the perturbation $w$ shifts the energy of the eigenstates with different mirror eigenvalues in opposite directions, and thus a gap closure in the LL spectrum interchanges "valence" and "conduction" LLs with different mirror eigenvalues. Assuming, in the flavor of $\boldsymbol{k} \cdot \boldsymbol{p}$ theory, that the LLs are gapped at other momenta, this gap closing corresponds to a 1D topological phase transition, in which the Berry phase of the LL (considered as a band in the $k_z$ direction) changes by $\pi$. This change corresponds to pumping a charge of $e/2$ to the surface of the sample parallel to the plane of NSNL [4]. For this reason, a change in the Hall response of the metallic surface states is expected to occur at the critical values of the magnetic field, corresponding to the gap closure in the LL spectrum.

### 2. In-plane field

Now we consider the case of an in-plane magnetic field $\boldsymbol{B} = (B, 0, 0)$ with $B > 0$. The corresponding Peierls substitution is

$$(\hbar k_x, \hbar k_y, \hbar k_z) \mapsto (\hbar k_x, -\mathrm{i}\hbar\partial_y + eA_y, -\mathrm{i}\hbar\partial_z + eA_z)$$
$$\equiv (\hbar k_x, \Pi_y, \Pi_z) \quad (40)$$

with the commutation relationship

$$[\Pi_y, \Pi_z] = \mathrm{i}\hbar e\left(\partial_z A_y - \partial_y A_z\right) = -\mathrm{i}\hbar eB. \quad (41)$$

In this case we express the $\Pi$ operators as

$$\Pi_y = \sqrt{\frac{e\hbar v_z B}{2v_y}}\left(a + a^\dagger\right), \ \ \Pi_z = \mathrm{i}\sqrt{\frac{e\hbar v_y B}{2v_z}}\left(a - a^\dagger\right). \quad (42)$$

Substituting this into the Hamiltonian of Eq. (20) and using a set of Dirac matrices

$$\gamma_1 = \tau^y, \qquad \gamma_2 = \tau^z\sigma^x, \qquad \gamma_3 = \tau^z\sigma^y,$$
$$\gamma_4 = \tau^x \qquad \gamma_5 = \tau^z\sigma^z \quad (43)$$

we arrive at the following Hamiltonian matrix

$$\mathcal{H}(k_x) = \begin{pmatrix} 0 & \sqrt{b}a & -\mathrm{i}\hbar v_x k_x & -w \\ \sqrt{b}a^\dagger & 0 & w & -\mathrm{i}\hbar v_x k_x \\ \mathrm{i}\hbar v_x k_x & w & 0 & -\sqrt{b}a \\ -w & \mathrm{i}\hbar v_x k_x & -\sqrt{b}a^\dagger & 0 \end{pmatrix}. \quad (44)$$

Applying the parametrization of Eq. (30) and setting $k_x = 0$, we find recurrence equations for the eigenstates of $\mathcal{H}(k_x = 0)$ with *zero energy*

$$\psi_{n+1}^{s,+} = -\psi_n^{-s,+}\sqrt{n+1}\frac{\sqrt{b}}{w} \quad (45)$$

$$\psi_{n+1}^{s,-} = \psi_n^{-s,-}\frac{1}{\sqrt{n+1}}\frac{w}{\sqrt{b}}. \quad (46)$$

We can construct four linearly independent solutions for this system, starting with a single non-trivial component, $\psi_0^{s,r} = 1$. The solutions generated for $r = +$ are not normalizable because of the cumulative divergent factor $\sqrt{n!}$, and are therefore non-physical. The normalized solutions for $r = -$ are

$$\left|\psi^{+,-}\right> = \mathrm{e}^{-\frac{w^2}{2b}}\left(\begin{matrix} 0 & \cosh\left(\frac{wa^\dagger}{\sqrt{b}}\right)|0\rangle & 0 & \sinh\left(\frac{wa^\dagger}{\sqrt{b}}\right)|0\rangle \end{matrix}\right)^\top$$

$$\left|\psi^{-,-}\right> = \mathrm{e}^{-\frac{w^2}{2b}}\left(\begin{matrix} 0 & \sinh\left(\frac{wa^\dagger}{\sqrt{b}}\right)|0\rangle & 0 & \cosh\left(\frac{wa^\dagger}{\sqrt{b}}\right)|0\rangle \end{matrix}\right)^\top$$

where we used

$$\sum_{n=0}^{+\infty}\frac{1}{n!}\frac{w^{2n}}{b^n} = \mathrm{e}^{w^2/b} \quad (47)$$

to obtain the normalization factor.

For $k_x \neq 0$ the first order correction to zero energy can be obtained by diagonalizing the matrix

$$M_{rs} = \hbar v_x k_x \left<\psi^{r,-}\right|(\tau^y \otimes \mathbb{1}_\sigma)\left|\psi^{s,-}\right>. \quad (48)$$

Evaluating the matrix leads to

$$M = \hbar v_x k_x \mathrm{e}^{-\frac{2w^2}{b}}\begin{pmatrix} 0 & -\mathrm{i} \\ \mathrm{i} & 0 \end{pmatrix} \quad (49)$$

and the first order correction to the energy becomes

$$\Delta\varepsilon(k_x) = \pm\hbar v_x k_x \mathrm{e}^{-\frac{2w^2}{b}}. \quad (50)$$

We see that the dispersion of the zeroth LLs is exponentially suppressed for weak magnetic fields $b \lesssim w$. This is confirmed by the numerical analysis presented in Fig. 2d.

We now show that the crossing of the lowest LLs is protected by system symmetries. Notice that the composition of the time-reversal and the glide plane

$$\Xi(\boldsymbol{k}) = \Theta \circ g(\boldsymbol{k}) \quad (51)$$

remains a symmetry of the system even in the presence of an in-plane magnetic field. Using the Dirac basis of Eq. (22), the glide symmetry is represented by

$$g(\boldsymbol{k}) = \mathrm{e}^{\mathrm{i}\varphi_{\boldsymbol{k}}}\gamma_{34} = \mathrm{e}^{\mathrm{i}\varphi_{\boldsymbol{k}}}\tau^z \quad (52)$$

where $\varphi_{\boldsymbol{k}}$ is a $\boldsymbol{k}$-dependent phase factor due to the non-symmorphic shift. A composition with $\Theta = \mathrm{i}\sigma^y\mathcal{K}$ gives a manifestly antiunitary operator

$$\Xi(\boldsymbol{k}) = \mathrm{e}^{-\mathrm{i}\varphi_{\boldsymbol{k}}}\tau^z\sigma^y\mathcal{K} \quad (53)$$

which squares to

$$\Xi\left[(\Theta \circ g)\boldsymbol{k}\right]\Xi(\boldsymbol{k}) = \mathrm{e}^{-\mathrm{i}\varphi_{(\Theta \circ g)\boldsymbol{k}}}\tau^z\sigma^y\mathcal{K}\mathrm{i}\mathrm{e}^{-\mathrm{i}\varphi_{\boldsymbol{k}}}\tau^z\sigma^y\mathcal{K}$$
$$= -\exp\left[\mathrm{i}\left(\varphi_{\boldsymbol{k}} - \varphi_{(\Theta \circ g)\boldsymbol{k}}\right)\right]. \quad (54)$$

For an applied in-plane magnetic field, only the momentum parallel to the field is a good quantum number. If this component is set to zero, $k_x = 0$, then (54) evaluates to $-\mathbb{1}$, hence, by Kramer's theorem, the twofold degeneracy is enforced at this point. (As a consistency check, remember that the case of $w = 0$ corresponds to a Dirac semimetal, where the zero-energy crossing of the chiral LLs is well-known.)

Let us briefly compare the response of NSNLs discussed here to that of *accidental* nodal loops (ANLs). ANLs in a mirror-invariant plane of a time-reversal symmetric system inevitably appear in pairs centered around points $\pm\boldsymbol{k}_0$, and the symmetry (51) relates LLs of one ANL to the LLs of the other, thus avoiding the two-fold degeneracy at zero energy. As a consequence, ANLs *do not* produce chiral LLs and therefore should have transport properties qualitatively different from those of NSNLs.

## D. NODAL CHAIN MATERIAL CANDIDATES

In this section we present a detailed discussion of the methods used to identify real material candidates for the nodal chain phase, and present the identified materials. We used two complementary approaches: first principles calculations and tight-binding modeling. The latter was done using the Wannier-based tight-binding models with long range hoppings, derived directly from first principles calculations, and also by creating a model with short range hoppings based purely on symmetry analysis and fitting it's parameters to the *ab initio* band structure.

### 1. Crystal structure and symmetry

The material candidates for the nodal chain realization identified in this work are all from the class of Iridium tetrafluoride ($\mathrm{IrF}_4$). This parent compound crystallizes in space group #43 ($Fdd2$) with the conventional unit cell parameters [15]

$$a = 9.64(1)\,\text{Å}, \quad b = 9.25(1)\,\text{Å}, \quad c = 5.67(1)\,\text{Å}. \quad (55)$$

The primitive unit cell contains two structural units.

In this crystal structure, shown in Fig. 3a-b, every Ir site is surrounded by an octahedral cage of six F atoms, four of which are shared with the neighboring octahedra, while the remaining two belong solely to one iridium

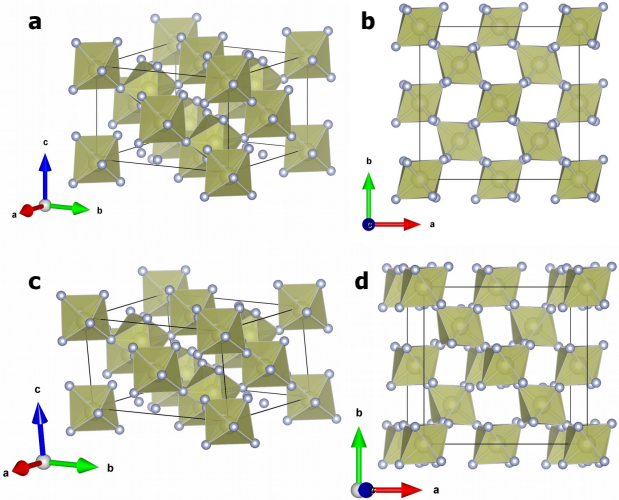

FIG. 3. **Crystal structure of unstrained and strained IrF$_4$.** (a) Crystal structure of unstrained IrF$_4$ in space group #43. (b) Same as (a) shown along the [001] direction. (c) Crystal structure of IrF$_4$ strained along the [011] direction. The symmetry is reduced to that of space group #9 with only one glide plane. The nodal chain reduces to NSNLs under this strain. (d) Same as (c) shown along the [001] direction. The crystal structures were plotted using VESTA 3 [14].

site. These octahedra form a bipartite lattice indicated in color in Fig. 3a-b of the main text. Appropriate straining of the crystal shown in Fig. 3c-d will transform the nodal chain into a single NSNL.

Other representatives of the IrF$_4$ material class were considered. The lattice constants of XY$_4$ (X=Ir, Ta, Re; Y=F, Cl, Br, I) were obtained by optimizing numerically both the lattice constants and the atomic positions of the compounds, starting from the IrF$_4$ structure. The resultant lattice parameters are listed in Table IV. Band structures of several such compounds are presented in Fig. 6. Nodal chains are clearly visible in all of these materials.

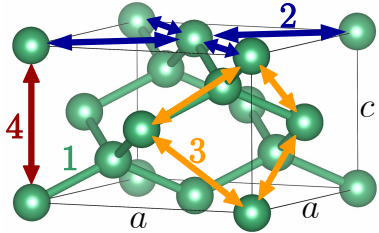

FIG. 4. **Construction of the tight-binding model.** The numbers indicate sites related by corresponding hoppings in Hamiltonian of Eq. (56).

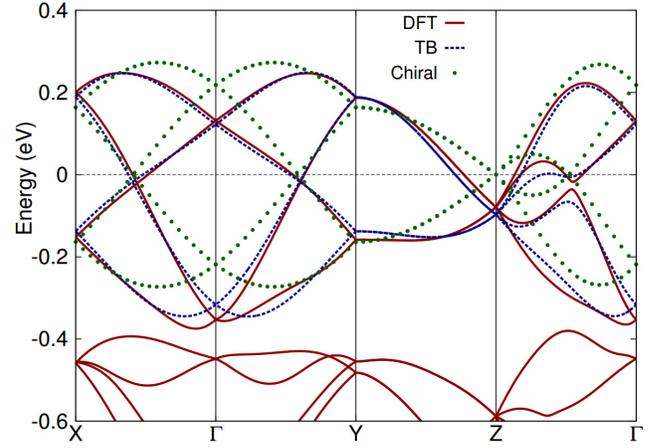

FIG. 5. **Band structure of IrF$_4$.** The solid red and dotted green lines correspond to the band structure calculated from first-principles and from the chiral symmetric model, respectively, and were already presented in Fig. 3f of the main text. The dashed blue line represents the tight-binding model constructed in subsection D 3.

### 2. First-principles calculations

The electronic structure calculations have been performed within the framework of density functional theory (DFT) [16], using generalized gradient approximation (GGA) [17] with the projector augmented-wave [18] implemented in VASP (Vienna ab-initio simulation program) [19, 20]. Spin-orbit coupling was included into consideration via the pseudopotentials. For all the materials presented the band structures were obtained using a 500 eV energy cutoff for the plane waves and the energy precision of $10^{-8}$ eV. The Brillouin zone was sampled using a $10 \times 10 \times 10$ Gamma-centered mesh. The lattice constants and atom positions listed in Table IV are fully relaxed untill the total energy is converged to $10^{-7}$ eV and the forces on atoms are below $10^{-3}$ eV/Å.

Wannier90 [21–23] code was used to construct a Wannier-based tight binding model. The $20\times20$ model was obtained by projecting the band structure around the Fermi level onto the five $d$-orbitals from all four Ir atoms in the unit cell. The surface band structures and densities of states were calculated using this tight-binding model and the iterative Green's function technique of Ref. [24], which was implemented in Wannier_tools package [25].

### 3. Tight-binding model

We now discuss a short-ranged symmetry-based tight-binding model of IrF$_4$. The DFT calculations predict the set of bands with $5d$-Ir character to be separated from the $5s$-Ir bands above and the $2p$-F bands below. The local octahedral crystal field splits the $5d$-Ir orbitals into a partially filled lower-lying $t_{2g}$ manifold and an empty higher energy $e_g$ states. The spin-orbit coupling further

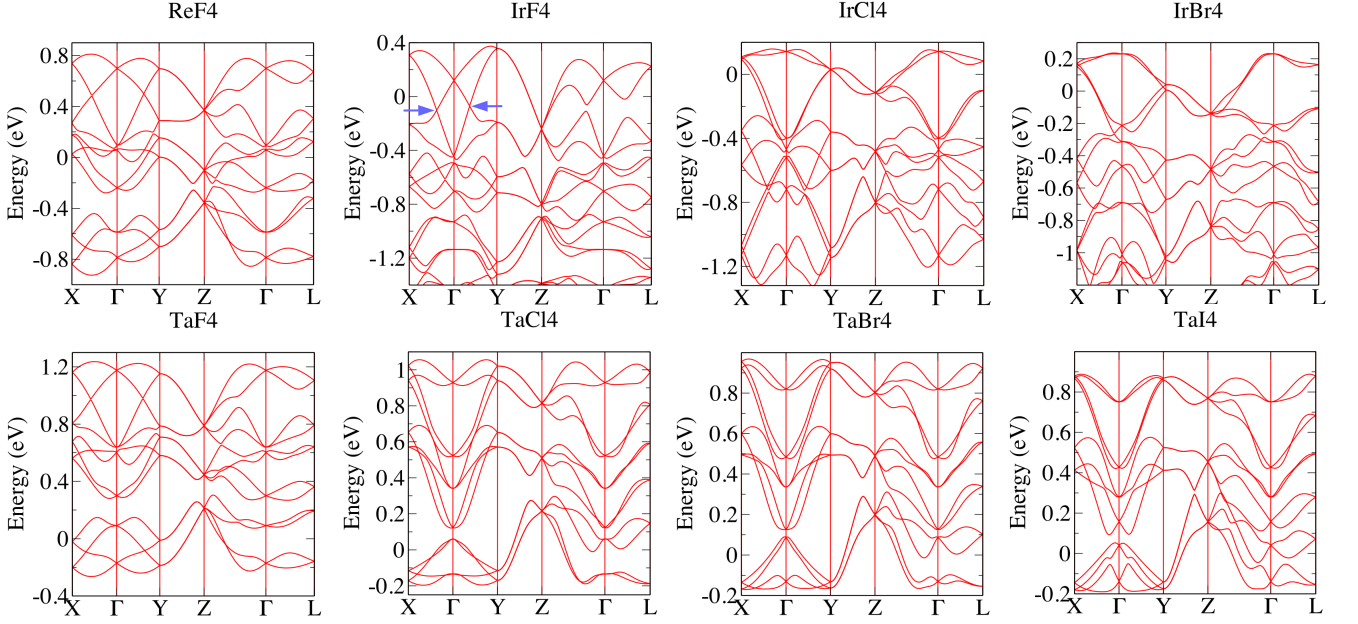

FIG. 6. **Band structure of IrF₄ material class.** The symmetry imposed nodal loops are visible in all materials. The blue arrows indicate the nodal loops of $IrF_4$ that are discussed in the main text.

splits the $t_{2g}$ orbitals into a filled $J_{\text{eff}} = \frac{3}{2}$ quadruplet, and a half-filled $J_{\text{eff}} = \frac{1}{2}$ doublet, which are well separated in energy. Since there are two Ir sites per primitive unit cell, it is possible to formulate an effective four-band tight-binding model. To simplify the analysis, we assume a perfect $C_{4z}$ screw rotational symmetry, which is only mildly broken in IrF₄. This increases the space group from the face-centred orthogonal #43 ($Fdd2$) to the body-centred tetragonal #109 ($I4_1md$).

We construct a tight-binding model compatible with the enlarged group that reproduces the central four bands of IrF₄ to better than 4% fraction of the overall bandwidth (determined as root-mean-square average of the difference between the original bands and the fit). It contains five parameters: $t_1, T_1, t_2, T_3, t_4$, where the small (capital) symbols stand for spin independent (dependent) hopping amplitudes, and the subscript denotes the relative position of the sites between which the hopping process takes place in accord with the caption of Fig 4.

The resultant tight-binding Hamiltonian can be written compactly, using a set of Pauli matrices $\boldsymbol{\sigma} = (\sigma^x, \sigma^y, \sigma^z)$ for the spin degrees of freedom, and another set of Pauli matrices $\boldsymbol{\tau} = (\tau^x, \tau^y, \tau^z)$ for the sublattice degrees of freedom. The electron annihilation operator at site $i$ with position vector $\boldsymbol{r}_i$ is given by $c_i = (c_{i\uparrow}, c_{i\downarrow})^{\top}$.

The real space Hamiltonian reads

$$
\begin{aligned}
\mathcal{H} = &\sum_{\{i,j\}\in 1} c_i^\dagger \left( t_1 \mathbb{1} + \mathrm{i} T_1 \lceil \boldsymbol{r}_{ij} \times \boldsymbol{e}_z \rceil \cdot \boldsymbol{\sigma} \right) c_j \\
&+ t_2 \sum_{\{i,j\}\in 2} c_i^\dagger c_j + t_4 \sum_{\{i,j\}\in 4} c_i^\dagger c_j \\
&+ \mathrm{i} T_3 \sum_{\{i,j\}\in 3} c_i^\dagger \lceil (\boldsymbol{e}_z \cdot \boldsymbol{r}_{ij})(\boldsymbol{e}_z \times \boldsymbol{d}_{ij}) \rceil \cdot \boldsymbol{\sigma} \, c_j
\end{aligned}
\tag{56}
$$

where $\boldsymbol{d}_{ij} = \boldsymbol{r}_{ik} + \boldsymbol{r}_{jk}$, $k$ is the common nearest neighbor of sites $\{i,j\} \in 3$, and $\lceil \boldsymbol{v} \rceil = \boldsymbol{v}/|\boldsymbol{v}|$ is the unit vector in a specified direction.

We remark that further hopping terms are possible and that symmetries allow for other spin-orbit terms [26], but we found the fitted amplitudes for such processes to be small compared to the terms already included in Eq. (56). The fitted values of the TB parameters and the corresponding value of the chemical potential in eV are

$$
\begin{aligned}
t_1 &= 0.0548, \quad T_1 = -0.0577, \quad t_2 = -0.0153 \\
T_3 &= 0.0071, \quad t_4 = -0.0068, \quad \mu = 0.0179.
\end{aligned}
\tag{57}
$$

Note that the hoppings between the two sublattices $(t_1, T_1)$ strongly dominates over intra-sublattice hoppings $(t_2, T_3, t_4)$. This indicates that the chiral symmetry of the system is only weakly broken, Indeed, neglecting the intra-sublattice hoppings leads to an ideal chiral symmetry of the model, while still producing reasonable band structure, shown in Fig. 5. The parameters of the chiral tight-binding model are

$$
\begin{aligned}
t_1 &= 0.0548, \quad T_1 = -0.0577, \quad t_2 = 0 \\
T_3 &= 0, \quad t_4 = 0, \quad \mu = 0.
\end{aligned}
\tag{58}
$$

| | | | |
|---|---|---|---|
| ReF$_4$ | 9.710 Å | 9.050 Å | 5.630 Å |
| Re | 0.0052 | 0.0052 | 0.9948 |
| F$_1$ | 0.9777 | 0.0242 | 0.2909 |
| F$_2$ | 0.4506 | 0.5023 | 0.8193 |
| IrF$_4$ | 9.764 Å | 9.479 Å | 5.778 Å |
| Ir | 0.0052 | 0.0052 | -0.0052 |
| F$_1$ | 0.9777 | 0.0242 | 0.2909 |
| F$_2$ | 0.4506 | 0.5023 | 0.8193 |
| IrCl$_4$ | 11.082 Å | 10.947 Å | 7.541 Å |
| Ir | 0.0153 | 0.0153 | -0.0153 |
| Cl$_1$ | -0.0039 | 0.0201 | 0.2961 |
| Cl$_2$ | 0.4547 | 0.4737 | 0.8212 |
| IrBr$_4$ | 11.772 Å | 11.506 Å | 8.021 Å |
| Ir | 0.0153 | 0.0153 | -0.0153 |
| Br$_1$ | -0.0125 | 0.0278 | 0.2981 |
| Br$_2$ | 0.4481 | 0.4813 | 0.8175 |
| TaF$_4$ | 10.134 Å | 9.887 Å | 5.861 Å |
| Ta | 0.0015 | 0.0015 | 0.9985 |
| F$_1$ | 0.0013 | 0.0142 | 0.2818 |
| F$_2$ | 0.4589 | 0.4841 | 0.8322 |
| TaCl$_4$ | 11.481 Å | 11.468 Å | 7.862 Å |
| Ta | 0.0203 | 0.0203 | -0.0203 |
| Cl$_1$ | 0.0140 | 0.0147 | 0.2900 |
| Cl$_2$ | 0.4550 | 0.4560 | 0.8292 |
| TaBr$_4$ | 12.153 Å | 12.102 Å | 8.425 Å |
| Ta | 0.0227 | 0.0227 | -0.0227 |
| Br$_1$ | 0.0119 | 0.0146 | 0.2929 |
| Br$_2$ | 0.4533 | 0.4575 | 0.8266 |
| TaI$_4$ | 13.149 Å | 13.088 Å | 9.046 Å |
| Ta | 0.0212 | 0.0212 | 0.9788 |
| I$_1$ | 0.0097 | 0.0129 | 0.2951 |
| I$_2$ | 0.4559 | 0.4603 | 0.8237 |

TABLE IV. **Crystal structure of IrF$_4$ material class XY$_4$ with space group #43.** The numbers next to the chemical formula indicate the conventional unit cell parameters $a, b, c$, respectively. Listed below are the coordinates of atoms. The Wyckoff positions are $8a$ for the X atom and $16a$ for both inequivalent Y atoms. In all cases, the crystals contain two formula units per primitive unit cell.

The chiral symmetry (CHS) operator is $\mathcal{C} = \tau^z$. Comparison of the DFT results with the two approximations is shown in Fig. 5.

## E. TOPOLOGICAL INVARIANT ON THE CURVED PLANE

In the main text we introduced a $\mathbb{Z}_2$ topological invariant of a curved plane, the projection of which onto the (100) surface is shown with a magenta line in Fig. 3f of the main text. The invariant was computed using the method of Refs. [27, 28] as implemented in the Z2Pack software package (https://z2pack.ethz.ch) [29]. Its value was found to be non-trivial, suggesting an odd number of Kramers pairs of edge states to cross the magenta line of Fig. 3f at some energies. The corresponding surface spectral function shown indeed exhibits a single Kramers pair of edge (surface) states, as shown in Fig. 7a.

The value of the $\mathbb{Z}_2$ invariant is also constrained to be non-trivial by crystalline symmetries. To see this, we apply an established technique of Refs. [27, 28] to compute the value of topological invariants by tracking the changes in Wannier charge center (WCC) positions (or, alternatively, a Wilson loop) on cuts of the Brillouin zone. In particular, we consider a cut $\mathcal{P} = (k_x, k_\ell)$, shown as a magenta plane in Fig. 7b. The cut is formed by $k_x$ and a curved line $k_\ell$, and it exhibits a gapped spectrum. The projection of this cut onto the (100) surface is shown with a magenta line in Fig. 3f of the main text. Importantly, this cut $\mathcal{P}$ of the Brillouin zone is symmetric under the glide $g_x$, and contains four inequivalent time-reversal invariant momenta (TRIMs), shown as dark points on $\mathcal{P}$.

We now construct the WCCs $\ell(k_x)$ by carrying out the non-Abelian parallel transport of Bloch states along $k_\ell$ at different values of $k_x$ (some of such parallel transport paths are shown as yellow dashed lines in Fig. 7b) (see Refs. [21, 27] for details of the procedure). The $k_\ell$ lines that connect TRIMs at $k_x = $ A or $k_x = $ C are mapped onto themselves by time-reversal symmetry, and hence the WCCs at these values of $k_x$ should appear in Kramers degenerate pairs [27, 30]. At values of $k_x$ other than A and C the degenerate centers split, as shown in Fig. 7c. Since lines located at B $\pm k_x$ are related by $g_x$ containing vertical shift by half-unit cell, the WCCs at B $\pm k_x$ are themselves displaced by half unit cell along $\ell$, as indeed can be seen in Fig. 7c. Due to this shift the connectivity of the WCCs between $k_x = $ A and $k_x = $ C is constrained in such a way that the $\mathbb{Z}_2$ invariant is trivial (non-trivial) if there are $4n$ ($4n+2$) occupied bands [31], meaning that the WCCs do not exchange (do exchange) partners when $k_x$ goes from A to C. This proves the existence of the topological phase for the filling $4n + 2$.

## F. CHIRAL SYMMETRIC MODEL

Consider the Hamiltonian of Eq. (56) with parameters of Eq. (58). As discussed in Sec. D 3, such a choice of parameters corresponds to admitting only inter-sublattice (NN) processes. Representing the spin degree of freedom by Pauli matrices $\boldsymbol{\sigma} = (\sigma^x, \sigma^y, \sigma^z)$ and the sublattice degree of freedom by $\boldsymbol{\tau} = (\tau^x, \tau^y, \tau^z)$, the chiral symmetry (CHS) operator $\mathcal{C} = \mathbb{1}_\sigma \otimes \tau^z$ fulfills

$$\mathcal{C}^2 = \mathbb{1} \qquad \text{and} \qquad \{\mathcal{C}, \mathcal{H}(\boldsymbol{k})\} = 0. \qquad (59)$$

The symmetry manifests itself in the particle-hole symmetric spectrum of $\mathcal{H}(\boldsymbol{k})$ in Fig. 3c of the main text.

In the sublattice basis, the chiral-symmetric Hamiltonian has the block-off-diagonal form

$$\mathcal{H}(\boldsymbol{k}) = \begin{pmatrix} \mathbb{0} & M(\boldsymbol{k}) \\ M^\dagger(\boldsymbol{k}) & \mathbb{0} \end{pmatrix}. \qquad (60)$$

At momenta, where a spectral gap exists between the conduction and valence bands, $\det M(\boldsymbol{k}) \neq 0$. We can

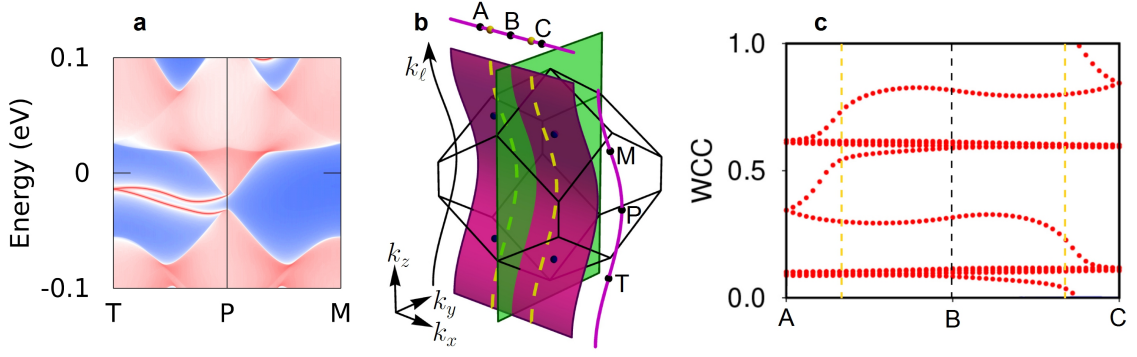

FIG. 7. **Determining the $\mathbb{Z}_2$ invariant on the curved plane.** (a) The surface spectral function along the magenta line in Fig. 3f of the main text reveals a non-trivial $\mathbb{Z}_2$ invariant. Points T, M correspond to time-reversal invariant momenta (TRIMs), and P corresponds to the point where the magenta line crosses the surface Brillouin zone boundary. (b) The curved cut $\mathcal{P}$ of the Brillouin zone is shown. The projection of $\mathcal{P}$ on the right (with points T,P,M) corresponds to the magenta line in Fig. 3f, while the projection on the top (with points A,B,C) does not correspond to any physical surface. The dashed yellow lines indicate two paths in momentum space related by glide reflection $g_x$. The dark blue points on the sheet correspond to bulk TRIMs. (c) Wannier charge centers (WCC) along vertical threads of the magenta plane $\bar{\ell}(k_x)$ computed from *ab initio* for IrF$_4$. The vertical shift associated with $g_x$ leads to glide symmetry of the WCC plot with respect to the central dashed line. This enforces the $\mathbb{Z}_2$ invariant to be non-trivial, since there are $4n + 2$ occupied bands.

therefore define a winding number $z(c)$ for any closed path $c$, along which the spectrum is gapped

$$z(c) = \frac{1}{2\pi} \operatorname{Im} \oint_c d\boldsymbol{k} \cdot \boldsymbol{\nabla}_{\boldsymbol{k}} \left[ \log \det M(\boldsymbol{k}) \right] \qquad (61)$$

which presents a $\mathbb{Z}$-valued topological charge [32].

For a mirror-symmetric loop $c$, the invariants of Eqs. (21) and (61) are related by [4, 33, 34]

$$z(c) = z_2(c) \mod 2. \qquad (62)$$

Any two paths $c_{a,b}$ with $z(c_{a,b}) = \pm 1$ are topologically *distinct* in the chiral case, but the distinction is lost once the CHS is broken. Thus, if one-dimensional lines perpendicular to the mirror plane crossing the three-dimensional Brillouin zone of a chiral-symmetric system have distinct chiral charges, they need to be separated by a gapless line in the momentum space. If the chiral charge of the lines is different by 2, this line becomes gapped upon breaking the chiral symmetry (but preserving mirror/glide).

Furthermore, we show in the present section that in a crystal with a symmetry $\mathcal{S}$, the winding numbers $z(c)$ and $z(\mathcal{S}c)$ are related by

$$z(\mathcal{S}c) = \zeta z(c) + \frac{N_{\text{occ.}}}{2\pi} \oint_c d\boldsymbol{k} \cdot \boldsymbol{\nabla}_{\boldsymbol{k}} \left[ \arg(v_{\boldsymbol{k}}) \right] \qquad (63)$$

where $N_{\text{occ.}}$ is the dimension of $M(\boldsymbol{k})$ and the appropriate form of $v_{\boldsymbol{k}}$ and $\zeta$ is listed in Table V. The integral in (63) vanishes if $c$ is contractible (i.e. if it does not wind across the BZ). Relations (62) and (63) provide handy tools to explain the appearance of the nodal net in Fig. 4c of the main text.

To outline the derivation of (63), we discuss the case of a *unitary* symmetry $\mathcal{S}$ that *exchanges* the two crystal sublattices, corresponding to the first row of Table V. Such a symmetry is off-diagonal in the sublattice basis, i.e.

$$\mathcal{S}_1(\boldsymbol{k}) = U(\boldsymbol{k}) \otimes (a_{\boldsymbol{k}}\tau^x + b_{\boldsymbol{k}}\tau^y) \qquad (64)$$

The $\boldsymbol{k}$-dependence of $U(\boldsymbol{k}) \in \mathrm{U}(N_{\text{occ.}})$ and $a_{\boldsymbol{k}}, b_{\boldsymbol{k}} \in \mathbb{C}$ is assumed to be continuous. The unitarity of (64) implies $|a_{\boldsymbol{k}}|^2 + |b_{\boldsymbol{k}}|^2 = 1$ and $a_{\boldsymbol{k}}b_{\boldsymbol{k}}^* - b_{\boldsymbol{k}}a_{\boldsymbol{k}}^* = 0$. Operator (64) alters the upper right block of (60) as

$$M(\mathcal{S}\boldsymbol{k}) = v_{\boldsymbol{k}}U(\boldsymbol{k})M^\dagger(\boldsymbol{k})U^\dagger(\boldsymbol{k}). \qquad (65)$$

where $v_{\boldsymbol{k}} = (a_{\boldsymbol{k}}^* - \mathrm{i}b_{\boldsymbol{k}}^*)(a_{\boldsymbol{k}} - \mathrm{i}b_{\boldsymbol{k}}) \in \mathrm{U}(1)$. Consequently

$$\det M(\mathcal{S}_1\boldsymbol{k}) = (v_{\boldsymbol{k}})^{N_{\text{occ.}}} \left[ \det M(\boldsymbol{k}) \right]^*. \qquad (66)$$

from which we conclude the relation (63) with $\zeta$ and $v_{\boldsymbol{k}}$ given in the first row of Table V. We can similarly find $\zeta$ and $v_{\boldsymbol{k}}$ for symmetries that *do not* exchange the two sublattices and for *antiunitary symmetries* that (anti)commute with $\mathcal{C}$.

The presence of a nodal chain in Fig. 1c of the main text is compatible with relations of Eqs. (62) and (63). To see this, consider two loops $c_{1,2}$ in Fig. 8a related by $c_2 = -C_{2z}c_1 = g_x c_1$. The relation of Eq. (63) implies $z(c_2) = -z(c_1)$. Loops $c_{1,2}$ enclose a NL so that $z_2(c_2) = z_2(c_1) = 1$. The last two relations imply

$$|z(c_2) - z(c_1)| = 2. \qquad (67)$$

This means that for the CHS model deforming $c_1$ onto $c_2$ must be associated with encountering at least two NLs. This is indeed true – the red NL in Fig. 8a needs to be intersected twice. We can analogously consider loops $c_{3,4}$ in Fig. 8a related by $c_4 = -C_{2z}c_3 = g_y c_3$, and infer

$$|z(c_4) - z(c_3)| = 2. \qquad (68)$$

| sublattice | anti-/unitary | $\mathcal{S}(\boldsymbol{k})$ | | $\zeta$ | $v_{\boldsymbol{k}}$ | examples in IrF$_4$ |
|---|---|---|---|---|---|---|
| $\{\mathcal{S},\mathcal{C}\}=0$ | $\mathcal{S}\mathcal{H}(\boldsymbol{k})\mathcal{S}^{\dagger}=\mathcal{H}(\mathcal{S}\boldsymbol{k})$ | $U(\boldsymbol{k})\otimes(a_{\boldsymbol{k}}\tau^{x}+b_{\boldsymbol{k}}\tau^{y})$ | | $-$ | $(a_{\boldsymbol{k}}^{*}-\mathrm{i}b_{\boldsymbol{k}}^{*})(a_{\boldsymbol{k}}-\mathrm{i}b_{\boldsymbol{k}})$ | $g_{x},g_{y}$ |
| $[\mathcal{S},\mathcal{C}]=0$ | $\mathcal{S}\mathcal{H}(\boldsymbol{k})\mathcal{S}^{\dagger}=\mathcal{H}(\mathcal{S}\boldsymbol{k})$ | $U(\boldsymbol{k})\otimes(a_{\boldsymbol{k}}\mathbb{1}_{\tau}+b_{\boldsymbol{k}}\tau^{z})$ | | $+$ | $(a_{\boldsymbol{k}}^{*}-b_{\boldsymbol{k}}^{*})(a_{\boldsymbol{k}}+b_{\boldsymbol{k}})$ | $C_{2z}$ |
| $\{\mathcal{S},\mathcal{C}\}=0$ | $\mathcal{S}\mathcal{H}^{*}(\boldsymbol{k})\mathcal{S}^{\dagger}=\mathcal{H}(\mathcal{S}\boldsymbol{k})$ | $U(\boldsymbol{k})\otimes(a_{\boldsymbol{k}}\mathbb{1}_{\tau}+b_{\boldsymbol{k}}\tau^{z})\mathcal{K}$ | | $+$ | $(a_{\boldsymbol{k}}^{*}-\mathrm{i}b_{\boldsymbol{k}}^{*})(a_{\boldsymbol{k}}-\mathrm{i}b_{\boldsymbol{k}})$ | $\Theta\circ g_{x},\Theta\circ g_{y}$ |
| $[\mathcal{S},\mathcal{C}]=0$ | $\mathcal{S}\mathcal{H}^{*}(\boldsymbol{k})\mathcal{S}^{\dagger}=\mathcal{H}(\mathcal{S}\boldsymbol{k})$ | $U(\boldsymbol{k})\otimes(a_{\boldsymbol{k}}\mathbb{1}_{\tau}+b_{\boldsymbol{k}}\tau^{z})\mathcal{K}$ | | $-$ | $(a_{\boldsymbol{k}}^{*}-b_{\boldsymbol{k}}^{*})(a_{\boldsymbol{k}}+b_{\boldsymbol{k}})$ | $\Theta,\Theta\circ C_{2z}$ |

TABLE V. **Transformation of chiral winding number under crystal symmetries.** The four rows characterize the possible crystal symmetries $\mathcal{S}$ as being anti-/unitary and anti-/commuting with the chiral symmetry $\mathcal{C}$. The third column indicates the general form of operator $\mathcal{S}(\boldsymbol{k})$ where $\mathcal{K}$ is the complex conjugation. The columns $\zeta$ and $v_{\boldsymbol{k}}$ indicate the proper parameters in relation (63). The last column gives examples for all types of symmetries in the material class studied in section D.

Therefore, deforming $c_3$ onto $c_4$ is associated with encountering at least two Weyl lines in the CHS case. This time it is the blue NL in Fig. 8a that needs to be intersected twice to connect the two loops.

To explain the apperance of the additional NL imposed by CHS, consider a pair of horizontal loops $c_{5,6}$ in Fig. 8a related by $c_6 = -(g_x \circ \Theta)\,c_5$. The relation of Eq. (63) implies that $z(c_6) = -z(c_5)$, and we obtain

$$|z(c_6) - z(c_5)| = 2. \tag{69}$$

Hence, a deformation of $c_5$ into $c_6$ is accompanied by crossing two nodal lines. In this case, the presence of the nodal chain itself is not sufficient to satisfy this constraint. Thus, CHS imposes an *additional* NL located in the $k_z = 0$ plane and separating $c_{5,6}$. We indicate such possible NLs by dashed green lines in Fig. 8a.

A suitable choice of considered paths can explain the actual *topology* of the additional NL. First, note that Eq. (69) implies that the additional NL *cannot terminate* at the red NL, because that would correspond to just a *single* gap closing between $c_5$ and $c_6$, so that it would not be possible to change the topological invariant by 2. Instead, the NL necessarily has to *touch* the red NL, having a ring-like shape. This leaves us with only two possibilities indicated by the dashed green and magenta lines in Fig 8b. We now show that it is the first option has to be realized.

Consider a path $c_7 = C_{2z}c_5$, such that $z(c_7) = z(c_5)$. The paths $c_{5,7}$ can be *merged* into $c_8$ with $|z(c_8)| = 2$. If we finally consider a path $c_9 = -(g_x \circ \Theta)\,c_8$, we conclude using Eq. (63) that

$$|z(c_9) - z(c_8)| = 4. \tag{70}$$

Therefore, deforming $c_8$ onto $c_9$ is accompanied by crossing of *four* NLs. This implies that the topology of the nodal net *cannot* take the form (2) of Fig. 8b. More careful argumentation determines the topology of the nodal chain uniquely to that of Fig. 4c of the main text. We expect analogous findings for all eight space groups listed in Fig. 1 of the main text.

Let us finally tackle the distinction of the regions designated as "0" and "2" in Fig. 3h of the main text. The labels correspond to the winding numbers of straight paths traversing the Brillouin zone in the [100] direction. The two regions are related by $C_{2z}$. Since these loops are *non*-contractible, we have to take into account the integral term in Eq. (63). We adopt the tight-binding convention with phase factors $\mathrm{e}^{\mathrm{i}\boldsymbol{k}\cdot\boldsymbol{R}}$ where $\boldsymbol{R}$ are the Bravais vectors. With such a choice, $\mathcal{H}(\boldsymbol{k}) = \mathcal{H}(\boldsymbol{k} + \boldsymbol{G})$. The representation of $C_{2z}$ in this convention is

$$C_{2z}(\boldsymbol{k}) = -\mathrm{i}\sigma_z \otimes \begin{pmatrix} \mathrm{e}^{-\mathrm{i}(k_z - k_y)/2} & 0 \\ 0 & \mathrm{e}^{-\mathrm{i}(k_z + k_x)/2} \end{pmatrix}. \tag{71}$$

and the phase $v_{\boldsymbol{k}}$ in Eq. (63) becomes

$$v_{\boldsymbol{k}} = (a_{\boldsymbol{k}}^{*} - b_{\boldsymbol{k}}^{*})(a_{\boldsymbol{k}} + b_{\boldsymbol{k}}) = \mathrm{e}^{\mathrm{i}(k_x + k_y)/2} \tag{72}$$

which winds once for any one-dimensional path threading the Brillouin zone in the [100] direction. Since in our case $N_{\mathrm{occ.}} = 2$, we obtain from Eq. (63) that

$$c(k_y, k_z) = -c(-k_y, k_z) + 2 \tag{73}$$

compatible with the region labels in Fig. 3h of the main text.

Finally, note that the appearance of an additional nodal line in the $k_z = 0$ plane is compatible with the symmetries of the system in the CHS case. The little group of any point in the $k_z = 0$ plane contains at least two symmetry operations: the CHS and the product symmetry $C_{2z} \times \mathcal{T}$, where $\mathcal{T}$ is the time-reversal operator. Due to the presence of these two symmetries, the existence of double degeneracies in the $k_z = 0$ plane corresponds to vanishing of a general $2 \times 2$ Hamiltonian, subjected to the two symmetry constraints. As a result, one has a single equation to satisfy by tuning two parameters (in-plane momenta), and hence a general solution admits the appearance of nodal lines in the $k_z = 0$ plane.

### G. MINIMAL $\boldsymbol{k} \cdot \boldsymbol{p}$ MODELS

A minimal model Hamiltonian of a NSNL is given by Eq. (20) of Section B. That Hamiltonian, however, only includes $\boldsymbol{k}$-independent perturbations to a $\boldsymbol{k}$-linear Dirac Hamiltonian. A more complete $\boldsymbol{k} \cdot \boldsymbol{p}$ expansion should also include all possible perturbations linear in $\boldsymbol{k}$. The full list of such perturbations is given by the following 11 terms that can be included into the Hamiltonian of Eq. (20) of Section B:

$$\begin{aligned} &k_x\gamma_5, k_x\gamma_{12}, k_x\gamma_{15}, k_x\gamma_{25}, \\ &k_y\gamma_5, k_y\gamma_{12}, k_y\gamma_{15}, k_y\gamma_{25}, \\ &k_z\gamma_{23}, k_z\gamma_{13}, k_z\gamma_{35}. \end{aligned} \tag{74}$$

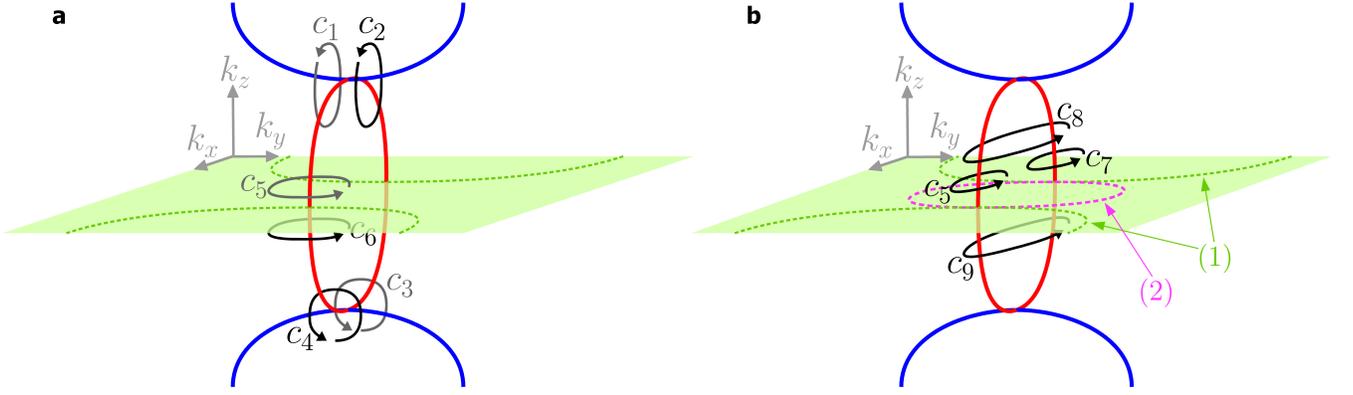

FIG. 8. **Paths considered in the discussion of the chiral symmetric model.** (a) The difference of $z(c_1)$ and $z(c_2)$ can be explained by the red NL separating them. Similarly, the blue NL explains the difference between $z(c_3)$ and $z(C_4)$. However, the nodal chain alone does not explain the difference between $z(c_5)$ and $z(c_6)$, and an additional NL in the $k_z = 0$ plane is imposed. (b) The invariants for paths $c_{8,9}$ are related by Eq. (70), which is consistent with the green NL (1), but not with the magenta NL (2). This uniquely fixes the topology of the nodal net.

A minimal Hamiltonian of a nodal chain is somewhat more complicated. This is because such a model cannot take the form of a $\boldsymbol{k} \cdot \boldsymbol{p}$ expansion, since finite order polynomials can not reproduce the periodic structure of the chain. Instead, the minimal model of a nodal chain can be obtained by combining a polynomial expansion in the momenta perpendicular to the axis of the chain with the fully periodic function of momentum parallel to the chain.

To derive such a nodal chain Hamiltonian, we start again with the model of Eq. (20) developed in Section B that contains a glide $g_z$ commuting with $\gamma_1, \gamma_2, \gamma_5$ and anticommuting with $\gamma_3, \gamma_4$ matrices. To obtain a nodal chain running in the $k_y$ direction, we have added a glide $g_x$ that commutes with $\gamma_2, \gamma_3, \gamma_4$ and anticommutes with $\gamma_1, \gamma_5$ matrices. Assuming the simplest possible $k_y$-dependence in the form of cosine and sine functions, the complete $(k_x, k_z)$-linear model of a nodal chain is given by combining the following 11 terms

$$
\begin{gathered}
k_x \gamma_1, \sin k_y \gamma_2, k_z \gamma_3, \\
\cos \tfrac{k_y}{2} \gamma_{34}, \sin \tfrac{k_y}{2} \gamma_{15}, k_x \gamma_{12}, k_z \gamma_{23}, \\
k_x \cos \tfrac{k_y}{2} \gamma_5, k_z \sin \tfrac{k_y}{2} \gamma_4, k_x \cos \tfrac{k_y}{2} \gamma_{25}, k_z \sin \tfrac{k_y}{2} \gamma_{24}.
\end{gathered}
\tag{75}
$$

Assuming further that the two NSNLs making up the nodal chain are related by symmetry (as is the case in space groups #109 and #122) the number of independent parameters in the minimal model can be reduced to six. In this case the Hamiltonian of a nodal chain is

$$
\begin{aligned}
\mathcal{H}_{\mathrm{NC}}^{\mathrm{sym.}}(\boldsymbol{k}) = {} & \hbar v_\perp \left(k_x \gamma_1 + k_z \gamma_3\right) + v_\parallel \sin k_y \gamma_2 \\
& + w \left(\cos \tfrac{k_y}{2} \gamma_{34} - \sin \tfrac{k_y}{2} \gamma_{15}\right) \\
& + a \left(k_x \gamma_{12} + k_z \gamma_{23}\right) \\
& + b \left(k_x \cos \tfrac{k_y}{2} \gamma_5 + k_z \sin \tfrac{k_y}{2} \gamma_4\right) \\
& + c \left(k_x \cos \tfrac{k_y}{2} \gamma_{25} - k_z \sin \tfrac{k_y}{2} \gamma_{24}\right)
\end{aligned}
\tag{76}
$$

where an additional condition $|w/v_\parallel| > \sqrt{2}$ needs to be imposed to guarantee a band ordering compatible with a nodal chain in the center of the band quadruplet. Note that in both Eqs. (75) and (76), we omitted possible terms proportional to the identity matrix.

Alternatively, one can construct a true $\boldsymbol{k} \cdot \boldsymbol{p}$ expansion around the touching point (TP) of the two NSNLs in the nodal chain. The model captures only the two bands that form the nodal chain. Up to second order in momentum the corresponding Hamiltonian is

$$
\begin{aligned}
\mathcal{H}_{\mathrm{TP}}(\boldsymbol{k}) = {} & \left(c_0 k_y + c_1 k_x^2 + c_2 k_y^2 + c_3 k_z^2\right) \sigma_z \\
& + k_x k_z \left(c_4 \sigma_x + c_5 \sigma_y\right)
\end{aligned}
\tag{77}
$$

where a representation, in which both glide operators $g_x$ and $g_z$ are proportional to $\sigma_z$ is chosen. Such a model has lost the periodicity in $k_y$, and depending on the choice of the parameters $c_i$ it may replace one or both closed nodal loops by open nodal lines of parabolic/hyperbolic shape. Note that we also omitted terms proportional to the identity matrix: $k_y \mathbb{1}, k_x^2 \mathbb{1}, k_y^2 \mathbb{1}, k_z^2 \mathbb{1}$.

[1] S. M. Young and C. L. Kane, Phys. Rev. Lett. **115**, 126803 (2015).

[2] H. Weng, Y. L. Liang, Q. Xu, R. Yu, Z. Fang, X. Dai, and Y. Kawazoe, Phys. Rev. B **92**, 045108 (2015).

[3] A. A. Burkov, M. D. Hook, and L. Balents, Phys. Rev. B **84**, 235126 (2012).

[4] R. D. King-Smith and D. Vanderbilt, Phys. Rev. B **47**, 1651(R) (1993).

[5] T. L. Hughes, E. Prodan, and B. A. Bernevig, Phys. Rev. B **83**, 245132 (2011).

[6] H. C. Po, H. Watanabe, M. P. Zaletel, and A. Vishwanath, Sci. Adv. **2**, e1501782 (2016).

[7] M. I. Aroyo, J. M. Perez-Mato, C. Capillas, E. Kroumova, S. Ivantchev, G. Madariaga, A. Kirov, and H. Wondratschek, Z. Krist **221**, 15 (2006).

[8] C. J. Bradley and A. P. Cracknell, *The Mathematical Theory of Symmetry in Solids* (Clarendon Press, Oxford, 1972).

[9] J. Ruan, S.-K. Jian, D. Zhang, H. Yao, H. Zhang, S.-C. Zhang, and D. Xingd, ArXiv e-prints (2016), arXiv:1603.01279.

[10] B.-J. Yang and N. Nagaosa, Nat. Commun. **5**, 4898 (2014).

[11] S. M. Young, S. Zaheer, J. C. Y. Teo, C. L. Kane, E. J. Mele, and A. M. Rappe, Phys. Rev. Lett. **108**, 140405 (2012).

[12] J. A. Steinberg, S. M. Young, S. Zaheer, C. L. Kane, E. J. Mele, and A. M. Rappe, Phys. Rev. Lett. **112**, 036403 (2014).

[13] T. Bzdušek, A. Rüegg, and M. Sigrist, Phys. Rev. B **91**, 165105 (2015).

[14] K. Momma and F. Izumi, J. Appl. Cryst. **44**, 1272 (2011).

[15] P. R. Rao, A. Tressaud, and N. Bartlett, Inorg. Nucl. Chem. **28**, 23 (1976).

[16] W. Kohn and L. J. Sham, Phys. Rev. **140**, A1133 (1965).

[17] J. P. Perdew, K. Burke, and M. Ernzerhof, Phys. Rev. Lett. **77**, 3865 (1996).

[18] P. E. Blöchl, Phys. Rev. B **50**, 17953 (1994).

[19] G. Kresse and J. Furthmüller, Phys. Rev. B **54**, 11169 (1996).

[20] G. Kresse and D. Joubert, Phys. Rev. B **59**, 1758 (1999).

[21] N. Marzari and D. Vanderbilt, Phys. Rev. B **56**, 12847 (1997).

[22] I. Souza, N. Marzari, and D. Vanderbilt, Phys. Rev. B **65**, 035109 (2001).

[23] A. A. Mostofi, J. R. Yates, Y.-S. Lee, I. Souza, D. Vanderbilt, and N. Marzari, Computer Physics Communications **178**, 685 (2008).

[24] M. P. L. Sancho, J. M. L. Sancho, J. M. L. Sancho, and J. Rubio, Journal of Physics F: Metal Physics **15**, 851 (1985).

[25] Q. S. Wu, https://github.com/quanshengwu/wannier_tools (2015).

[26] T. Moriya, Phys. Rev. **120**, 91 (1960).

[27] A. A. Soluyanov and D. Vanderbilt, Phys. Rev. B **83**, 235401 (2011).

[28] R. Yu, X. L. Qi, A. Bernevig, Z. Fang, and X. Dai, Phys. Rev. B **84**, 075119 (2011).

[29] D. Gresch, G. Autés, O. Yazyev, M. Troyer, D. Vanderbilt, B. A. Bernevig, and A. A. Soluyanov, in preparation (2016).

[30] L. Fu and C. L. Kane, Phys. Rev. B **74**, 195312 (2006).

[31] A. Alexandradinata, Z. Wang, and B. A. Bernevig, Phys. Rev. X **6**, 021008 (2016).

[32] A. P. Schnyder, S. Ryu, A. Furusaki, and A. W. W. Ludwig, Phys. Rev. B **78**, 195125 (2008).

[33] J. Zak, Phys. Rev. Lett. **62**, 2747 (1989).

[34] S. Ryu, A. P. Schnyder, A. Furusaki, and A. W. W. Ludwig, New J. Phys. **12**, 065010 (2010).



\end{document}